\documentclass[12pt]{article}
\usepackage{latexsym,amssymb,amsfonts,amsmath,graphicx}
\usepackage{amsmath}
\usepackage{amsfonts}
\usepackage{amssymb}
\usepackage{graphicx}
\usepackage{color}
\newcommand{\oo}{\overline}

\newcommand{\ba}{\begin{array}}
\newcommand{\ea}{\end{array}}
\newcommand{\be}{\begin{equation}}
\newcommand{\ee}{\end{equation}}
\newcommand{\bea}{\begin{eqnarray}}
\newcommand{\eea}{\end{eqnarray}}
\newcommand{\beaa}{\begin{eqnarray*}}
\newcommand{\eeaa}{\end{eqnarray*}}

\newcommand{\basa}{\begin{assumption}}
\newcommand{\easa}{\end{assumption}}

\newcommand{\bas}{\begin{assum}}
\newcommand{\eas}{\end{assum}}

\begin{document}
\title{Boolean delay equations on networks: An application to 
economic damage propagation}
\author{Barbara Coluzzi$^a$, Michael Ghil$^{a,b,c}$, 
St\'ephane Hallegatte$^{d,e}$, \\and G\'erard Weisbuch$^{f}$}

\maketitle

\begin{center}
a) {\em Environmental Research and Teaching Institute,
Ecole Normale Sup\'{e}rieure}, 24, rue Lhomond, 
F-75231 Paris Cedex 05, France.\\

b) {\em Geosciences Departement and
Laboratoire de M\'{e}t\'{e}orologie Dynamique
(CNRS and IPSL), \'Ecole Normale Sup\'{e}rieure},
24, rue Lhomond, F-75231 Paris Cedex 05, France.\\

c) {\em Department of Atmospheric and Oceanic Sciences
and Institute of Geophysics and Planetary Physics,
University of California}, Los Angeles, CA 90095-1565, USA.\\

d) {\em Centre International de Recherche sur l'Environnement et le 
D\'eveloppement}, 4bis avenue de ls Belle-Gabrielle, 94736
Nogent-sur-Marne Cedex, France.\\

e) {\em \'Ecole Nationale de la M\'et\'eorologie}, M\'et\'eo France, France.\\

f) {\em Laboratoire de Physique Statistique}, \'Ecole Normale
Sup\'erieure, 24, rue Lhomond, F-75231 Paris Cedex 05, France.\\

\end{center}

\begin{abstract}

We introduce economic models based on Boolean Delay Equations:
this formalism makes easier to take into account the complexity of the 
interactions between firms and is particularly appropriate for studying 
the propagation of an initial damage due to a catastrophe. 
Here we concentrate on simple cases, which allow to understand 
the effects of multiple concurrent production paths as well as the 
presence of stochasticity in the path time lengths or in the network 
structure. 

In absence of flexibility, the shortening of production of a single
firm in an isolated network with multiple connections usually ends up by 
attaining a finite fraction of the firms or the whole economy, whereas the 
interactions with the outside allow a partial recovering of the activity, 
giving rise to periodic solutions with waves of damage which propagate across 
the structure. The damage propagation speed is strongly dependent 
upon the topology. The existence of multiple concurrent production paths does 
not necessarily imply a slowing down of the propagation, which can be as 
fast as the shortest path.
\end{abstract}

\newpage

\section{Introduction}

In most economic models, the production system is modeled as a unique
representative producer (e.g., with the Cobb-Douglas function of the
Solow model) or as a set of sectors, in which there is a unique representative
producer by sector (e.g., in General Equilibrium Models). But the real
production system can best be seen as a network composed of firms, producing
different goods and services, and connected by suppliers and customer links. 
In such a network, the firm $j$ supplies a fraction of its production to
other firms, that use this production as an input for their own production
function (see [Fig. 1], for instance). 

Introducing the role of networks in the economic system can lead to complex 
endogenous economic dynamics \cite{Heletal04}. But the network formalism is 
also well adapted to study the cascade effects generated by exogenous events, 
positive in the case of new orders from the market 
\cite{RoLe81,RoLe86,RoLe89,LeHaSa07,Baketal93,OkHeSo04,BaSa09} 
and negative in the case of financial crisis \cite{DGaetal04}, random local 
strikes affecting production \cite{SoFuAo01,WeBa07}, or natural disasters
\cite{He07,Ha08,HeHa08}. In classical economic models, where the production 
system is represented as a unique representative producer or as a small
set of representative producers, indeed, exogenous events effects on the
numerous firms have to be averaged over all firms, or over all firms
of each sector. Because such effects are often highly heterogeneous, and
because consequences and responses are highly nonlinear, this averaging
process can bias the analysis, and makes it impossible to assess the 
consequences of an exogenous shock.
 
The case of natural disasters is particularly interesting, because disaster
impacts are very heterogeneous and affect especially strongly a small set
of firms \cite{Ti97,WeTiDa02}. In such cases, the total economic impact of the
catastrophe can be much higher than the direct impacts of the event, because
indirect effects through supply chains can be large. For instance,
an earthquake that destroys a bridge can cause losses that are much larger
than the value of the bridge, because impacts on transportation costs and
duration can impair production in many firms.

Such results have been reported using input-output models at the sector level
\cite{LiSaHa07,Ok04,RoLi05,Ha08}, or specific network models \cite{Choetal01}.
To account for heterogeneity at the firm level, an appropriate input-output 
formalism has been developed in \cite{He07,Ha08,HeHa08}, and used to analyze
disaster consequences. This approach has demonstrated that the shape and the
structure of the network play an important role in disaster vulnerability, 
justifying the introduction of network effects in the economic assessment of 
natural disasters. The purpose of the present manuscript is to simplify as
much as possible the formalism, within the framework of Boolean Delay
Equations (BDEs), to be able to go beyond the simple observation that the
network structure has an influence on disaster consequences, and to analyze
this dependency.

BDEs are semi-discrete dynamical systems, whose discrete variables 
evolve in continuous time; they have been introduced about 25 years ago 
\cite{DeeGhil84,Mull84,GhilMull85}, and they are related to the 
kinetic logic of Thomas \cite{Th79}: nevertheless, in BDEs, the memory of the 
system can contain more and more information as the time goes on, allowing
for solutions of increasing complexity, which hence display deterministic 
chaotic 
behavior, as it has been recently observed experimentally \cite{Zhetal09}. 
Apart from their intriguing mathematical properties, BDEs represent an useful 
tool in the modeling of complex systems, characterized by threshold 
behavior, multiple feedbacks, and distinct time delays. They have been in 
particular successfully applied to the study of climate dynamics \cite{SG01}, 
of earthquake physics \cite{ZKG03a,ZKG03b}, and of real life problems 
\cite{Oketal03,GC05} (see \cite{GhilZalCol08} for a recent review). The 
present work is a first step towards the application of BDEs to phenomena in 
the economic realm and, at the same time, it deals with the clarification of 
the role played by stochasticity in these systems.

By using BDEs, we plan to take into account the importance of the topology of 
the network, and of the different delays involved in the production paths
\cite{RoLe81,RoLe86,RoLe89,LeHaSa07,OkHeSo04,BaSa09}, on 
the total losses due to the initial catastrophe. The manuscript will develop 
the role of the connectivity of the structure: how the multiplicity of
products needed by each industry affects the vulnerability of the whole 
industry, and what are the effects of multiple production paths from one firm 
to 
down-stream production firms. We are specially interested by long term 
production shortage spatio-temporal patterns generated by the generic 
asynchrony due to variable time lengths of concurrent production paths.

Even though we are using Boolean variables to describe the production of
individual firm, a zero value should not be interpreted in terms of a 
destruction of the production unity: we rather mean that some shortage has 
been generated through production interactions; similarly, a level of one 
simply implies that the firm has recovered from previous impairing of 
production. When taking instead the interpretation of the zero value as a 
complete destruction of the firm, our models could be interesting from the 
point of view of the study of damage spreading in an Operational Research 
framework. Finally, we will show that a macroeconomic quantity, measuring the 
density of fully active firms, allows to characterize the propagation of the 
consequences of the initial catastrophe in the various considered cases. 

\section{The models}
A realistic representation of the economy at the firm level would make 
possible to assess both the direct and the indirect losses due to
a catastrophic event, by taking into account also the backward and forward
propagations of the event's consequences, hence the failure avalanches and
the ripple effects which move across the chains of suppliers and producers 
in the network. For instance, the analysis of the impacts on the regional 
economy of the Northridge earthquake \cite{Ti97,GoRiDa98} makes it evident 
that a catastrophe can have very heterogeneous repercussions, and that the
indirect losses due in particular to damages to the transport 
infrastructure system can be definitely higher than the direct losses 
themselves. Similar conclusions are obtained by the study of the
consequences of the Loma Prieta and Northridge earthquakes \cite{WeTiDa02},
which underlines the importance of taking into account both the direct and
the indirect losses, too; moreover this work shows that the repercussions can 
be 
less extended on firms belonging to a market larger than the strictly local 
one, hence suggesting to consider also economic models with adaptation 
(see, e.g., \cite{Choetal01}). 

Whereas the first steps towards a model suitable for realistic
evaluations have been presented in \cite{He07,Ha08,HeHa08}, we introduce here
some quite simplified toy models, which we will study within the framework
of Boolean Delay Equations: the present approximation neglects some key 
ingredients, such as in particular the division in sectors of the economy, and 
the related observation that a given firm can usually choose between more 
providers of the same good, also in a strictly local economy. Nevertheless, 
with these over-simplifications, the numerically observed behaviors can be 
understood in 
detail from the theoretical point of view, and we
find that our results are a good starting point for future analysis. 

We take the Boolean variable $x_i(t)=0$ to mean that firm $i$ at time $t$ 
is impaired and can not fully produce (respectively, $x_i(t)=1$ implies that 
it is not impaired): this can be due both to the firm being itself 
damaged or to the fact that it lacks of the necessary inputs, since some of 
its 
suppliers (or some of the suppliers of its suppliers and so on) were 
previously damaged. Therefore, this simple modeling takes into account the 
role of the chains of suppliers and producers in real economies.

We study networks of $N$ firms, assumed to be placed on the vertexes of a 
directed graph defined by the connectivity matrix $A$. In our notation, 
$A_{ij}=1$ if part of the output of firm $j$ (the supplier) is needed as input 
for firm $i$ (the customer) and $A_{ij}=0$ otherwise. As a first step for 
assessing the losses due to the propagation of the consequences of a 
catastrophic event, hence the vulnerability of an economic network to a 
natural disaster, we analyze the vulnerability of the present models to the 
initial damage of a single firm. Therefore we will assume, without
loss of generality, that the firm in the origin is initially destroyed 
during an interval of time of length $\tau_c$. We will look in particular at 
the 
dependence of the propagation of this event on the structure 
of the matrix $A$, hence on the resulting network topology, and at its 
dependence on the ensemble of the delays that exist in the economic 
system. 

We consider both isolated networks, the free models, and networks interacting 
with the outside, hence economies with adaptation, the forced models. In the 
first cases, the availability 
$S_{ji}$ of the good manufactured by firm $j$ at firm $i$ is simply delayed 
from production $x_j$ with a constant time length $\tau_{i,j}$ according to:
\begin{equation}
S_{ji}(t)=x_j(t-\tau_{i,j}) \hspace{.3in} \mbox{ free models.}
\label{free}
\end{equation}
In the presence of adaptation, which means that the economy is not locally
isolated but there is instead the availability of some external 
rescue input, the stock $S_{ji}$ is made again disposable after that the 
production of the firm $i$ has been impaired for a duration of time that we 
take, in a first approximation, still equal to $\tau_{i,j}$:
\begin{equation}
S_{ji}(t)=\oo{x}_i(t-{\tau}_{i,j})\vee x_j(t-{\tau}_{i,j})
\hspace{.3in} \mbox{ forced models} 
\label{forced},
\end{equation}
where $\vee$ is the OR operator and $\oo{(\cdot)}$ means the Boolean negation.
These equations give the truth table reported in [Tab.~1].

\begin{table}
\begin{center}
\begin{tabular}{||c|c|c|c||}
\hline
\hline
$x_i$ & $x_j$ & $S_{ji}$ & free models \\
\hline
\hline
0 & 0 & 0 & 
$j$ and $i$ inactive, the stock can not be reconstituted \\
0 & 1 & 1 & 
$j$ active and $i$ inactive, the good is stocked \\
1 & 0 & 0 & 
$j$ inactive and $i$ active, the stock is finished \\
1 & 1 & 1 & 
$j$ active and $i$ active, the stock is updated \\
\hline
\hline
$x_i$ & $x_j$ & $S_{ji}$ & forced models \\
\hline
\hline
0 & 0 & 1 & 
$j$ and $i$ inactive, the stock is supplied from outside \\
0 & 1 & 1 & 
$j$ active and $i$ inactive, the good is stocked \\
1 & 0 & 0 & 
$j$ inactive and $i$ active, the stock is finished \\
1 & 1 & 1 & 
$j$ active and $i$ active, the stock is updated \\
\hline
\hline
\end{tabular}
\end{center}
\caption{The output table of the stock $S_{ji}$ of a given product as Boolean 
function of the activities $x_i,x_j$ of the customer firm $i$ and of the 
supplier firm $j$, in the free models described by 
Eq.~(\ref{free}) and in the forced models described by Eq.~(\ref{forced}),
respectively.}
\end{table}

In our simple toy models, the production $x_i(t)$, of firm $i$ at time $t$, 
needs the presence of all the stocks of goods usually provided by the 
suppliers  $\{j: A_{ij} \neq 0\}$ to which the firm $i$ is connected in the 
network. Besides, we assume that $x_i(t)$ also depends upon the value 
$\mu_i(t)$ of an external coefficient, which is an independent Boolean 
variable that allows in particular for the initial destruction of
the firm in the origin. These constraints can be expressed, both for the free 
and the forced models, by the system of $N$ BDEs:
\begin{equation}
x_i(t)=\mu_i(t) \prod_{j=1}^N \oo{A}_{ij} \vee S_{ji}(t)
\hspace{.3in}
\mbox{ for } i=1,\dots,N.
\label{syst}
\end{equation}
Here the product $\prod$, which runs over all the sites of the network, 
means the Boolean operator AND, $\wedge$. The 
dependence on the delays $\{\tau_{i,j}\}$ is implicit, through the stocks,
given by Eq.~(\ref{free}) or Eq.~(\ref{forced}).

In order to get well defined solutions for this system, one has 
to fix the initial values of the set of variables $\{x_i(t)\}$ in 
the interval $[-\tau_{max},0]$, where $\tau_{max}$
is the largest possible delay. Moreover, one has to choose the behaviors of 
the external functions $\{\mu_i(t)\}$. We only study the simple
cases in which the economy starts undamaged and all the external functions take
the value one during the evolution, apart from the destruction of
the single firm in the origin, {\em i.e.} on the node $i=1$. Notice that
this firm is not definitely eliminated form the network: we assume instead
that it is forced to stop the activity from the time $t=0$ up to the time 
$\tau_c$; this means $\mu_1(t)=0$ for $t \in [0,\tau_c]$. 

We compare results for deterministic delays, chosen all equal to $\tau_0$, 
with those for the more realistic situation of random delays 
$\{ \tau_{i,j} \}$, independently and uniformly 
distributed in the interval $[\tau_{min},\tau_{max}]$, with:
\begin{equation}
{\cal P}(\tau_{i,j})=\frac{\tau_{min}}{\tau_{max}} 
\sum_{k=1}^{\tau_{max}/\tau_{min}}
\delta \left (\tau_{i,j}/\tau_{min}-k \right ),
\label{Ptauij}
\end{equation}
{\em i.e.} the random variables are multiples of the unity of time fixed by 
$\tau_{min}$. We usually take $\tau_0=\tau_{min}=1$ day and 
$\tau_{max}=10$ days. Though the possibility of defining BDE systems 
characterized by random delays with a given probability distribution was 
already stated in \cite{DeeGhil84}, the present study is the first example in 
this direction to our knowledge.

Still, in the first part of this work we look at a deterministic braid chain 
network, corresponding to a connectivity matrix $A$ with 
circulant structure (see [Fig. 1]), whereas in the second part we study 
a directed random graph (DRG). We will consider in particular the family of 
directed random graphs, $D(N,p)$ \cite{Ka90,LuSe09}, obtained by generalizing 
the rule of the well known Erd\H{o}s-R\'enyi undirected random graph (RG) 
\cite{ErRe59,ErRe60,ErRe61}, where 
each of the $N(N-1)$ directed links is present or absent with the 
same independent probability $p\in[0,1]$. Hence, the elements of the 
connectivity matrix $\{A_{ij} \}$ are random variables, assumed to be 
independently and identically distributed: 
\begin{equation}
{\cal P}(A_{ij})=
\left \{
\begin{array}{lcl}
\delta \left(A_{ij} \right) &\mbox{ for }& i=j \nonumber \\
p \delta \left( A_{ij} -1 \right ) + (1-p) \delta \left ( A_{ij} \right ) 
& \mbox{ for } & i \neq j\\ 
\end{array}
\right.
\label{PAij}
\end{equation}
The probability $p$ is straightaway related to the mean number of 
input-output connections ({\em i.e.} the in/out-degree), 
$z=\langle k_{in} \rangle = \langle k_{out} \rangle =(N-1) p \simeq Np$, of 
the resulting directed network.

In the considered context of BDEs with stochasticity, the delays 
$\{ \tau_{i,j} \}$ and the elements of the connectivity matrix $\{A_{ij}\}$ 
are 
quenched random variables: when defining the BDE system one fixes 
their values, with probabilities given by Eq.~(\ref{Ptauij}) and
Eq.~(\ref{PAij}), to a given disorder configuration, and they are assumed to
be constant on the time scale of the evolution we are interested in.

\begin{table}
\begin{center}
\begin{tabular}{||c||c|c|c|c||}
\hline
\hline
& Fr, $\tau_0$ & Fr, ${\cal P}(\tau_{i,j})$& Fo, $\tau_0$ & Fo, 
${\cal P}(\tau_{i,j})$  \\ 
\hline
\hline
CM, $n=1$ & 3.2 & 3.3 & 3.6 & 3.6  \\ 
\hline 
CM, $n>1$ & 3.4 & 3.5 & 3.7 & 3.8  \\ 
\hline
DRG & 4.2 & 4.3 & 4.4 & 4.4  \\ 
\hline  
\hline
\end{tabular}
\caption{We summarize here the main characteristics of the models studied
in the following, and we indicate in each box the section where the model
with all the corresponding details is discussed and the figures showing
our results are presented. We label Fr the free models described by 
Eq.~(\ref{free}) and Fo the forced models obeying to Eq.~(\ref{forced}); then, 
we look both at the case in which all the delays are deterministically taken 
equal to the same time unit $\tau_0$, and at the random delay one, with 
${\cal P}(\tau_{i,j})$ defined by Eq.~(\ref{Ptauij}); still, CM refers to the 
deterministic braid chain network obtained from a circulant matrix, and DRG to 
the directed random graph, with ${\cal P}(A_{ij})$ given by Eq.~(\ref{PAij}); 
finally, we distinguish between single ($n=1$), and multiple ($n>1$), 
input-output connections in the braid chain, whereas the models defined on the
DRG with average in/out-degree $z$ both smaller and larger then one are treated
in the same sections.}
\end{center}
\end{table}

With the aim of describing the solutions, and particularly their behaviors 
in the limit of a large number $N$ of variables, we find interesting
to introduce a macroeconomic observable, that is to say the average density of 
firms which maintain their production, $\rho(t)$. In the deterministic case it 
is given by:
\begin{equation}
\rho(t) \equiv \frac{1}{N} 
\sum_{i=1}^N  x_i(t).
\label{defrho0}
\end{equation}
In the presence of stochasticity, in the delays and/or in the elements of the 
connectivity matrix, the behaviors of the systems are better captured from the 
density averaged over disorder:
\begin{equation}
\langle \rho(t) \rangle \equiv \int \prod_{i,j} d\tau_{i,j} 
{\cal P}(\tau_{i,j})
\int \prod_{h,k} dA_{hk} {\cal P}(A_{hk}) \rho(t), 
\label{defrho1}
\end{equation}
that can be computed numerically by considering an enough large number, 
${\cal N}_s$, of different disorder configurations.
We also look at the density averaged over a time window of length 
corresponding to the smallest possible delay, {\em i.e.} $\tau=\tau_0$ in the 
deterministic case and $\tau=\tau_{min}$ in the stochastic one:
\begin{equation}
\rho_{av}(t) \equiv \int_{t}^{t+\tau} \rho(s)ds.
\label{defrhoav}
\end{equation}

Finally, in the considered simple toy models, the total number of impaired 
firms at time $t$, $\theta_{tot}(t)$, gives just the evaluation of the total 
losses due to the propagation at this time of the effects of the initial 
destruction of the single firm in the origin, hence it is also a measure of 
damage spreading:
\begin{equation}
\theta_{tot}(t) \equiv \sum_{i=1}^N \oo{x}_i(t).
\label{defztot}
\end{equation}
Accordingly, the density can be straightaway obtained as:
\begin{equation}
\rho(t)=1-\frac{1}{N}\theta_{tot}(t).
\label{rdaz}
\end{equation}
When taking random delays and/or a random network structure, we 
study the average $\langle \theta_{tot}(t) \rangle$, defined analogously to
the average density in Eq.~(\ref{defrho1}).

We summarize in [Tab. 2] the different cases that we consider.

\section{Circulant matrix}

\subsection{Topology}
Here we look at the network topology corresponding to a braid chain, which is 
obtained from a connectivity matrix $A$ (of size $N \cdot N$) with circulant 
structure. When taking periodic boundary conditions for the indexes 
($i\pm N \equiv i$), the elements can be written as: 
\begin{eqnarray}
A_{ij}= \left \{
\begin{array}{ll}
1 & \mbox{ for } j=i-1,i-2, \dots, i-n  \\
0 & \mbox{ otherwise}, \\
\end{array}
\right.
\end{eqnarray}
for some integer $n < N$. This models a directed braid chain, with the same
in/out-degree $n$ for all the nodes. Each firm is linked to $2n$ other firms, 
and needs as inputs part of the goods manufactured by the previous $n$ ones 
(its suppliers), whereas the produced outputs are used from the next $n$ ones 
(its customers). The resulting deterministic topology is strongly connected: 
starting from any node one finds at least one directed path along which the 
signal can propagate to any other node, and in fact there are multiple 
concurrent paths as soon as $n>1$. Most of these paths have the same time 
length in the purely deterministic case of equal delays, whereas they have 
usually different time lengths for random delays. In other words, the random 
delays allow to model the asynchrony in the concurrent production paths. We 
present in [Fig. 1] an example for $n=2$ in/out-degree.

\begin{figure}[ht]
\begin{center}
\includegraphics[width=.8\textwidth,angle=0]{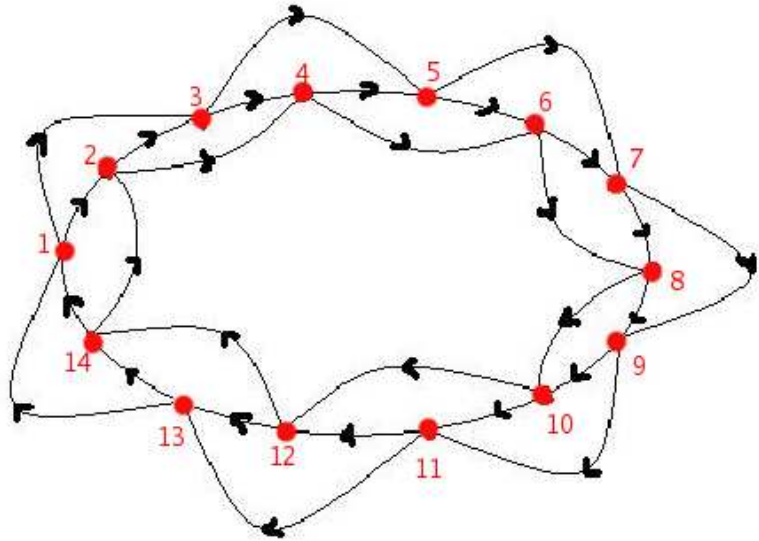}
\caption{The braid chain structure given by a circulant matrix with periodic 
boundary conditions in the case of $n=2$ in/out-degree. The size of the 
network is $N=14$, and the nodes are arbitrarily ordered clockwise. In the 
deterministic case of equal delays also the choice of the origin (the $i=1$ 
node) is completely arbitrary. Note that instead, in the presence of 
stochasticity, the different concurrent paths connecting two given nodes do 
not usually have the same duration in time, although they might 
have the same spatial length in terms of number of steps along the chain.}
\end{center}
\end{figure}

When $A$ is a circulant matrix, the equation for the variable $x_i$ of the 
system (\ref{syst}) can be simplified into:
\begin{eqnarray}
x_i(t)&=&\mu_i(t) \left[ \prod_{j=1}^{n}
 x_{i-j}(t-\tau_{i,i-j}) \right] \label{freecirc} \\
x_i(t)&=&\mu_i(t) \left[ \prod_{j=1}^{n}
\oo{x}_{i}(t-\tau_{i,i-j}) \vee x_{i-j}(t-\tau_{i,i-j}) \right], 
\label{forcedcirc}
\end{eqnarray}
for the free models defined by Eq.~(\ref{free}) and the forced models 
defined by Eq.~(\ref{forced}), respectively. The considered systems give an
extremely simplified description of the input-output structure of real
economic network; nevertheless, braid chain structures 
turn out to be an interesting approximation when interpreting each 
node $i$, with its production $x_i$, as corresponding to a whole sector of the
economy.

\subsection{The free model with $n=1$ and equal delays}
The free model on a braid chain network, with a single input-output 
connection for node, $n=1$, and deterministically taken delays
all equal to the same time unit $\tau_0$, is an example of conservative system 
of BDEs \cite{DeeGhil84,Mull84,GhilMull85,GhilZalCol08}. Its dynamics is
periodic from the beginning for all the initial states, without any transient. 
The model is described by the set of equations:
\begin{equation}
x_i(t)=\mu_i(t)\:x_{i-1}(t-\tau_0) \hspace{.3in} i=1,\dots,N,
\label{hpde}
\end{equation}
where we fix initial values $x_i(t)=1 \: \forall i$ for $t\in [-\tau_0,0)$, 
and we represent the initial destruction of the single firm
on node $i=1$, for a duration $\tau_c$, by the choice of the external 
coefficient: $\mu_1(t)=0$ for $t\in [0,\tau_c)$.
The solution displays a wave of nodes taking the value zero the one after
the other, which propagates periodically across the chain, as in the example 
given in [Fig. 2a]. 

\begin{figure}
\begin{center}
\includegraphics[width=.45\textwidth,angle=0]{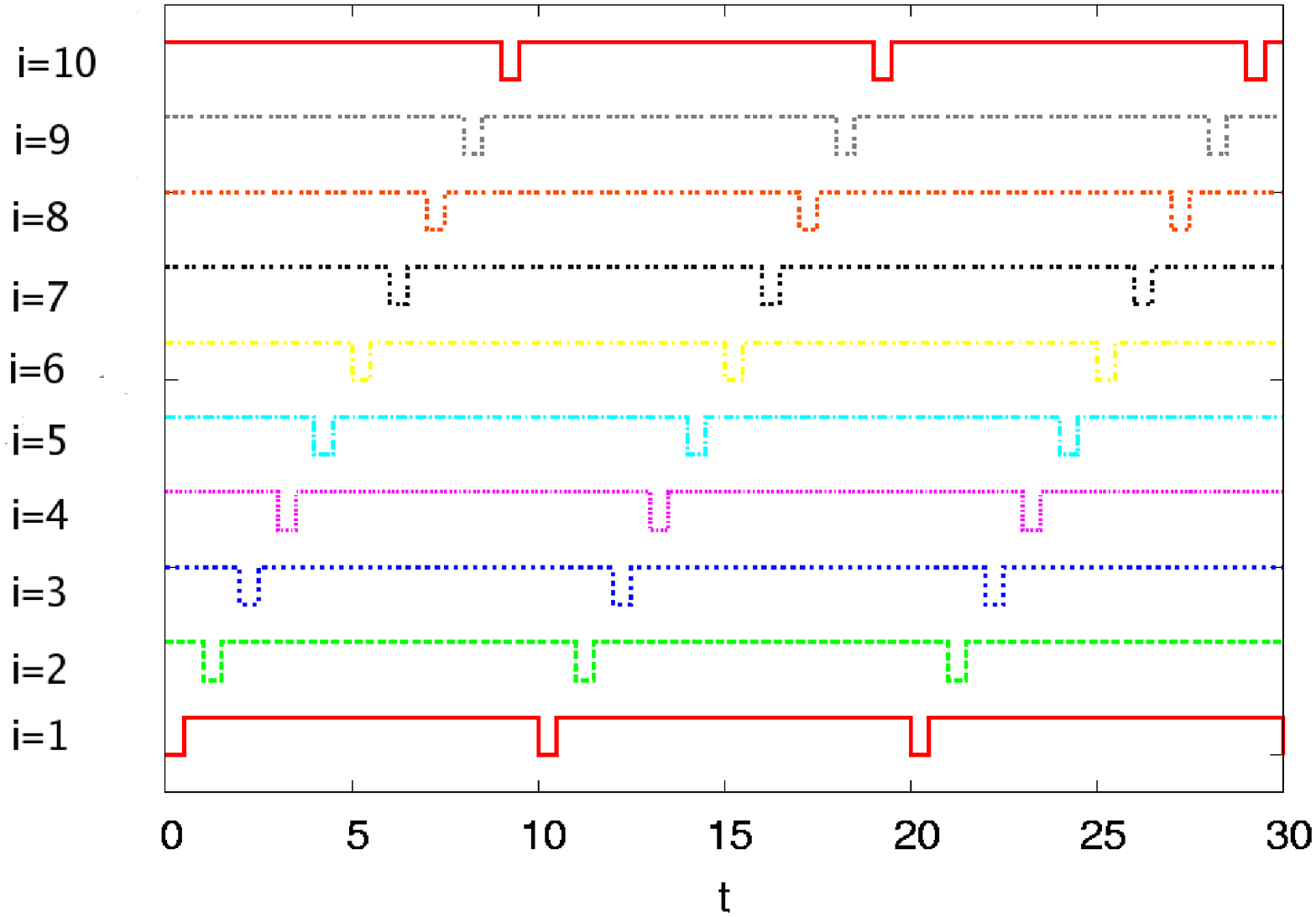}
\includegraphics[width=.45\textwidth,angle=0]{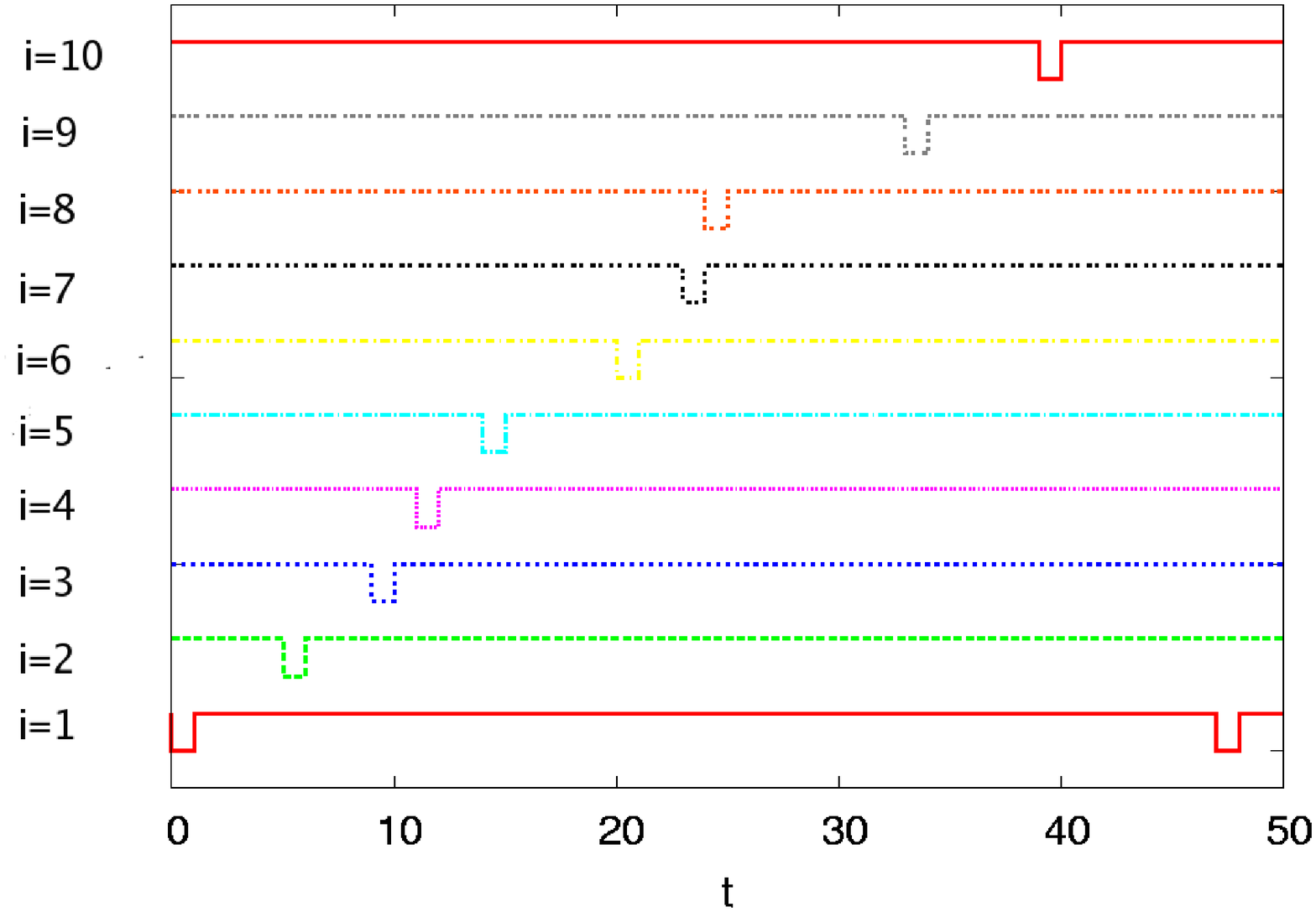}
\caption{Time evolution of the state of the 10 nodes of the free
model on a braid chain, with $n=1$, after an initial perturbation of 
node $i=1$. Each Boolean variable $x_i(t)$ is plotted as function of $t$ and it
is a piecewise constant function varying between 0 and 1. We moreover shift
the solutions $x_i(t)$ along the $y$ axis, as a function of the node index 
$i$, for visualizing the propagating wave in the network. On the left, the 
delays are deterministically taken all equal to $\tau_0=1$ day, and the 
duration of the initial perturbation is $\tau_c= \tau_0/2=0.5$ day.
On the right, we look at a particular configuration of random delays, 
$\{\tau_{i,i-1}\}$, uniformly distributed between $\tau_{min}$= 1 day and 
$\tau_{max}$=10 days, yielding a period of 47 days. Here the duration of the
initial perturbation is $\tau_c= \tau_{min}=1$ day.}
\end{center}
\end{figure}

This simple system is analogous to the model introduced in \cite{GhilZalCol08} 
as a first step towards a BDE equivalent of hyperbolic partial differential 
equations. In that work, we were discretizing the unidimensional space 
lattice, and in the analogy the node index $i$ plays the role of the 
discretized coordinate. We studied the evolution starting from an initial
state where only the variable associated to the lattice point in 
the origin was equal to one, and we found correspondingly a propagating wave 
of points taking the value one in the spatio-temporal pattern.

In fact, when the duration of the initial perturbation, $\tau_c$, is smaller 
than the unit of time, $\tau_0$, the role of the external coefficients 
can be absorbed into the initial conditions. The evolution shows in any case 
no transient and the period of the solution is ${\cal \pi}=N \tau_0$. For 
values of $\tau_c$ multiple of $\tau_0$, the density of firms which maintain 
their production takes the constant value $\rho(t)=1-{\tau_c}/(N{\tau_0})$, 
whereas more generally it has period $\tau_0$. 

This dynamics implies that, on the one hand, the damage does not spread,
but, on the other hand, the activity never recovers completely: each impaired
firm gets back to the unaffected state after the time $\tau_c$, but 
contemporaneously its production shortage reaches its customer, with a 
constant delay $\tau_0$. This unrealistic result is linked to one of our 
simplifying assumptions, namely the discretization of firm production 
capacity, with no possibility for overproduction or production rescheduling.

\subsection{The free model with $n=1$ and random delays}
In the free model on a braid chain network with $n=1$, the effect of 
stochasticity in the delays can be straightaway worked out explicitly:
\begin{eqnarray}
x_i(t)&=&
x_{i-1}(t-\tau_{i,i-1})=
\nonumber \\ &=& 
x_{i-2}(t-\tau_{i,i-1}-\tau_{i-1,i-2})= \dots =\nonumber \\
& =& 
x_i \left ( t - \sum_{k=0}^{N-1} \tau_{i-k,i-k-1} \right ).
\end{eqnarray}
Correspondingly, one finds that the solution is still periodic from 
the beginning for all the initial states, and the system is therefore once 
again conservative. 

In the considered case of an initial perturbation of node $i=1$ of duration 
$\tau_c$, we get in particular a periodically propagating wave 
of nodes taking the value zero the one after the other in the spatio-temporal 
pattern, as in the deterministic case. The difference is that the propagation 
time of the perturbation, from a supplier $i-1$ to its customer $i$, is now 
given by the quenched random variable $\tau_{i,i-1}$, and the period $\pi$ of 
the solution, for a given disorder configuration, is equal to the sum of these 
delays along the whole chain (see [Fig. 2b]), with in average 
$\langle \pi \rangle = N \langle \tau_{i,i-1} \rangle = N (\tau_{max}+1)/2$.

\begin{figure}[htpb]
\begin{center}
\includegraphics[width=.45\textwidth,angle=0]{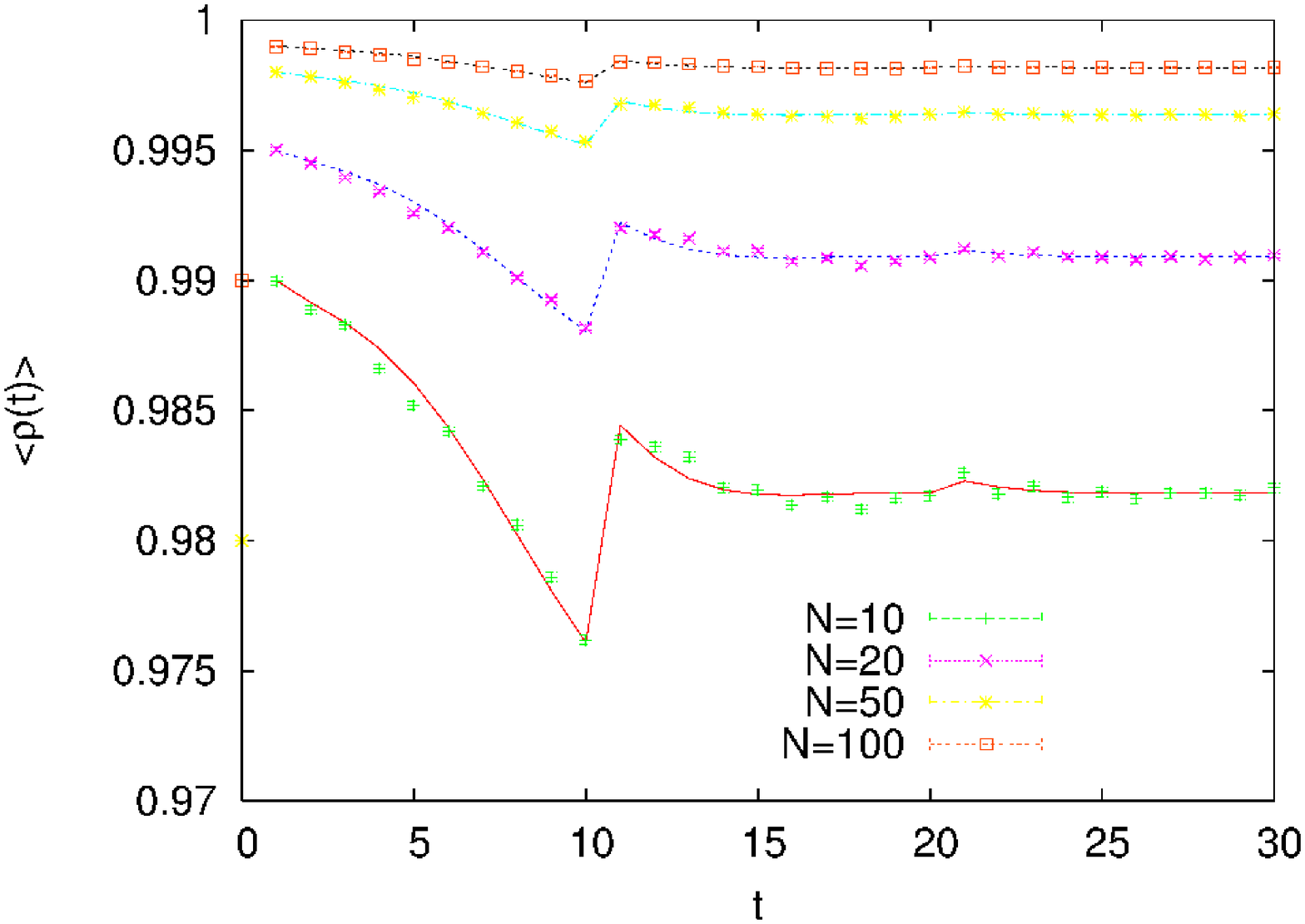}
\includegraphics[width=.45\textwidth,angle=0]{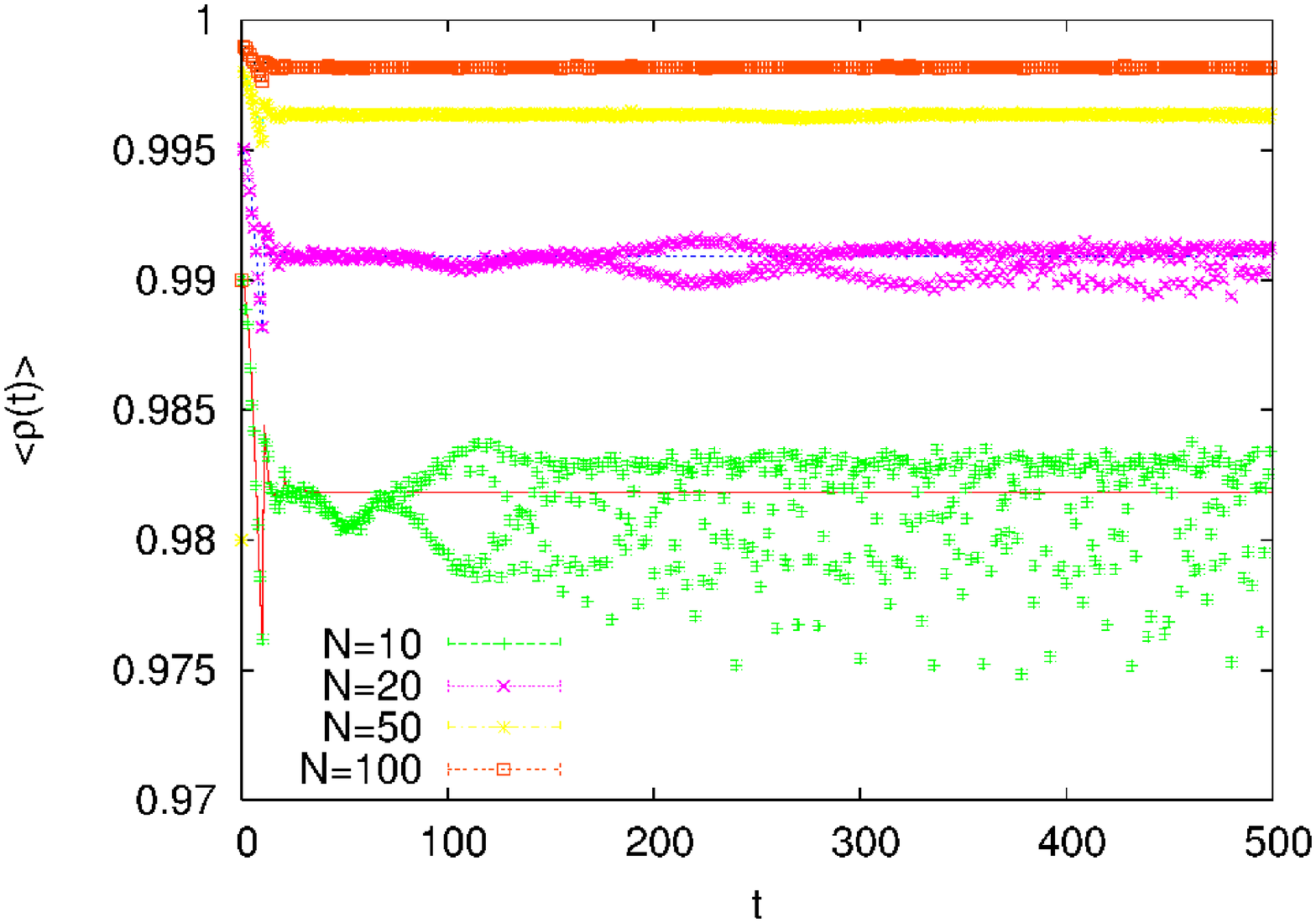}
\caption{Time plot of the average fraction of fully active firms, 
$\langle \rho(t) \rangle$, in the free model on a braid chain, with $n=1$, and 
stochasticity in the delays. Here $\tau_{max}=10$ days, as in [Fig. 2b],
and we compare the behaviors for network sizes $N=10$, 20, 50 and 100 with the 
expected ones (see Appendix~A). On the left we present the details at short 
times, and on the right the whole considered time window. The data are 
averaged over ${\cal N}_s$ different random configurations of the 
delays, a number which is taken enough large to give errors of the order of 
the point size in the plot.}
\end{center}
\end{figure}

The density averaged over the disorder, $\langle \rho(t) \rangle$, can be 
computed analytically by applying the central limit theorem (see Appendix~A). 
In [Fig. 3], we compare the expected behavior, given by
Eq.~(\ref{ztotcmn1}), with the numerical results, for different network sizes 
$N$: at short times, one can observe the jumps at $t=\tau_{max}$; at long 
times, there are corrections to the expected behavior, since in the 
considered sums of random delays the same variables appear more than once;
nevertheless, the mean asymptotic value is in agreement with Eq.~(\ref{asym}). 

The case $\tau_c \neq \tau_{min}$ can be understood within the same framework, 
and one finds that, for large $N$ values, the average density does only depend 
on the ratio of the duration of the initial damage, $\tau_c$, to the size of 
the network $N$ itself:
\begin{equation}
\langle \rho_{av}(t)\rangle \simeq 1-\frac{\tau_c}{N} 
\langle \theta_{tot}(\tau_c=1,t) \rangle.
\label{risn1r}
\end{equation}
Here we are looking at $\langle \rho_{av}(t) \rangle$, 
assumed to be averaged also over a time window of length $\tau_{min}=1$ day: 
in fact, for $\tau_c$ not multiple of $\tau_{min}$,
the density $\langle \rho(t) \rangle$ approximately oscillates
between the expected solution for $[\tau_c/\tau_{min}]$ and for 
$[\tau_c/\tau_{min}]+1$ (where $[y]$ means the integer part of $y$),
with period $\tau_{min}$: this result becomes correct in the asymptotic 
large time limit.

\subsection{The free model with $n>1$ and equal delays}
Generally, the BDE systems which describe the free models on a braid chain,
with $n$ input-output connections, can 
be reduced to closed sets of $n$ equations; these equations involve the 
products of only $n$ variables, corresponding to nodes in consecutive 
positions along the chain, but each variable appears more than once, at 
different times. 

In particular, in the deterministic case of delays all equal to $\tau_0$, 
and taking $N$ multiple of $n$ for simplicity, one has:
\begin{eqnarray}
x_i(t)& = &
\prod_{k=N/n}^{N-n+1} x_{i+n-1} (t - k \tau_0 ) \cdot
 x_{i+n-2} ( t - k \tau_0 ) \cdot 
...  \cdot x_i ( t - k \tau_0 ) \nonumber \\
x_{i+1}(t)& = &
\left [ \prod_{k=N/n}^{N-n+2} x_{i+n-1} ( t - k \tau_0) \cdot
 x_{i+n-2} ( t - k \tau_0 ) \cdot
... \cdot  x_{i+1} ( t - k \tau_0 ) \right ] \cdot \nonumber \\ 
& \cdot&
\left [ \prod_{k=N/n+1}^{N-n+2} x_{i} ( t - k \tau_0 ) \right ] \nonumber \\
...&=&... \nonumber \\
x_{i+n-1}(t)& = &
\left [ \prod_{k=N/n}^{N-n+2} x_{i+n-1} ( t - k \tau_0 ) \right ] \cdot 
\nonumber \\ &\cdot&
\left [ \prod_{k=N/n+1}^{N-n+2}  x_{i+n-2} ( t - k \tau_0 ) \cdot
... \cdot  x_{i+1} ( t - k \tau_0 ) \cdot  x_{i} ( t - k \tau_0 ) \right ]. 
\label{semplif}
\end{eqnarray}
In this expression they are contained all the possible paths along which the 
information, here the damage due to the initial perturbation, can 
propagate across the network. The reduction is analogous to the one
of systems of differential equations to systems of higher order equations 
depending upon a smaller number of variables.  

As soon as the number of connections is $n>1$, the dynamics is dissipative 
\cite{DeeGhil84,Mull84,GhilMull85,GhilZalCol08}: the asymptotically 
stable solution is the state in which the whole economy is attained 
by the consequences of the initial damage, $x_i \equiv 0 \: \forall i$, 
and it is finally reached for any duration of this starting perturbation  
$\tau_c \ge \tau_0$ (see [Fig. 4]). This result can be qualitatively explained 
by noticing that the present economic models lack of flexibility, since the 
$n$ different inputs to a given firm are considered to be linked by AND 
logical operators. Besides, the system is isolated and the deterministic 
topology of the braid chain is strongly connected.

\begin{figure}
\begin{center}
\includegraphics[width=.45\textwidth,angle=0]{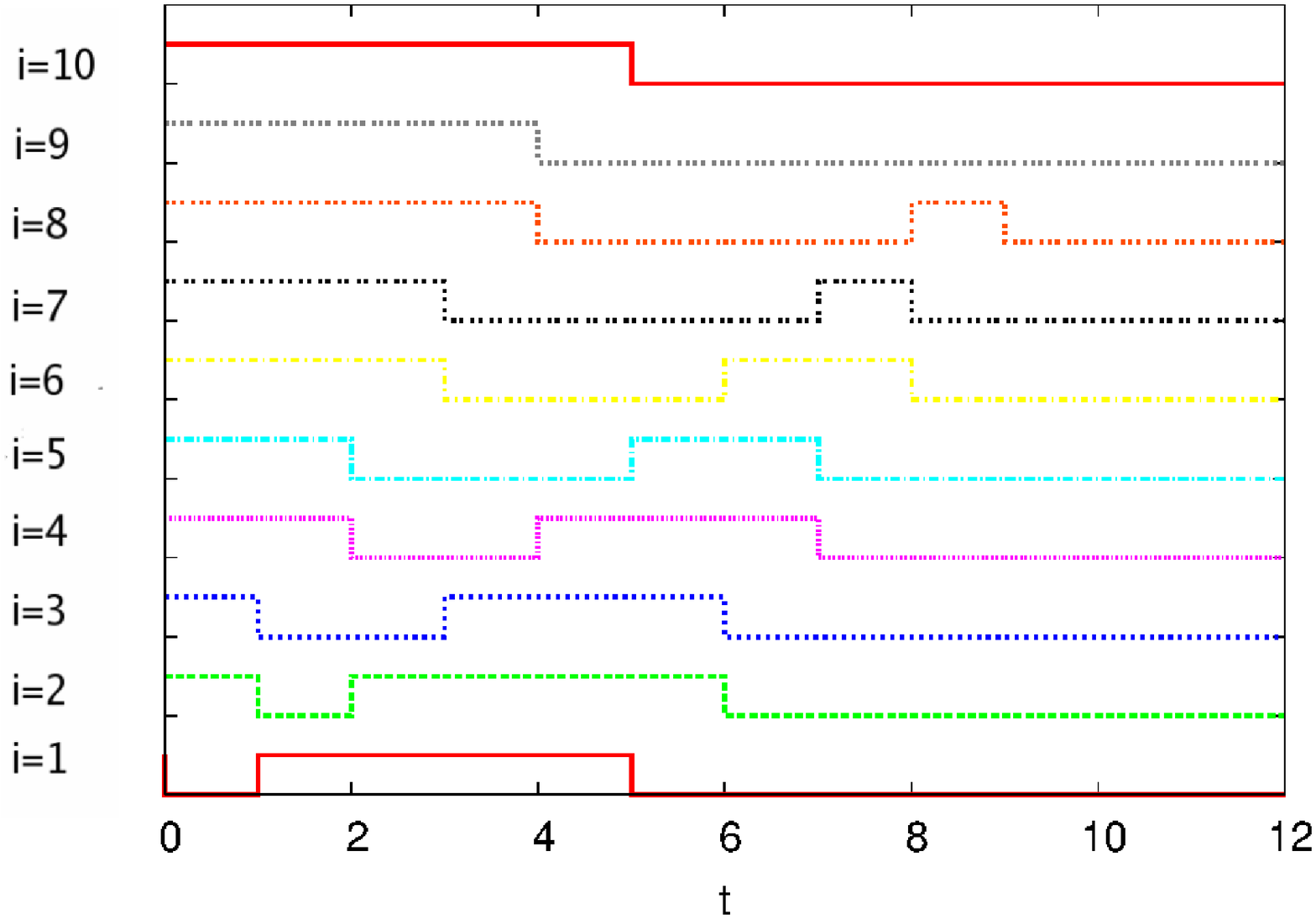}
\includegraphics[width=.45\textwidth,angle=0]{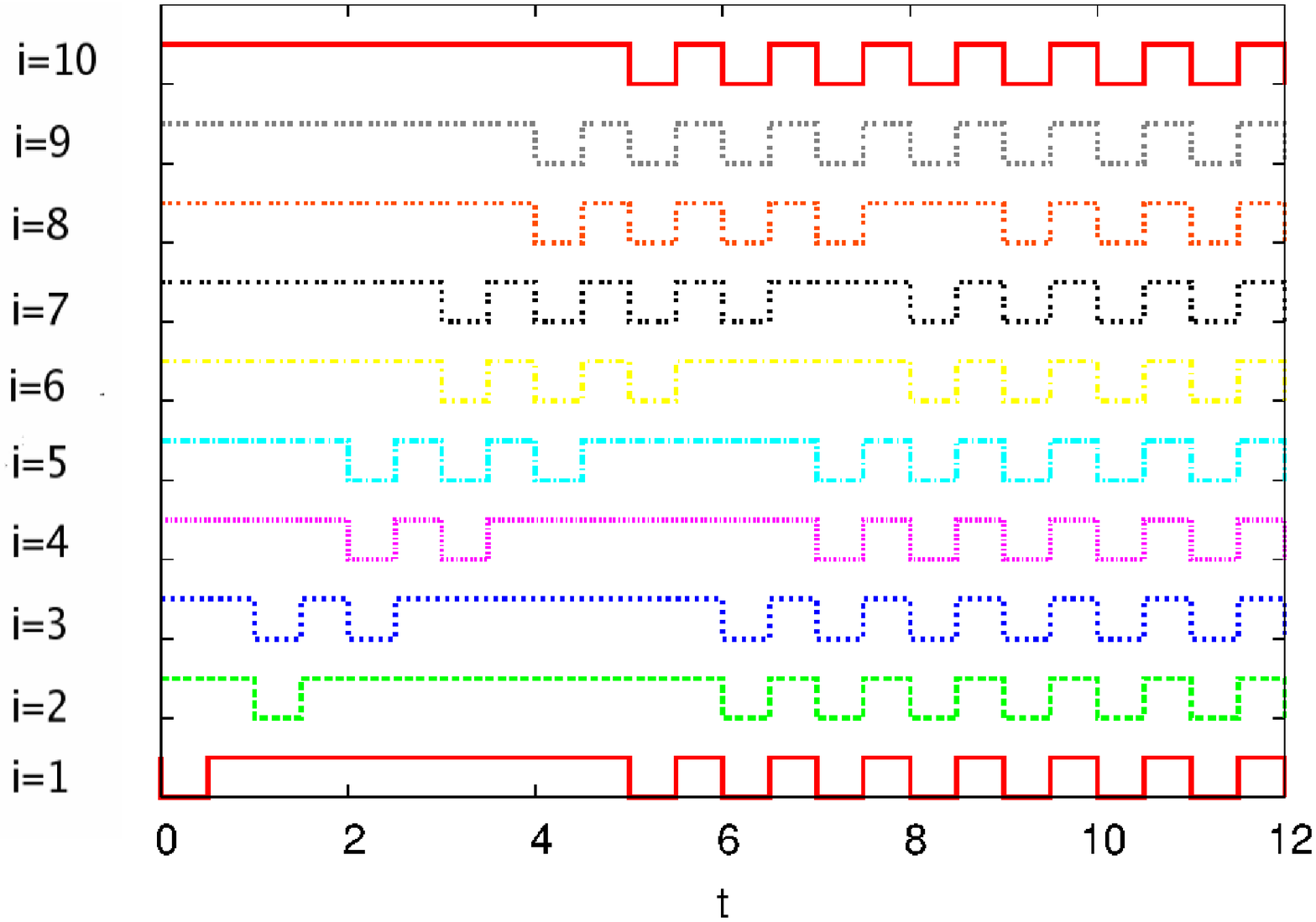}
\caption{Time evolution of the state of the 10 nodes of the free model
on a braid chain, with $n=2$, after an initial perturbation of node 1. We plot 
$x_i(t)$ as a function of $t$, the solutions corresponding to different node 
index $i$ being shifted along the $y$ axis in order to visualize the 
spatio-temporal pattern. Here all the delays are equal to $\tau_0=1$ day. On 
the left, the duration of the initial perturbation is $\tau_c= \tau_0=1$ 
day, whereas on the right we take $\tau_c=\tau_{0}/2=0.5$ days. See the text 
for details.}
\end{center}
\end{figure}

We analyze the dynamics of the model starting from a situation in which the 
impaired firms occupy nodes in consecutive positions along the braid chain:  
let us say that in the interval $[t,t+1)$ (in unity of $\tau_0$) there are 
$\theta_{tot}(t)=k$ impaired firms, {\em i.e.}, $x_{i}(t)=0$ for 
$i \in [i_m(t), i_M(t)]$, where $i_M(t)=i_m(t)+k-1$ and $k \ll N$. At this 
point, it is helpful to imagine a clock, with the hour-hand marking 
the position of the impaired firm closest to the origin, $i_m(t)$,
and the minute-hand marking the position of the one farthest from the origin, 
$i_M(t)$, at time $t$. One finds that both the hour-hand and 
the minute-hand move with constant velocities, given respectively by 1 and 
$n$, and therefore (as soon as $n>1$), the width of the set of firms which
are unable to fully produce increases with constant velocity, given by $n-1$, 
or in other words the damage spreads to a total number of $n-1$ more firms at 
each subsequent time step.

In fact, at each time step $t'=t+1$, the firms in the next 
$n$ positions after $i_M(t)$ will be impaired, since at least one of the 
stocks that they need is finished, hence $i_M(t+1)=i_M(t)+n$. The first firm 
in the sequence is instead the only one which recovers its unaffected state, 
all its suppliers being previously fully active, hence $i_m(t+1)=i_m(t)+1$. 
Correspondingly, one gets just $\theta_{tot}(t')=\theta_{tot}(t)+(n-1)$. 
The same argument can be applied again, starting from the new sequence, 
and so on. 

Following this analysis, the evolution will stop at the time $t_{trans}$, when 
the minute-hand reaches the hour-hand with one turn of advantage: the system 
has attained the asymptotically stable steady state 
$x_i \equiv 0 \: \forall i$, and the activity can not fully recover at all. 
Notice that the result is unchanged if, in the last time step of the 
transient, the 
minute-hand surpasses the hour-hand, since $x_i=0$ describes a shortage in the 
production of firm $i$ aside from the number of stocks of goods which are not 
available. Moreover, during the transient, each firm can recover at most once.

Summarizing, $\theta_{tot}(t)$ is linearly increasing with $t$ up to reach the 
system size $N$; correspondingly, for an initial perturbation destroying 
the production of one firm for a duration $\tau_c=\tau_0=1$ day, 
$\rho(t)=1-{1}/{N}-({n-1})t/{N}$ for $t \lesssim 
t_{trans}$ and $\rho(t)=0$ for $t\ge t_{trans}$, where the length of the 
transient is straightaway given by:
\begin{equation}
t_{trans}\simeq\frac{N-1}{n-1} \tau_0.
\label{ttransndet}
\end{equation}

When generalizing to durations of the initial event 
$\tau_c  \neq \tau_0$, it is important to distinguish between:
\begin{itemize}
\item $\tau_c > \tau_0$: Here the clock-argument turns out to be valid after 
the first $\sim \tau_c/\tau_0$ time steps; therefore, at large times, the 
damage spreading velocity is constant and equal to $n-1$, and the asymptotic 
stable state is the one where all the firms are impaired. 
\item $\tau_c < \tau_0$: As shown by data in [Fig. 4b], here
one can say that it is the behavior of the initially damaged firm,
\begin{equation}
x_1(t)=\left \{
\begin{array}{ccc}
0 & \mbox{ for } & t \in [0,\tau_c) \\
1 & \mbox{ for } & t \in [\tau_c, 1), 
\end{array}
\right.
\label{firstfirm}
\end{equation}
that spreads across the network, with constant velocity $n-1$;
the asymptotic solution is periodic of period equal to the unit of time, 
$\tau_0$, with all the firms synchronously impaired only in 
the first $\tau_c$ part of each period; the density $\rho_{av}(t)$, 
averaged over the period, is linearly decreasing with time, 
in this case up to the asymptotic constant value $\tau_c/\tau_0$, reached after
a transient of length given again by Eq.~(\ref{ttransndet}).
\end{itemize}

These results are confirmed by the analysis that we present in Appendix~B,
where $\theta_{tot}(t)$ is computed explicitly.

\subsection{The free model with $n>1$ and random delays}
We now consider the free model on a braid chain, with multiple 
$n>1$ input-output connections, in the presence of stochasticity in the 
delays. Here the $\{ \tau_{i,j} \}$ are quenched independent random variables, 
taken uniformly distributed in the interval $[\tau_{min},\tau_{max}]$ 
accordingly to Eq.~(\ref{Ptauij}), and the unit of time is given by 
$\tau_{min}$, that we choose again equal to 1 day. 

As usual, we look in particular at the behavior of the system after the 
destruction of the single firm in the origin for a duration $\tau_c$. 
For $\tau_c \ge \tau_{max}$, we expect 
the damage to spread all over the network. In fact, the damage can not spread 
more slowly than in the peculiar case in which all the delays are equal to 
$\tau_{max}$: it follows that the state $x_i \equiv 0 \: \forall i$ is 
asymptotically stable also in the presence of stochasticity, and that it is 
surely reached, after a transient $t_{trans}$.

Nevertheless, now the concurrent paths, with the same space length along the 
chain, usually do no longer have the same time length, since in the sums of 
the involved delays appear different random variables. The presence of
asynchrony has multiple consequences (see [Fig. 5]): the reduced set of 
equations (\ref{semplif}) does generally contain a number of terms 
increasing with the path space lengths; during the dynamics, 
there is often no more a sequence of consecutive positions along the 
chain all occupied by impaired firms; as soon as $\tau_c<\tau_{max}$, 
each firm can be impaired and recover more than once; 
most intriguingly, because of this reason, the dynamics 
can be ruled by a faster time scale than the average delay. 

\begin{figure}[htb]
\begin{center}
\includegraphics[width=.45\textwidth,angle=0]{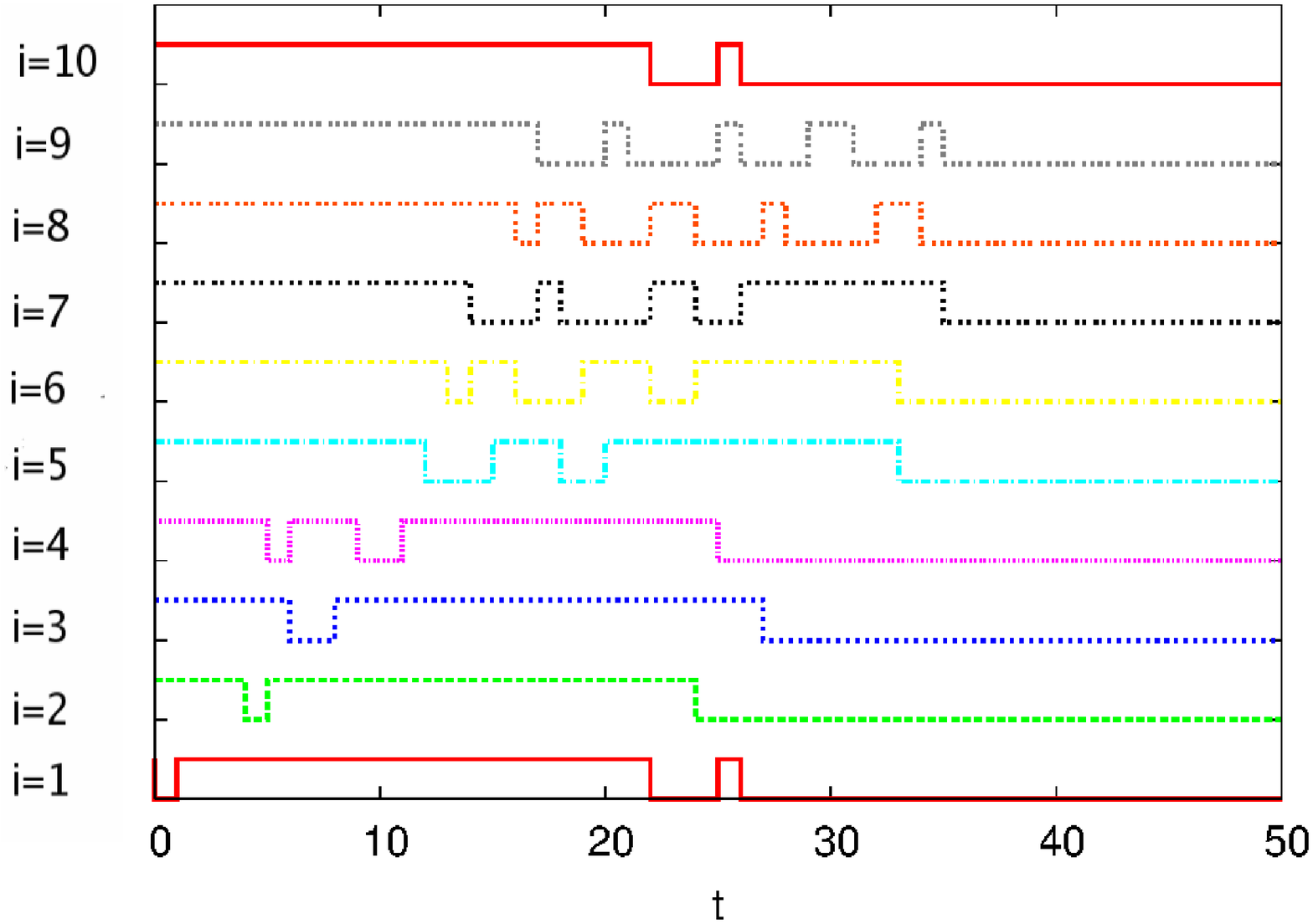}
\includegraphics[width=.45\textwidth,angle=0]{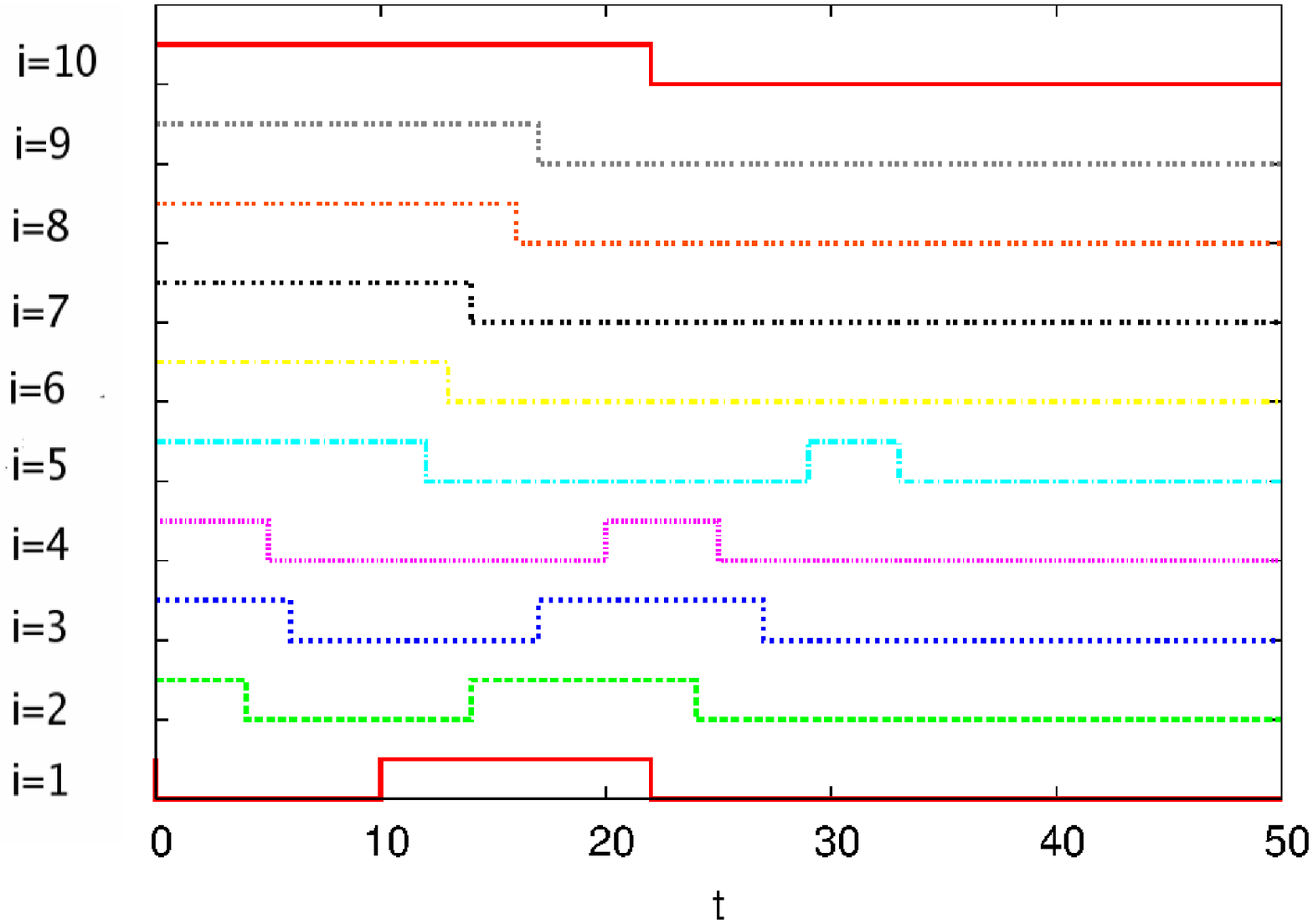}
\caption{Time evolution of the state of the 10 nodes of the free model 
on a braid chain, with $n=2$, after an initial perturbation of 
node 1. As in the previous cases, we plot $x_i(t)$ as a function of $t$ 
and we shift the solutions corresponding to different node index $i$
along the $y$ axis in order to visualize the spatio-temporal pattern. We look 
at a particular configuration of random delays, $\{\tau_{i,j}\}$, 
uniformly distributed between $\tau_{min}$= 1 day and $\tau_{max}$=10 days. 
The duration of the initial perturbation is $\tau_c=\tau_{min}$ on the left, 
and $\tau_c=\tau_{max}$ on the right, which yield transients of length 35 and 
33, respectively.}
\end{center}
\end{figure}

In fact, since the delays are independent identically distributed random 
variables along the whole chain, in the limit $1 \ll t \ll t_{trans}$, 
one can argue that the damage spreads with constant average velocity 
$\alpha$, where $\alpha_{min} \le \alpha \le \alpha_{max}$: the lower
(respectively upper) limit is easily obtained by taking
all the delays equal to $\tau_{min}$ (respectively to $\tau_{max}$), hence:
\begin{equation}
\frac{n-1}{\tau_{max}} \le \alpha \le n-1.
\end{equation} 
Nevertheless, as we are going to discuss, the upper limit is usually
a definitely better approximation to the average velocity than
the lower one.
 
In order to understand this point, we use once again the analogy 
with a clock, where here the hour-hand marks the position 
$i_m(t)$ of the impaired firm nearest to the origin and the minute-hand the 
position $i_M(t)$ of the impaired firm farthest from the origin: the key 
ingredients are that the network is braid chain structured and that the firm 
is affected by the consequences of the initial event, being unable
to fully produce, as soon as a single one of the $n$ stocks that it needs is 
unavailable. Therefore, in the long time limit, 
for $1 \ll \tau_{max} \ll n \ll N$, the average velocity of the 
hour-hand is negligible, $\langle i_m(t)\rangle \sim const$, and the most of 
the region between the origin and the minute-hand position, $i_M(t)$, is 
occupied by impaired firms, $\langle \theta_{tot}(t)\rangle \sim 
\langle i_M(t) \rangle-const$. In other words, the long term 
dynamics is dominated by the hare, moving more quickly than the average runner.

In this limit, one can moreover evaluate how far does the hare go in one time 
step, of length $\tau_{min}=1$, hence the approximated average velocity 
$\alpha^*$ of the signal: from a given supplier $j$, the damage can spread up 
to the farthest customer $i$ with $\tau_{i,j}=1$ delay value; at most, it can 
move up to $i=j+n$, with probability ${\cal P}(\tau_{j+n,j}=1)=
1/\tau_{max}$. Generally, the probability that it moves of a distance at most 
$k$ is obtained by taking $\tau_{h,j}>1$ for $h=j+n,j+n-1,\dots, j+k+1$ and  
$\tau_{j+k,j}=1$. To get the average distance, 
one has to sum up each possible value $k=0,1...n$, multiplied
for its corresponding probability, implying that $i=j+k$ is the farthest 
reachable point; hence it follows:
\begin{equation}
\alpha^* = \sum_{k=0}^n 
\frac{k}{\tau_{max}}\left ( 1-\frac{1}{\tau_{max}} \right )^{n-k} 
\simeq n-(\tau_{max}-1),
\label{hare}
\end{equation}
where we used the expansions:
\begin{eqnarray}
\sum_{k=0}^{\infty} \epsilon^k&=&\frac{1}{1-\epsilon}, \nonumber \\
\sum_{k=0}^{\infty} k \: \epsilon^k&=&\epsilon \: \frac{d}{d\epsilon}
\frac{1}{(1-\epsilon)},
\end{eqnarray}
with $\epsilon=(1-1/\tau_{max})$. 

This result is still an underestimation of the effective average velocity in 
the considered limit, since there are corrections of order $\epsilon^n$ which 
make $\alpha$ larger, and we are moreover neglecting that the 
signal can go still faster on more than one time step. However, notice
that, in a large region of the parameter values, one finds 
$\alpha^*>(n-1)(\tau_{max}+1)/\tau_{max}$,
thereby confirming that the dynamics is usually dominated by a faster time 
scale than the one corresponding to the average delay. 

A different approach is worked out in Appendix~B, where the average number of 
impaired firms is explicitly computed as a function of the probability for the 
signal to have propagated of $k$ position in $t$ time steps: this analysis 
confirms the expectation of a linear behavior on a large time window, and
it allows to predict the effective slope.

\begin{figure}
\begin{center}
\includegraphics[width=.6\textwidth,angle=0]{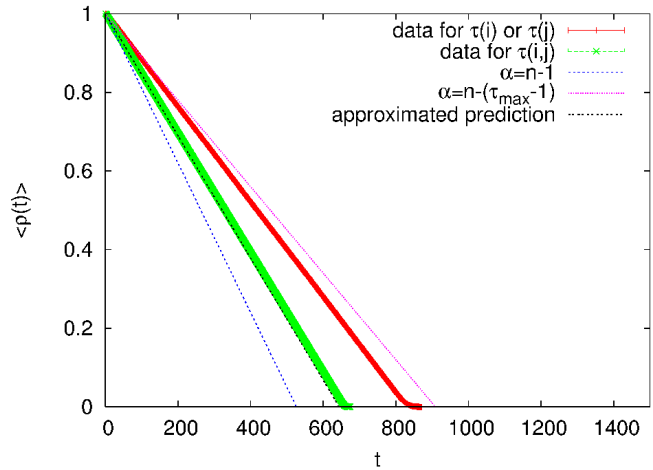}
\caption{The average density of fully active firms, $\langle \rho(t) \rangle$,
as a function of time, in the free model on a braid chain, after an initial 
perturbation of node 1 of duration 
$\tau_c=\tau_{min}$. Here the network size is $N=10000$, and the
number of input-output connections is $n=20$; the average is taken 
over ${\cal N}_s=100$ different configurations of the disorder. We study 
delays depending only upon the customers, $\tau_{i,j}=\tau(i)\: \: \forall j$, 
only upon the suppliers, $\tau_{i,j}=\tau(j)\: \: \forall i$, and upon both of 
them, $\tau_{i,j}=\tau(i,j)$; in all of
the cases they are independent random variables,
identically uniformly distributed according to Eq.~(\ref{Ptauij}), 
with $\tau_{min}=1$ day and $\tau_{max}=10$ days. We compare the data 
with the lower ($\alpha=n-1$), with the upper 
($\alpha=\alpha^*=n-(\tau_{max}-1)$) limits on the behavior, and
with the prediction obtained by the approach explained
in Appendix~B. See the text for details.}
\end{center}
\end{figure}

We present in [Fig. 6] our results on the average density of fully active 
firms, $\langle \rho(t) \rangle$, which is in fact linearly decreasing with 
time on a large window. The obtained behavior is well in agreement with our
expectation, since the data lie between the lower limit 
$\rho_{min}(t) \sim 1-(n-1)t/N$, corresponding to the peculiar case 
in which all the delays are equal to $\tau_{min}$, and the 
upper limit $\rho_{max}(t) \sim 1-\alpha^*t/N$, evaluated using 
the hare argument. Moreover, the approach worked out
in Appendix~B allows to get results on $\langle \rho(t) \rangle$
which are nearly indistinguishable from numerical data. 

Here we also consider delays depending only upon 
the customers, $\tau_{i,j}=\tau(i) \:\: \forall j$, or only upon the 
suppliers, 
$\tau_{i,j}=\tau(j) \: \forall i$,  where $\{ \tau(i)\}$ 
(respectively $\{ \tau(j)\}$) are, as usual, independent identically 
distributed 
random variables, which follow the law Eq.~(\ref{Ptauij}): we get the same, 
slightly slower, average signal velocity in these last two cases. This result 
can be qualitatively explained by the observation that here there is a smaller
number of 
propagation paths of different time lengths. 

We found the same kind of almost linear decay, with velocity quite
correctly approximated by $\alpha^*$, also for different values of the 
model parameters. We moreover checked that, as soon as $\tau_c \ge \tau_{min}$,
the average density shows a dependence on $\tau_c$ only on the first 
$\sim \tau_c$ time steps. Finally, we studied the behavior of the densities 
$\rho(t)$ obtained from single different disorder configurations: this
quantity does usually slightly fluctuate around the average value. 
In particular, the probability distributions of the transient length 
$t_{trans}$ is Gaussian-shaped, becoming rapidly peaked for increasing network 
size $N$: this means that most of the configurations of the delays gives a 
solution which reaches the asymptotic one, $x_i \equiv 0 \: \forall i$, 
in about the same time, of order $t_{trans} \sim \alpha^*/N$. In
other words, the density and $t_{trans}$ are self-averaging quantities,
whose values are independent on the particular disorder configuration
in the $N \rightarrow \infty$ limit.

Anyway, from the point of view of the dependence of the asymptotic solution
upon the disorder, it is interesting to notice that the 
steady state $x_i \equiv 0 \: \forall i$ is surely reached only 
for $\tau_c \ge \tau_{max}$: otherwise, there are configurations of the delays 
in which the smallest variable is $\min_{i,j} \{ \tau_{i,j} \}=
\tau^* > \tau_c$ 
and, for the same argument that we discuss in the deterministic case when 
$\tau_c < \tau_0$, the asymptotic solutions in these cases 
turn out to be periodic of period $\tau^*$, with all the firms simultaneously 
impaired in the first part of length $\tau_c$ of the period.
 
Nevertheless, since the probability of obtaining a 
configuration of the delays in which $\min_{i,j} \{ \tau_{i,j} \}=\tau^*$, 
${\cal P}_{\tau^*}$, approaches rapidly zero for increasing $N$ values,
in the limit of large network size $N$, periodic asymptotic 
solutions are usually found only when the duration of 
the initial damage $\tau_c$ is smaller than the smallest possible 
nearest neighbor time length, $\tau_{min}=1$. 
In fact, this probability ${\cal P}_{\tau^*}$ can 
be straightaway evaluated:
\begin{equation}
{\cal P}_{\tau^*}= 
\left (1-\frac{\tau^*-1}{\tau_{max}} \right)^{N^2} 
\mbox{ for }  \tau_{min} \le \tau^* \le \tau_{max},\: \tau^* \in \mathbb{N}. 
\label{probtozero}
\end{equation}

\subsection{The forced models with $n=1$}
We now turn onto the study of the forced models on a braid chain, defined 
accordingly to Eq.~(\ref{forcedcirc}). When the in/out degree is
$n=1$, and the delays are deterministically chosen all equal to $\tau_0$, 
the BDE system (\ref{syst}) can be reduced to the equation:
\begin{equation}
x_i(t)=\left \{\sum_{k=0}^{N-1} \oo{x}_{i-k} \left [t-(k+1) \tau_0 \right ]
\right \} \vee
x_i(t-N \tau_0),
\label{n1not}
\end{equation}
where also the sum in the parenthesis is to be interpreted in the sense of the 
OR Boolean operator, $\vee$. 
For randomly distributed delays the equation is slightly more complex:
\begin{equation}
x_i(t)= \left[ \sum_{k=0}^{N-1} \oo{x}_{i-k} \left 
(t-\sum_{j=0}^k\tau_{i-j,i-j-1} \right ) \right ] \vee
x_i \left (t- \sum_{j=0}^{N-1} \tau_{i-j,i-j-1} \right ),
\label{n1notr}
\end{equation}
since the time lengths of the paths are here given by the sums of the 
corresponding random variables.

\begin{figure}
\begin{center}
\includegraphics[width=.45\textwidth,angle=0]{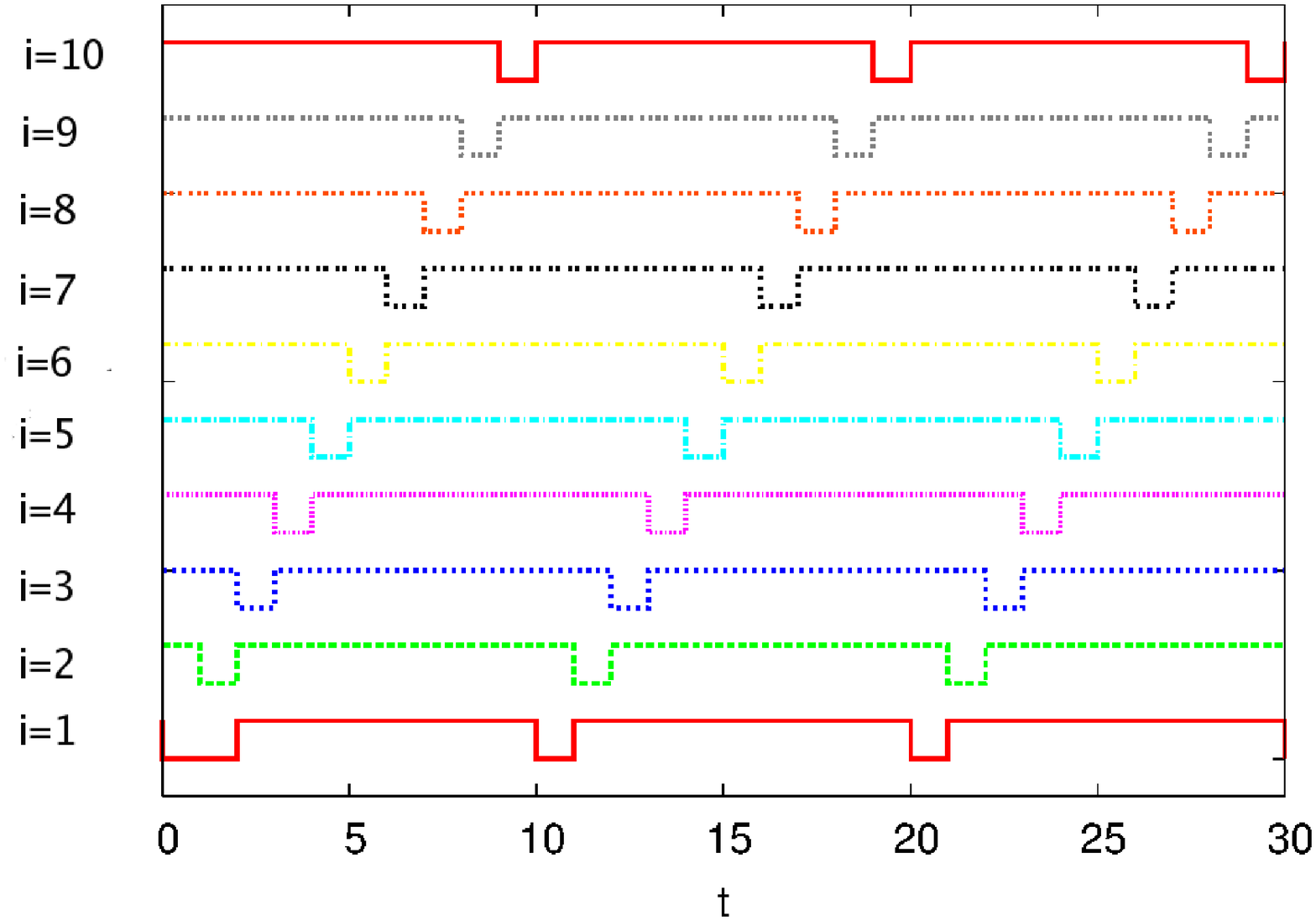}
\includegraphics[width=.45\textwidth,angle=0]{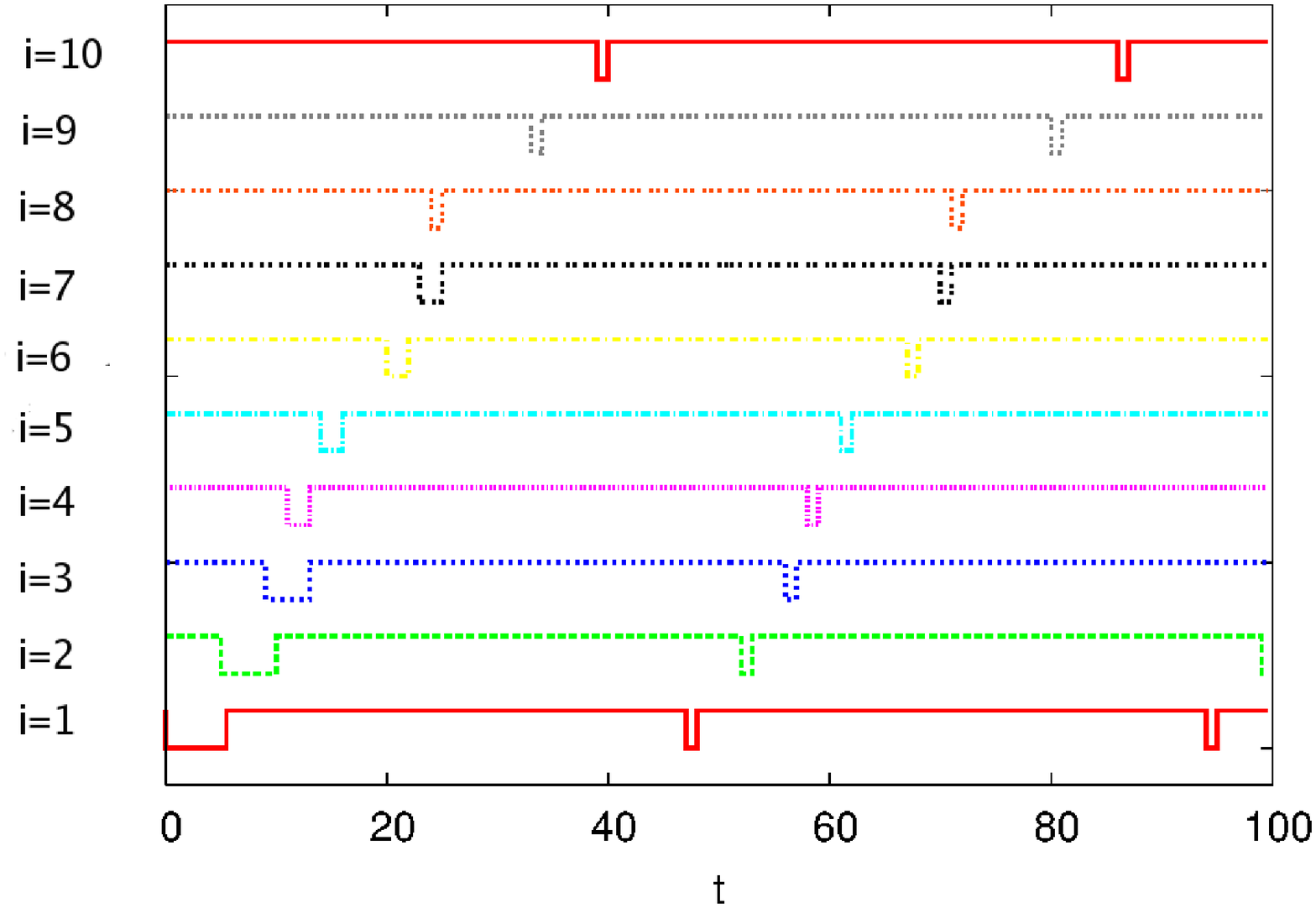}
\caption{Time evolution of the state of the 10 nodes of forced
models on a braid chain, with $n=1$, after an initial perturbation of 
node 1. We plot $x_i(t)$ as a function of $t$ and we shift the solutions
corresponding to different node index $i$ along the $y$ axis in order to 
visualize the spatio-temporal pattern. On the left we present data for the 
deterministic case of delays all equal to $\tau_0=1$ day, whereas on the right 
we look at a particular configuration of random delays 
$\{ \tau_{i,j} \}$, uniformly distributed between $\tau_{min}=1$ day and 
$\tau_{max}=10$ days, with $\min_{i} \tau_{i,i-1}=\tau^*=\tau_{min}$.
The duration of the initial damage is taken to be 
$\tau_c=2\tau_{0}=2$ days and $\tau_c= \langle \tau_{i,j} \rangle=5.5$ days,
respectively. See the text for details.}
\end{center}
\end{figure}

One can distinguish between: 
\begin{itemize}
\item A duration of the initial damage $\tau_c$ smaller than 
the smallest nearest neighbor propagation path, which means:
\begin{itemize}
\item $\tau_c \le \tau_{0}$ in the deterministic model; 
\item $\tau_c \le \min_{i,j} \{ \tau_{i,j} \}=\tau^*$ in the stochastic one,
where the probability to have $ \tau^* > \tau_{min}$ is given by
Eq.~(\ref{probtozero}), and becomes negligible in the limit of large system 
size.
\end{itemize}
Here the systems are conservative and the solutions are the same as in 
the corresponding free models (see [Fig. 2]). In particular, they are periodic 
from the beginning, with a wave of impaired firms which propagates in the 
spatio-temporal pattern. The periods are given by the sums of 
the delays along the whole chain, hence:
\begin{itemize}
\item $\pi=N \tau_0$ in the deterministic case;
\item $\pi=\sum_{i=1}^{N} \tau_{i,i-1}$, with 
$\langle \pi \rangle = N (\tau_{max}+1)/2$,
in the stochastic one.
\end{itemize} 
\item A duration of the initial damage $\tau_c$ larger
than the smallest nearest neighbor propagation path ($\tau_c > \tau_0$ 
or $\tau_c > \tau^*$ in the deterministic or stochastic model, respectively).
As shown in [Fig. 7], here the solutions are not periodic from the very 
beginning. In detail, when the firm in position $i$ is reached by the 
wave of damage, its activity is affected for a duration:
\begin{itemize}
\item no longer than $\tau_0$ in the deterministic case, 
\item no longer than $\tau_{i,i-1}$ in the stochastic one.
\end{itemize}
The asymptotic solutions are surely reached after a 
first cycle in which each firm is impaired and recovers once. Moreover:
\begin{itemize}
\item they are the same as for $\tau_c=\tau_0$ in the deterministic model, 
\item they are the same as for 
$\tau_c = \tau^*$ (usually equal to $\tau_{min}$ in the large size limit), 
in the stochastic one.
\end{itemize} 
\end{itemize}

\subsection{The deterministic forced model with $n>1$}
The behavior of the forced models on a braid chain, with in/out-degree $n>1$,
can be understood in detail when the delays are deterministically taken all 
equal to $\tau_0$. Here, Eq.~(\ref{forcedcirc}) simplifies into:
\begin{equation}
x_i(t)=\mu_i(t) \oo{x}_i(t-\tau_{0}) \vee \left[ \prod_{j=1}^{n}
x_{i-j}(t-\tau_{0}) \right].
\label{forcedcircd}
\end{equation}
Therefore, the firm $i$ maintains its production, as usual, if all its 
suppliers were fully active at the previous time step, but also if some 
of them were impaired and $i$ itself was already impaired: the production 
is recovered after one time step of length $\tau_0$. 

\begin{figure}
\begin{center}
\includegraphics[width=.45\textwidth,angle=0]{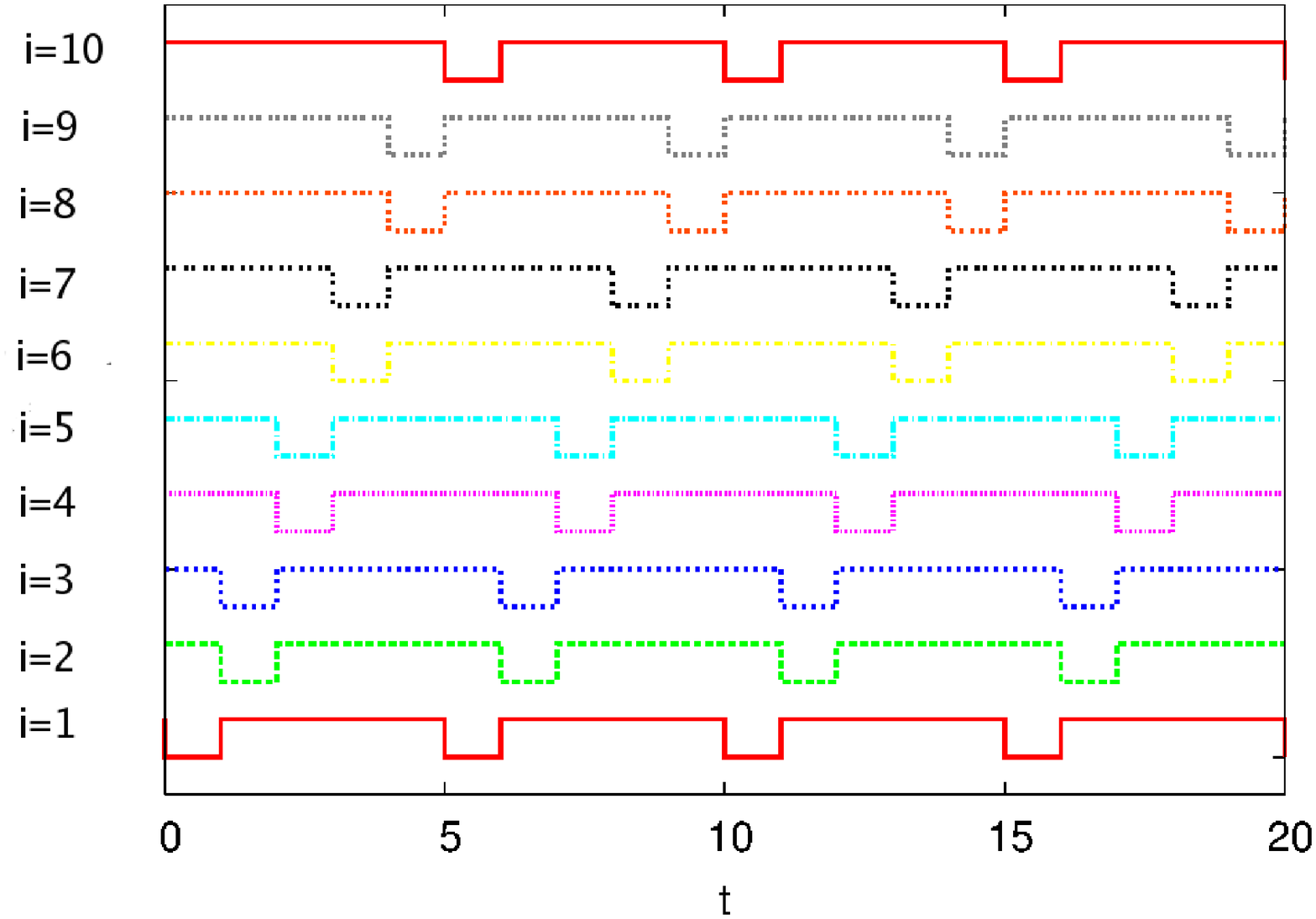}
\includegraphics[width=.45\textwidth,angle=0]{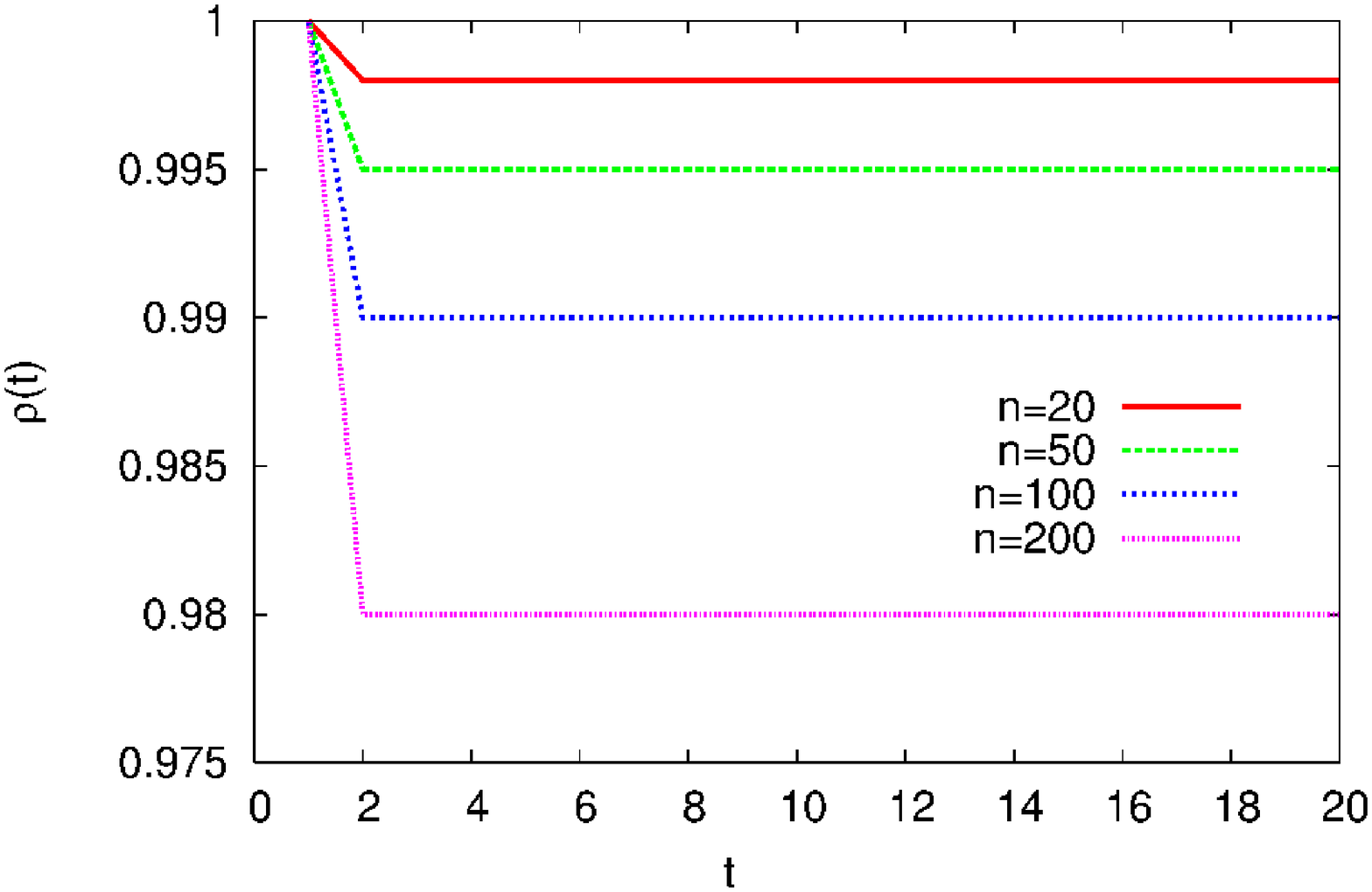}
\caption{On the left, we present the time evolution of the state of the 10 
nodes of the deterministic forced model on a braid chain, with $n=2$, after 
an initial perturbation of node 1. As in the previous considered cases, we 
plot $x_i(t)$ as a function of $t$ by shifting the solutions corresponding to 
different node index $i$ along the $y$ axis. On the right, we show the 
density of fully active firms as a function of time in the same model, for a 
network size $N=10000$, and different values of the input-output connections 
$n$. In both of the cases we take a duration of the initial damage 
$\tau_c=\tau_0=1$ day.}
\end{center}
\end{figure}

Let us take for simplicity $N$ multiple of $n$, and let us start by
considering a duration of the initial damage $\tau_c \ge \tau_0$: 
then, after the first time step, there are 
$n$ firms in consecutive positions along the chain which activity is 
simultaneously impaired; at the subsequent time step, these firms recover 
but the damage propagates to the next $n$ ones.
 
In other words, in this model, as soon as $t>1$, both the hour-hand and the 
minute-hand move with the same constant velocity $n$: the damage does not 
spread, but the activity never recovers completely. Correspondingly, 
the asymptotic solution is periodic of period $N/n$ and the density is just 
$\rho(t)=1-n/N$, as shown by the data presented in [Fig. 8]. 

Though here there is no complete breakdown of the economy, this simple 
case makes evident that, in the considered models, the presence of more 
connections does lead to a less
favorable final outcome, with more impaired firms. This result can be
explained by the fact that more connections do not lead to risk sharing,
since having one impaired supplier is enough to stop a firm production. This 
assumption amounts to say that each supplier of a firm provides a different 
type of goods or services, and that one supplier cannot compensate for the 
loss of another supplier. In such a situation, a firm that is connected to 
more suppliers has higher risks of being indirectly affected by a shock.

When $\tau_c<\tau_0$, one still finds a propagating wave of $n$ impaired 
firms in consecutive position along the chain; nevertheless, here their 
activity is simultaneously impaired only in the first part of length 
$\tau_c$ of the time step. The situation is similar to the one encountered
in the same case in the free models with $n=1$, {\em i.e.} it is the behavior
of the initially attained firm (see Eq.~(\ref{firstfirm})) which propagates
along the network.

\subsection{The stochastic forced model with $n>1$}

\begin{figure}[htpb]
\begin{center}
\includegraphics[width=.6\textwidth,angle=0]{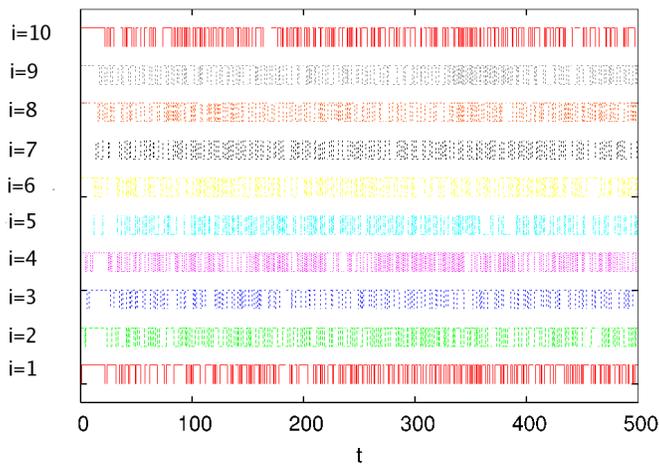}
\caption{Time evolution of the state of the 10 nodes of the stochastic 
forced model on a braid chain, with $n=2$, after an initial perturbation of 
node 1. We plot $x_i(t)$ as a function of $t$ by shifting the solutions 
corresponding to different node index $i$ along the $y$ axis in order to 
visualize the spatio-temporal pattern. As usual, we look at a particular 
configuration of random delays, $\{\tau_{i,j}\}$, uniformly 
distributed between $\tau_{min}$= 1 day and $\tau_{max}$=10 days. The duration 
of the initial perturbation is $\tau_c=\tau_{min}$. Notice that the solution 
does not display any periodicity in the considered time window.}  
\end{center}
\end{figure}

In the presence of stochasticity in the delays, the dynamics of the forced
models on a braid chain, with $n>1$ input-output connections, turns out 
instead to be quite complex. Here we take as usual random independent delays, 
uniformly identically distributed according to Eq.~(\ref{Ptauij}), with 
$\tau_{max}=10$, in the unit of time given by $\tau_{min}=1$ day. The solution 
for a typical disorder configuration, after an initial damage of 
duration $\tau_c=\tau_{min}$, is shown in [Fig. 9]: intriguingly, despite of 
the relatively small network size considered, $N=10$, and of the small $n=2$ 
value, one does not observe any periodicity in the considered time window. 

Since the delays are integer multiple of $\tau_{min}$, the asymptotic solution 
of the system (\ref{syst}), for a given disorder configuration, is surely
constant or periodic because of general results on BDEs 
\cite{DeeGhil84,Mull84,GhilMull85,GhilZalCol08}. Nevertheless, in the
same works it is shown that, for irrationally related delays, BDEs can have 
solutions of increasing complexity, which display a number of jumps in time 
intervals of the same length increasing with time; this peculiar behavior
is observed in particular in the case of 
linear systems such as the simple example:
\begin{equation}
x(t)=x(t-\tau) \bigtriangledown x(t-1).
\end{equation}
Here, $\tau \in (0,1)$ is irrational, and $\bigtriangledown$ means the XOR 
Boolean operator, hence $x(t)$ is true ($x(t)=1$) only if a single one
among the two variables $x(t-\tau)$ and $x(t-1)$ is true. It can moreover
be shown that these aperiodic solutions can be approximated with the
desired accuracy, for increasingly long times, by the periodic solutions of 
nearby BDEs systems with rationally related delays, which approximate 
better and better the irrational ones.

Though a careful analysis would be necessary in order to extend these
results to the present case, we notice that Eq.~(\ref{forcedcirc}), describing 
the model, is equivalent to:
\begin{equation}
\oo{x}_i(t)=\mu_i(t) \left \{
\sum_{j=1}^n {x}_{i}(t_{ij}) 
\bigtriangledown \oo{x}_{i-j}(t_{ij})
\bigtriangledown  \left [
\oo{x}_{i}(t_{ij}) \cdot x_{i-j}(t_{ij}) \right ]
\right \},
\end{equation}
where we label $t_{ij}=t-\tau_{i,i-j}$. The equivalence is due to the Boolean 
relation $\oo{a\vee b}=\oo{a}\wedge\oo{b}=
\oo{a}\bigtriangledown \oo{b} \bigtriangledown (\oo{a} \cdot \oo{b})$,
and it makes clear that the considered system is partially linear. This 
suggests that the long periods observed in the behavior of the solutions can 
be related to the known results on the existence of small, partially linear 
BDEs, with rationally related delays, which approximate the increasingly 
complex solutions of systems with irrationally related delays. 

In order to better characterize our numerical findings, we find interesting
to study various quantities:
\begin{itemize}
\item The length of the transient, $T_{trans}$, rigorously defined just as  
the time elapsed before that periodicity settles in. In 
the previously considered stochastic free models, $T_{trans}$ is the same as 
the time $t_{trans}$, which the average density takes to reach the asymptotic 
zero value. Instead, as we are going to discuss, in the present case one
finds $T_{trans} \gg t_{trans}$, and it is therefore important to distinguish
between the different meanings of the two transients.
\item The period $\pi$ of the asymptotic solution itself, possibly equal
to zero if it is constant.
\item The period-averaged asymptotic density of fully active firms, 
$\rho_{\infty}$, defined as
\begin{equation}
\rho_{\infty}=\frac{1}{\pi}\int_{T_{trans}}^{T_{trans}+\pi}\rho(t)dt.
\end{equation}
\end{itemize} 
   
\begin{figure}[htpb]
\begin{center}
\includegraphics[width=.3\textwidth,angle=0]{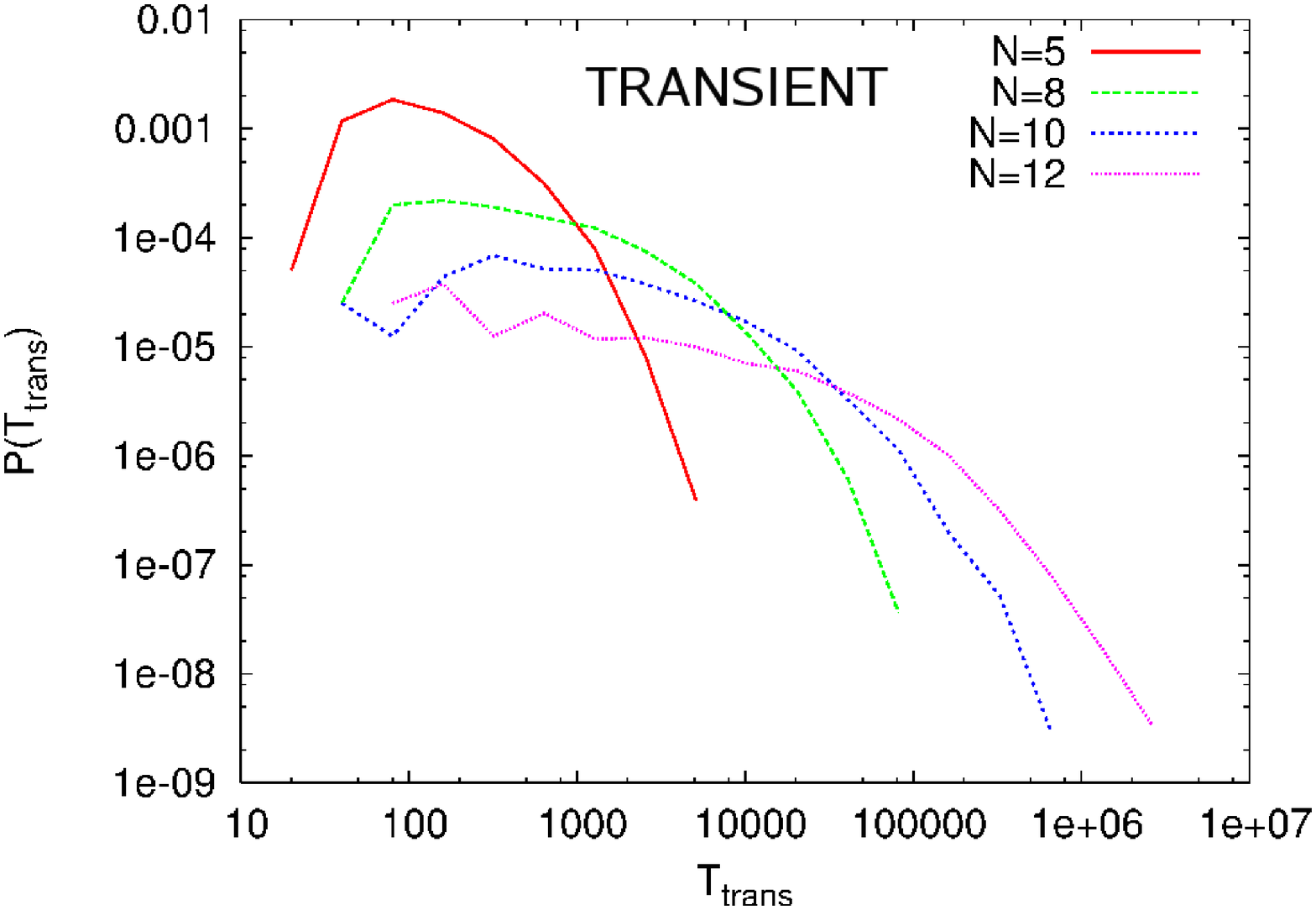}
\includegraphics[width=.3\textwidth,angle=0]{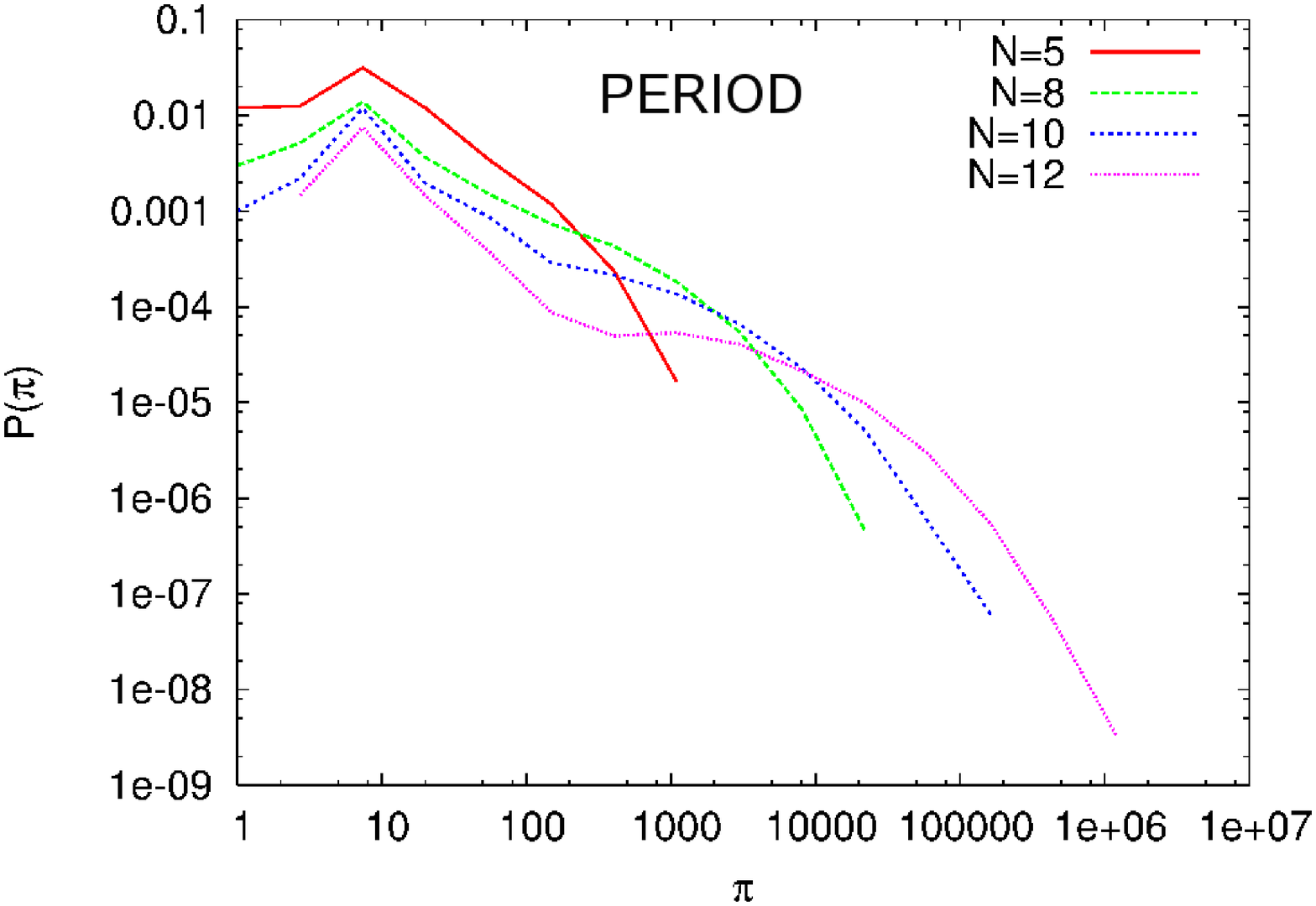}
\includegraphics[width=.3\textwidth,angle=0]{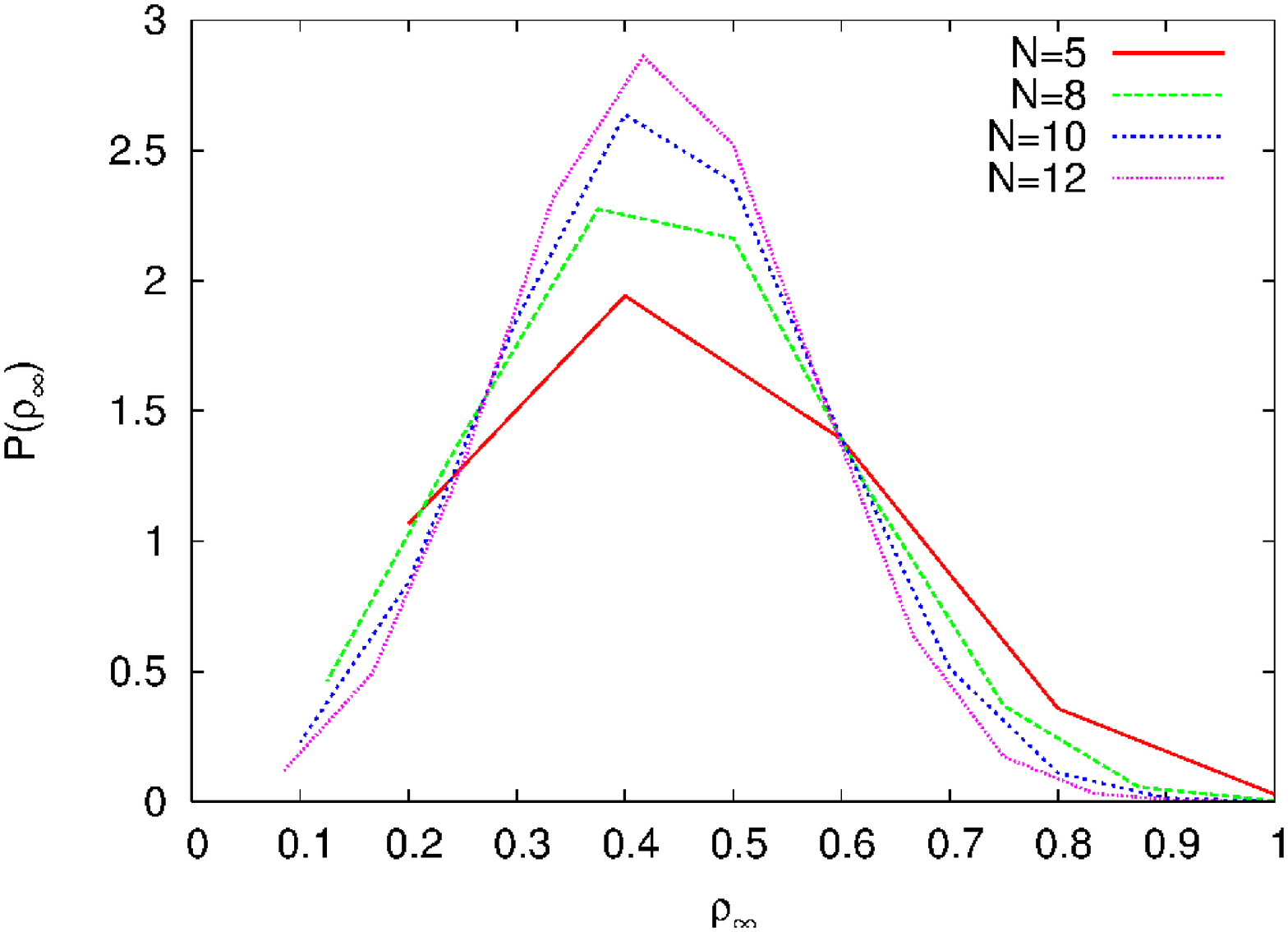}
\caption{Data on the forced model on a braid chain with random delays, 
uniformly distributed between $\tau_{min}=1$ day and $\tau_{max}=10$ days. 
Here we study the solutions of small systems of $N \sim 10$ equations with 
$n=2$, after an initial perturbation of node 1 of duration $\tau_c=\tau_{min}$.
The results are obtained from ${\cal N}_s=1000$ different disorder 
configurations. On the left we plot the probability distribution of the 
transient, $P(T_{trans})$, in the center the probability distribution of the 
period, $P(\pi)$, and on the right the probability distribution 
of the period-averaged asymptotic density of fully active firms, 
$P(\rho_{\infty})$. Both $P(T_{trans})$ and $P(\pi)$ are presented
in log-log scale, in order to make evident that they are not negligible
also at very large values. See the text for details.}  
\end{center}
\end{figure}

These quantities can be straightaway exactly measured up to relatively small 
system sizes $N \sim 12$. We present in [Fig. 10] our results on their 
probability distributions, $P(T_{trans})$, $P(\pi)$, and $P(\rho_{\infty})$ 
respectively, as obtained by considering ${\cal N}_s=1000$ different 
configurations of the random delays. 

The most striking feature displayed by the data is that both the transient, 
$T_{trans}$, and the period, $\pi$, of the solution can increase very fast 
with the network size $N$. In order to make it evident, we have plotted the 
resulting probability distributions in log-log scale
(see [Fig. 10a] and [Fig. 10b]): for instance, $P(T_{trans})$ is not
negligible for values of the transient as large as $T_{trans}\sim 10^6$ (in
units of $\tau_{min}$), in systems of $N=12$ nodes. In detail, we found that 
the average transient diverges exponentially with $N$, 
$\langle T_{trans} \rangle \propto \exp(const N)$, and that, at least for the 
considered $N$ values, the average period $\langle \pi \rangle$ displays a 
similar behavior. 

The data on the asymptotic density of fully active firms, $\rho_{\infty}$, 
averaged over the disorder dependent period $\pi$, are shown in [Fig. 10c]: its
probability distribution, $P(\rho_{\infty})$, turns out to be bell-shaped, 
and it becomes more peaked around the mean value for increasing network 
size $N$, therefore suggesting that the quite complex dynamics does not imply 
the unpredictability of the behavior of macroscopic intensive quantities, at 
least in the large network size limit. 

In fact, whereas to predict the details of the solutions for
given disorder configurations of the random delays seems to be a quite hard
task, one can use an approach similar to the one of Section~3.5 to obtain
the expected behavior of various average quantities.

For simplicity, we limit the analysis to the case in which the duration of the 
initial damage is not smaller than the smallest nearest neighbor 
propagation path, $\tau_c \ge \tau_{min}$. In the present extension of
the clock argument, we assume that the minute-hand marks the position $i_M(t)$ 
of the firm farthest from the origin, {\em i.e.} from the node where the 
initial damage acted, which has already been impaired at least for 
a duration $\tau_{min}$ at time $t$. We recall that the network is braid chain 
structured and that the production of a given firm, apart from the initially 
damaged one, is impaired for the first time as soon as one of the $n$ stocks 
that it needs is unavailable. For these reasons, in the limit 
$1 \ll \tau_{max} \ll n \ll N$, at large $t$, the signal is propagating across 
the chain with velocity still well approximated by Eq.~(\ref{hare}):
\begin{equation} 
\langle i_{M}(t) \rangle \sim \alpha^* t=[n-(\tau_{max}-1)]t. 
\end{equation}

The main difference with the previous stochastic free model is that here, 
because of the external rescue inputs, only a fraction $s^*$ of the firms 
between the origin and $\langle i_M(t) \rangle$ remains in average impaired 
during the time step of length $\tau_{min}$. Since the delays are taken 
independently and identically distributed along the whole chain, at large 
enough times this fraction is constant, and the damage spreads, up to invade 
the whole network, according to:
\begin{equation} 
\langle \theta_{tot}(t) \rangle \sim s^* \alpha^* t=s^*[n-(\tau_{max}-1)]t, 
\mbox{ for } 1 \ll t {\lesssim} t_{trans}.  
\end{equation}
Correspondingly, the average density of fully active firms is again linearly 
decreasing with $t$, up to the time $t_{trans}$, at which it reaches a nearly 
constant asymptotic value: 
\begin{equation}
\langle \rho(t) \rangle \sim 
\left \{
\begin{array}{lcl}
1- ({s^*}/{N}) \alpha^* t, & \mbox{ for }&
1 \ll t {\lesssim} t_{trans} \\
1-{s^*}, &\mbox{ for } & t \ge t_{trans}. 
\end{array}
\right.
\end{equation}

Hence, the effective transient, $t_{trans}$, relates to dynamics of 
$\langle \rho(t) \rangle$; in fact, it is the time at which the
minute-hand reaches again the origin after a whole tour, and it is
therefore of the same order as the time before that the density became zero, 
since the system attained the asymptotically stable configuration, 
$x_i \equiv 0 \: \forall i$, in the stochastic free model:
\begin{equation}
t_{trans}\sim \frac{N}{\alpha^*}=\frac{N}{n-(\tau_{max}-1)}. 
\label{ttransforc}
\end{equation}
Therefore, whereas in the free model one has $t_{trans}=T_{trans}$, in the 
forced case it is important to distinguish between the 
transient $T_{trans}$, rigorously defined as the time elapsed before that
periodicity settles in, that, as we showed numerically, is in average 
exponentially increasing with the network size $N$, and the definitely smaller 
effective transient $t_{trans}$, which can be more generally defined as the 
time at which macroscopic observables attain nearly constant average values. 

\begin{figure}[ht]
\begin{center}
\includegraphics[width=.8\textwidth,angle=0]{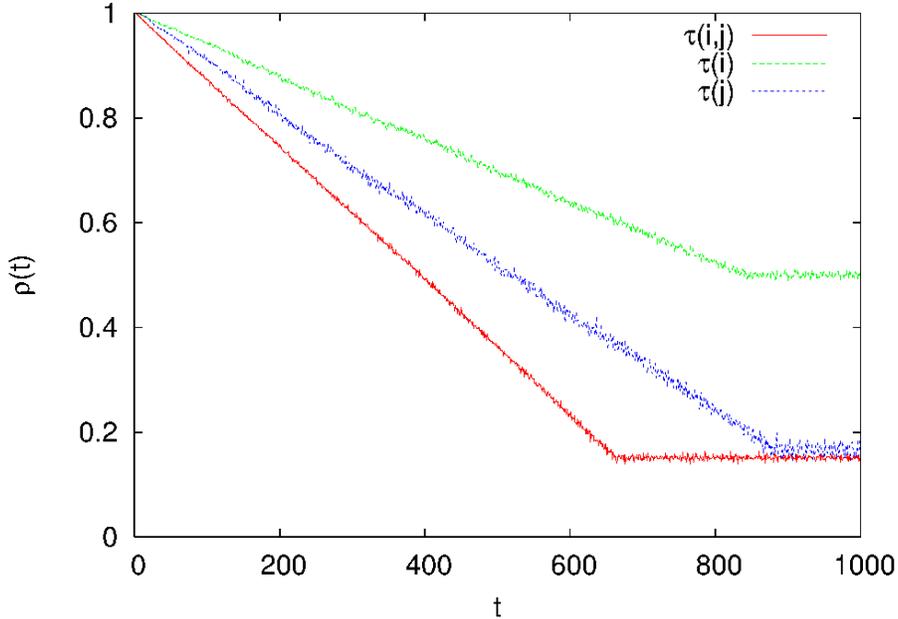}
\caption{The density of fully active firms $\rho(t)$ as a function of time, in 
the forced model on a braid chain with random delays, uniformly distributed 
between $\tau_{min}=1$ day and $\tau_{max}=10$ days, after an initial 
perturbation of node 1 of duration $\tau_c=\tau_{min}$. The network size is 
$N=10000$ and the in/out-degree is $n=20$. The results are for a single 
(typical) disorder configuration. We study delays depending only upon the 
customers ($\tau_{i,j}=\tau(i) \:\: \forall j$), delays depending only upon 
the 
suppliers ($\tau_{i,j}=\tau(j)\:\: \forall i$), and delays depending upon both 
of them ($\tau_{i,j}=\tau(i,j)$). See the text for details.}
\end{center}
\end{figure}

We present in [Fig. 11] the observed behavior of the density of fully active
firm as a function of time, after an initial perturbation of node 1 of
duration $\tau_c=\tau_{min}$, for a large network size $N=10000$ and 
in/out-degree $n=20$. In agreement with the picture emerging from 
the previous discussion, we find a nearly linear decay on a quite large time 
window, followed by small oscillations around the asymptotic value, which is 
reached in a time $t_{trans}$ of the same order of the estimation of the 
effective transient, $t_{trans} \sim 900$ (in units of $\tau_{min}$), that
one gets from Eq.~(\ref{ttransforc}). From this point of view, we also
recall that our evaluation of $\alpha^*$ is an underestimation
of the damage spreading velocity, hence the corresponding value of
$t_{trans}$ is to be considered an upper limit.

The data are for a single, typical, disorder configuration, in order to make 
evident that the fluctuations of $\rho(t)$ around the average are small, as it 
can be expected for a macroscopic intensive quantity: we checked in particular 
that both the effective transient length, $t_{trans}$, and the asymptotic mean 
value of the density, $1-s^*$, are usually the same for different choices of 
the random variables. 

Besides considering delays depending both upon the customers and the
suppliers, we present, in the same [Fig. 11], data on the system with
delays depending only upon the customers, $\tau_{i,j}=\tau(i)\: \: \forall j$, 
or 
only upon the suppliers, $\tau_{i,j}=\tau(j) \: \: \forall i$: the last two 
cases 
are characterized by compatible values of the effective transient,
$t_{trans}$, which is slightly larger than in the first one. This can be 
explaining by noticing that, as in the stochastic free model, 
when $\tau_{i,j}$ depend on both of the indexes, there are more 
concurrent paths of different time lengths in the system, hence the
average signal propagation velocity is higher.

The nearly constant asymptotic mean value of the density, $1-s^*$, is instead 
definitely larger when the delays depend only upon the customers, whereas the
other two considered cases give very close results. In fact, 
if $\tau_{i,j}=\tau(i)\: \: \forall j$, Eq.~(\ref{forcedcirc}) becomes:
\begin{equation}
x_i(t)=\mu_i(t) \oo{x}_i[t-\tau(i)] \vee \left\{ \prod_{j=1}^{n}
x_{i-j}[t-\tau(i)] \right\}.
\label{forcedcircs}
\end{equation}
Since the activity of firm $i$ is impaired as soon as a single
one of the stocks of products that it needs is lacking, the fact
that the delays in production are assumed to be independent on the
particular customer implies that there are much less combinations of the 
delays which can result in shortening the production of a given firm, 
{\em i.e.} the average fraction $s^*$ of contemporaneously impaired firms, 
in the region reached by the spreading of the initial perturbation, is 
smaller.

Moreover, in this particular case, one can make a further step in the 
analysis:
in fact, Eq.~(\ref{forcedcircs}) implies that, after the transient
$T_{trans}$, when averaging the dynamics on an enough large time window, 
each firm is roughly impaired for one half of the time;
for large $N \gg n$, this turns out to be equivalent to say that there
is in average one half of impaired firm at each time step, {\em i.e.}
$s^* \simeq 0.5$. This is just the result on the mean asymptotic value of the
density observed in [Fig. 11], which we checked to be independent upon 
the considered $n$. We also found that the solutions for
 $\tau_{i,j}=\tau(i)\: \: \forall j$
usually display a more regular behavior than in the other situations.

\section{Random graph}

\subsection{Topology}
We are interested in the topology induced by a directed 
random graph, which is obtained when the elements of the 
connectivity matrix $A$ are chosen accordingly to Eq.~(\ref{PAij}):
this is a quite well known random structure \cite{Ka90,LuSe09}, hence we start 
by summing up some of the main results.

Random graphs have been initially introduced by Erd\H{o}s and R\'enyi about
50 years ago \cite{ErRe59,ErRe60,ErRe61}, and they have been extensively
studied more recently, when also a number of related models have been
considered \cite{Bo98,Wa99,AlBa02,Ne03}. One gets an undirected 
Erd\H{o}s-R\'enyi random graph, $G(N,p)$, by taking the edges, connecting
each possible pair $(i,j)$ of the $N$ nodes, independently identically
distributed with probability $p$. In our notation, the matrix $A$ is 
symmetric, since the event $A_{ij}=1$ implies the event $A_{ji}=1$, and 
vice-versa. 

The total number of pairs of nodes is $N(N-1)/2$, and each edge 
contribute to the degree of 2 nodes, therefore the average degree
is straightaway $z=\langle k \rangle=(N-1)p\simeq N p$. More in detail, the 
probability distribution of the degree $k$ is binomial:
\begin{equation}
{\cal P}(k)=\left (
\begin{array}{c}
N-1\\
k\\
\end{array}
\right )
p^k (1-p)^{N-1-k} \simeq \frac{z^{k}}{k!} {e^{-z}},
\label{poisson}
\end{equation}
and it converges to a Poisson distribution with average value $z$ in the large
$N$ limit. One moreover defines the connected components ${\cal S}_c$ 
of the graph as the ensembles of nodes such that, from each node 
$i\in {\cal S}_c$, there is 
at least one path, across nodes belonging to the component itself, 
which reach each other possible node $j$ in the same connected component:
\begin{equation}
\forall (i,j) \in {\cal S}_c \Rightarrow \exists \: 
{\cal C}_{ij} : A_{ih_1} \cdot A_{h_1h_2}
\cdot ... \cdot A_{h_l j} \neq 0 \hspace{.3in} h_1, h_2, ...,h_l 
\in {\cal S}_c.
\end{equation} 

Therefore, the average size of connected components $S_c$ can 
be evaluated by starting from a randomly chosen node, and then looking at the 
number of its first neighbors $z_1$, of its second neighbors $z_2$, and so on.
For a Poisson distribution, which is a quite peculiar case, one simply has:
\begin{eqnarray}
z_1&=&\sum_{k=0}^{\infty} k \frac{z^k}{k!}e^{-z}=z \nonumber \\
z_2&=&\sum_{k=0}^{\infty} k(k-1) \frac{z^k}{k!}e^{-z}=z^2 \nonumber \\
z_l&=&\frac{z_2}{z_1} z_{l-1}=\left ( \frac{z_2}{z_1} \right )^{l-1} z_1=z^l,
\label{zallat}
\end{eqnarray}
hence one obtains:
\begin{equation}
S_c=\sum_{l=0}^{\infty}z^l=\frac{1}{1-z} \mbox{ for } z<1.
\label{serie}
\end{equation}
Correspondingly, one finds that, in the $N\rightarrow \infty$ limit, 
$\langle S_c \rangle$ diverges for $z \nearrow z_c=1$. Above this
``critical'' value it appears a giant connected component,
${\cal S}_{gc}$, which contains a finite fraction of the nodes, 
$S_{gc}=s_{gc}N$, such as it is usually observed in real networks. This 
``phase transition'' was already enhanced in the pioneering papers by 
Erd\H{o}s and R\'enyi \cite{ErRe59,ErRe60,ErRe61}, and was subsequently
studied in great detail both from the mathematical point of view \cite{Bo98}, 
and from the physical point of view \cite{AlBa02}. 

A simple argument for evaluating $s_{gc}$ for $z>z_c$ runs as 
follows \cite{Ka90,BaDoSo03}: let us assume that they still exist finite size 
connected components for $z>z_c$; then, it is $s_{gc}=1-v(z)$, where $v(z)$ is 
the probability for a random node of being in a finite size connected 
component. Therefore, $v(z)$ is the probability that all the nodes which can 
be reached from a random one is finite, but each node generates a Poisson 
branching process (the number of its first neighbors is Poisson distributed),
hence $v(z)$ must be the solution of the transcendental equation:
\begin{equation}
v(z)=\sum_{k=0}^{\infty} \frac{[zv(z)]^k}{k!}e^{-z}=e^{z[v(z)-1]}
\label{my_v}
\end{equation}
This result can also be derived in the 
framework of the probability generating functions \cite{NeStWa01}, which is 
well suitable for applications to more general degree distributions and to 
directed network, as recalled in Appendix~C. 

In fact, the directed random graph that we are considering, $D(N,p)$, is a 
simple generalization of the Erd\H{o}s-R\'enyi model: each directed link is 
chosen independently with the same probability $p$ among the $N(N-1)$ possible 
ones. Hence the average in/out-degree 
$z=\langle k_{in} \rangle =\langle k_{out}
\rangle $ is again given by $z=(N-1)p\simeq Np$, and ${\cal P}(k_{in})=
{\cal P}(k_{out})$ are still described by Eq.~(\ref{poisson}).
Notice that, here, the average in/out-degree $z$ corresponds to the 
connectivity
$n$ in the deterministic braid chain structure previously considered, and
furthermore that the total average degree of the node is $2z$, {\em i.e.}, if 
we were to transform the directed random graph in an undirected one, by 
interpreting each link as an edge, we would get an RG with average degree 
equal to twice the average in/out-degree of the starting DRG. 

\begin{figure}[htpb]
\begin{center}
\includegraphics[width=.6\textwidth,angle=0]{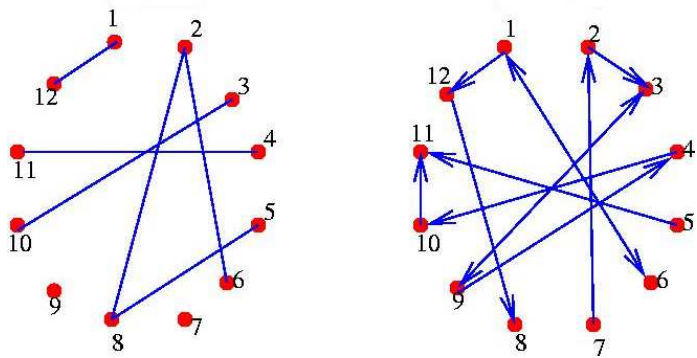}
\end{center}
\caption{A possible realization of a random graph (on the left) and 
of a directed random graph (on the right) with the same small
network size of $N=12$ nodes, and the same $p \sim 1/N$ (hence $z \sim 1$) 
value. The number of the edges in the RG is one half of the
number of the directed links in the DRG.}
\end{figure}

We present in [Fig. 12] an example of the kind of structures which can be
obtained: we compare in particular an RG and a DRG of equal small size 
$N=12$, with the same $p \sim 1/N$ (hence $z \sim 1$)
value. This figure makes evident the completely different topology
with respect to the brain chain, where for the same connectivity value
$n=1$ one would find a single connected component with directed links
$i \rightarrow i+1$ (see [Fig. 1]). Instead, in the shown example,
looking at the case of the undirected random graph for
simplicity, we observe two isolated nodes, three connected component of size 
2, and one connected component of size 4. 

The figure also makes evident that in the directed case 
\cite{BaDoSo03,NeStWa01,DoMeSa01,Broetal00}, since the existence 
of a path connecting $i$ to $j$ does not usually imply the existence of a path 
connecting $j$ to $i$, for each given node one can define:
\begin{itemize}
\item the out-component, which is the set of all nodes
that can be reached from it;
\item the in-component, which is the set of all nodes
from which it can be reached;
\item the strongly connected component, which is the set of all the nodes
that can be reached from it $and$ from which it can be reached, hence
the intersection of the in-component and the out-component;
\item the weakly connected component, which is the set of all the nodes
that can be reached from it $or$ from which it can be reached,
hence the union of the in-component and the out-component; this corresponds
to the connected component of the node in the graph obtained
by disregarding the directionality.
\end{itemize}

\begin{figure}[htpb]
\begin{center}
\includegraphics[width=.6\textwidth,angle=0]{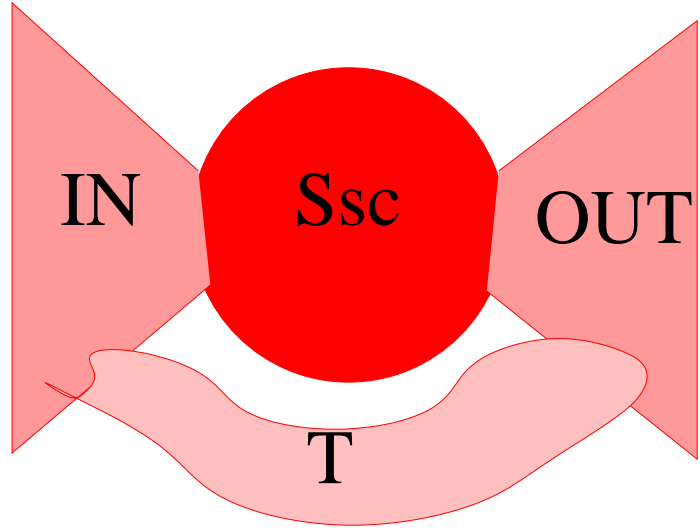}
\end{center}
\caption{A sketch of the bow tie topology structure characteristic of
directed network \cite{DoMeSa01}: with respect to the notation used in the 
text, the two bows correspond to the giant components 
${\cal I}\setminus{\cal S}_{sc}$ and ${\cal O}\setminus{\cal S}_{sc}$ 
respectively, whereas the tie represents the giant strongly connected 
component ${\cal S}_{sc}={\cal I}\cap{\cal O}$.
Here we consider the most general case in which 
${\cal W}={\cal I}\cup{\cal O}\cup{\cal T}$, where ${\cal T}$ contains in 
particular the paths linking the two bows without passing across the tie, 
which have been observed in real network such as the web \cite{Broetal00}.}
\end{figure}

The analogous of the ``phase transition'' in an undirected random graph can now
be characterized by the possible formation: 
of a giant in-component 
${\cal I}$, containing $I=s_IN$ nodes; of a giant out-component, 
${\cal O}$, containing $O=s_ON$ nodes; of a giant strongly connected 
component, 
${\cal S}_{sc}={\cal I}\cap{\cal O}$, containing $S_{sc}=s_{sc}N$ nodes; 
of a giant weakly connected component ${\cal W}$, containing $W=s_wN$ nodes.
In fact, one usually expects to observe two different transitions, 
{\em i.e.} two different
abrupt changes in the properties of the system in the large $N$ limit:
at the lower average in/out-degree $z^w_c$, corresponding to the critical 
average degree in the undirected graph, where it appears the giant weakly 
connected component, and at the higher average in/out-degree $z^d_c$, which
gives the effective transition in the directed network, where
the other giant components appear contemporaneously. 

For $z\ge z^d_c$ the resulting picture is captured by the bow tie structure 
sketched in [Fig. 13], which has been observed in many different real networks
\cite{Broetal00,Dietal09}: notice that the appearance of the giant in-component
(respectively out-component) corresponds to the divergence of the
number of nodes which can be reached from a given one (respectively
from which a given one can be reached), {\em i.e.} a node is in the
giant in-component if its out-component diverges, and vice-versa.
Moreover, above the transition, it can be necessary to introduce one more giant
component in order to fully characterize the topology, since one can have
${\cal W} \supset ({\cal I}\cup {\cal O})$: this is in particular
the case when there are directed paths between ${\cal I}$ and ${\cal O}$
which do not pass across the strongly connected component ${\cal S}_{sc}$,
as it has been found in the structure of the web \cite{Broetal00}, which can
be seen as a directed graph characterized by power law in/out-degree 
distributions.

The formalism of the probability generating functions for directed random 
graphs \cite{NeStWa01,DoMeSa01} turns out to be particularly simple when the 
in/out-degree distribution factorizes, 
${\cal P}_{k_{in},k_{out}}={\cal P}_{k_{in}}{\cal P}_{k_{out}}$,
since the in and out components are independent in the large 
network size $N$ limit, so that one has simply 
$s_{sc}=s_{I}s_{O}$. 
For the considered case of ${\cal P}(k_{in})={\cal P}(k_{out})$ given by 
Eq.~(\ref{poisson}), the transition does still occur at the 
critical  average in/out-degree $z^d_c=1$, and the giant 
in-component and out-component have the same size:
\begin{equation}
s_{I}=s_O=1-v(z),
\end{equation}
where $v(z)$ is once more given by the solution of Eq.~(\ref{my_v}).
Then, $s_{sc}=(1-v(z))^2$, whereas the fraction of nodes in 
${\cal I}\setminus {\cal S}_{sc}$, and equivalently in
${\cal O}\setminus {\cal S}_{sc}$, is equal to $v(z)(1-v(z))$:
the three regions in the bow tie have roughly the same size in the 
thermodynamic limit, as it has been observed in real networks 
\cite{Broetal00}, for $v \sim 1/2$, hence $z \sim \log 4$. 
Moreover, one simply has $z^w_c=z^d_c/2$: since when $v(z)$ is solution of
Eq.~(\ref{my_v}) one has $v(2z)=v^2(z)$, it follows that, in the region
$z>z^d_c$, $s_w=1-v^2(z)=s_I+s_O-s_{sc}$, which confirms that here
${\cal W} = {\cal I}\cup {\cal O}$.

\begin{figure}[htpb]
\begin{center}
\includegraphics[width=.6\textwidth,angle=0]{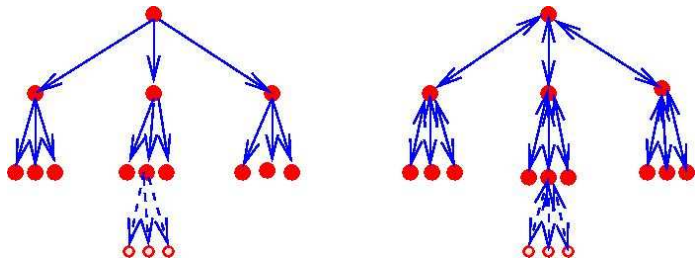}
\end{center}
\caption{The local tree-like structure characteristic of
Erd\H{o}s-R\'enyi random graphs: on the left we consider 
the damage spreading on a directed random graph $D(N,p)$, with
$z=\langle k_{out} \rangle = \langle k_{in} \rangle=3$, and on the
right the one in the undirected network with the same $p$ value, 
$G(N,p)$.}
\end{figure}

The other key ingredient of the topology of Erd\H{o}s-R\'enyi random graph 
is that their structure, for not too large $p$-values, is tree-like:
this is rigorously correct for the finite size connected components,
both below and above the critical point, since it can be shown
that here the probability of closed loops approaches zero in the
limit $N \rightarrow \infty$, but it is locally valid also within the
giant connected components, where the closed loops can be neglected in a first 
approximation. We compare in [Fig. 14] these kind of structures
for a DRG and for an RG with the same $z=3$ value: notice that
in the first case one has $z_2=\langle k_{in} \rangle \langle k_{out} \rangle=
z^2$, where in the second case one gets the same results, $z_2=z^2$,  
from Eq.~(\ref{zallat}). In other words, in random directed network the
probability to have both the link $i \rightarrow j$ and the link
$j \rightarrow i$ is $ \sim p^2$, and it is therefore negligible in the
region $p \ll 1$ that we are studying, whereas in undirected random graphs
we have to consider the edges emerging from the node apart the
one along which the signal arrived, in order to make a meaningful
computation. 

As we are going to discuss, this difference is evident in the results
on the damage spreading at short time in the BDE models on random graphs, 
that can be predicted quite accurately on the basis
of the local tree-like topology. Moreover, when considering randomly 
distributed delays, though also here two paths with the same space length 
along the network do not usually have the same time length, because
of the local tree-like topology, the number of 
different concurrent paths connecting two given nodes does not increase as 
fast with the average in/out-degree as in the previously considered 
deterministic braid chain. 

Finally, we note that on the behavior of $D(N,p)$ there are detailed rigorous 
mathematical results \cite{Ka90} (see also \cite{LuSe09} and references 
therein), which would require a definitely deeper analysis for being 
summarized.

\subsection{The free models with equal delays}
We introduce the free models, defined by Eq.~(\ref{free}), with
all the delays equal to the same time unit $\tau_0=1$ day,
on the directed random graph, $D(N,p)$, and we study as usual the consequences 
of a damage initially destroying a single firm for a 
duration $\tau_c$. For simplicity, we limit the analysis to the case 
$\tau_c=\tau_0$. 

In the presence of stochasticity in the network structure, the BDE system
(\ref{syst}) can no more be reduced, and one has in principle to solve a set 
of $N$ equations in $N$ variables, which can be rewritten in the form:
\begin{equation}
x_i(t)=\mu_i(t) \prod_{j\in {\cal I}(i)} x_j(t-\tau_0) 
\hspace{.3in} i=1,...,N.
\label{newform}
\end{equation}
Here the product, which means as before the AND Boolean operators $\wedge$, 
runs over all the values of the index $j$ labeling firms which are suppliers 
of $i$, hence belonging to the set of nodes 
${\cal I}(i)= \{ j : A_{ij} \neq 0 \}$, which corresponds to the definition
of the node in-component ${\cal I}(i)$; 
the in-degree of the node is the total number of suppliers of firm $i$,
{\em i.e.} $I(i)=k_{in}(i)$.
Analogously, the out-component of the node, 
${\cal O}(i)=\{ j: A_{ji} \neq 0 \}$, 
corresponds to the firms which are customers of $i$, which total number is 
$k_{out}(i)$.

Therefore, Eq.~(\ref{newform}) makes evident that the properties of the
BDE system strongly depend on the probability distributions of the 
in/out-degree, which in the considered case are independent, and described by 
Eq.~({\ref{poisson}), with mean values 
$\langle k_{in}(i) \rangle = \langle k_{out}(i) \rangle=(N-1)p=z$. 
Hence, for increasing $z$ values, the properties of 
the solutions will reflect the appearance of giant in/out-connected components 
in the graph, and one will find different kinds of solution for average 
in/out-degree lower or higher than the critical value $z^d_c=z_c=1$.  

In detail, the damage spreading in this simple BDE model can be understood
with the same argument used for evaluating the average size of the connected
components in the graph: at $t=1$, the signal propagates from the node $i$, 
occupied by the initially damaged firm, to its first neighbors, which
average number is $z_1=z$; at $t=2$, it reaches the second neighbors, 
which average number is $z_2=z^2$, and so on. From
Eq.~(\ref{zallat}) it follows that at time $t$ the damage did spread
up to reach in average $z^t$ nodes, hence:
\begin{equation}
\langle \theta_{tot}(t) \rangle \simeq z^t \mbox{ for }
t \ll \log N /\log z, \hspace{.5in} \mbox{ (DRG)}.
\label{st_drg}
\end{equation}
This implies that the average number of firms which production is impaired 
increases with time only if $z>z^d_c=1$, {\em i.e.} only if the graph is above 
its transition point. 

The argument does make use of the local tree-like
structure, since we are implicitly assuming that the probability for two
node reached at a given time step of being themselves connected by a different
path, {\em i.e.} the probability of closed loops, can be neglected. In fact,
this argument breaks down roughly at the time where the signal  
propagated to the whole connected component to which the initial node $i$ 
belongs, and one can find different situations:
\begin{itemize}
\item $i$ is in a connected component containing a small number of nodes, 
$S_c=0(1)$; in this case one expects that there are no loops, hence 
the system can completely recover from the initial damage, {\em i.e.} 
$\theta_{tot}(t)=0$ in the large time limit; a randomly
chosen initial node $i$ will belong almost surely to a connected 
component containing a finite number of nodes, in the 
$N\rightarrow \infty$ limit, for $z<z^d_c$, and with probability 
$v(z)$ for $z>z^d_c$;
\item the graph is above the critical point and $i$ belong to a giant
connected component; because of the
presence of closed loops, here a finite fraction
of the firms will be impaired in the asymptotic
solution, and the economic network does never recover completely
from the initial destruction event. 
\end{itemize}

Such results can be explained by noticing that, in our model, there are 
firms which have no ``clients'' in the network. This surprising possibility 
simply arises from the fact that some firms have only final customers as 
clients, not other firms. For these ``final-demand'' firms, being unable to 
produce does not have any consequences on the rest of the productive system, 
only on household well-being. Since $z$ measures the average number of clients
which are in the network, in the limit of a large total number $N$ of firms, 
the phase-transition just corresponds to the fact that, for $z<1$, the 
most of the firms are in this peculiar situation, or have a few 
customer firms which are themselves in this peculiar situation, 
and so on, so that the initial damage usually does not propagate;
conversely, as soon as $z$ is larger than one, a finite fraction $1-v(z)$ of 
the firms have clients in the network, which have themselves other clients
in the network, and so on.

To make more quantitative the analysis, in the large $N$ limit, one can say
that in directed random network above the transition point, in order to observe
damage spreading up to reach a finite fraction of the network, the initially
attained firm has to belong to the giant in-component, $i \in {\cal I}$:
this occurs with probability $1-v(z)$. Moreover, in the same limit, one 
expects that finally all the firms in the giant out-component ${\cal O}$ will 
be impaired, which means a fraction $s_O=1-v(z)$ of the whole network. 
Correspondingly, we find:
\begin{equation}
\langle \theta_{tot} (t) \rangle \simeq 
\left \{
\begin{array}{lcl}
z^t &\mbox{ for }& t \ll t_{trans} \\
\left [1-v(z) \right]^2 N &\mbox{ for }& t \gg t_{trans},\: N \gg 1\\ 
\end{array}
\right.
\end{equation}
where the time that $\langle \theta_{tot} (t) \rangle $ takes before
reaching the asymptotic constant average value is quite short also for
large $N$ values, since it is of order 
$t_{trans} \sim \log([1-v(z)]N)/\log z$.

Interestingly, $[1-v(z)]^2$ is also the fraction of nodes which
are in the giant strongly connected component, ${\cal S}_{sc}$, of the network;
nevertheless, one should note the different meaning of the two quantities:
when the initial node belongs to ${\cal I}$, the total number of firms
which activity is finally impaired is $[1-v(z)]N=S_{O}$, definitely
larger than $S_{sc}$ for small $v(z)$ values. This point can be better
understood by looking at the Erd\H{o}s R\'enyi undirected random graphs with 
the same average degree $z>1$: here, the initially
damaged firm $i$ has to belong to the giant connected component
${\cal S}_{gc}$, and the firms which activity is finally
impaired are the ones in ${\cal S}_{gc}$ as well, hence we find again that
$\langle \theta_{tot}(t) \rangle$ approaches $[1-v(z)]^2N$ at large times,
though it is clearly $S_{sc}=1-v(z)$.

Summarizing, both in the DRG and in the RG, the average asymptotic density of 
fully active firms, when the networks is above the transition point, turns out 
to be:
\begin{equation}
\langle \rho_{asym}(z) \rangle =1- [1-v(z)]^2=2v(z)-v^2(z),
\label{phasetransition}
\end{equation}
where we recall that $v(z)$ is the solution of Eq,~(\ref{my_v})). 
Instead, one important difference between the behavior 
of the considered
free BDE systems with equal delays on directed and undirected random graphs 
concerns the short time dynamics: as it is shown in [Fig. 14], in the 
undirected case the signal does also propagate back to the node from which it 
is arrived and, correspondingly, to correctly describe the 
behavior of $\langle \theta_{tot}(t) \rangle$, Eq.~(\ref{st_drg}) should
be replaced by:
\begin{equation}
\langle \theta_{tot}(t) \rangle = \sum_{l=0}^{[t/2]} z^{t-2l} 
\mbox{ for } t \ll \log N/\log z, \hspace{.5in} \mbox{(RG)},
\label{st_rg}
\end{equation}
where as usual we denote $[y]$ the largest integer smaller than $y$. 

\begin{figure}[htpb]
\begin{center}
\includegraphics[width=.45\textwidth,angle=0]{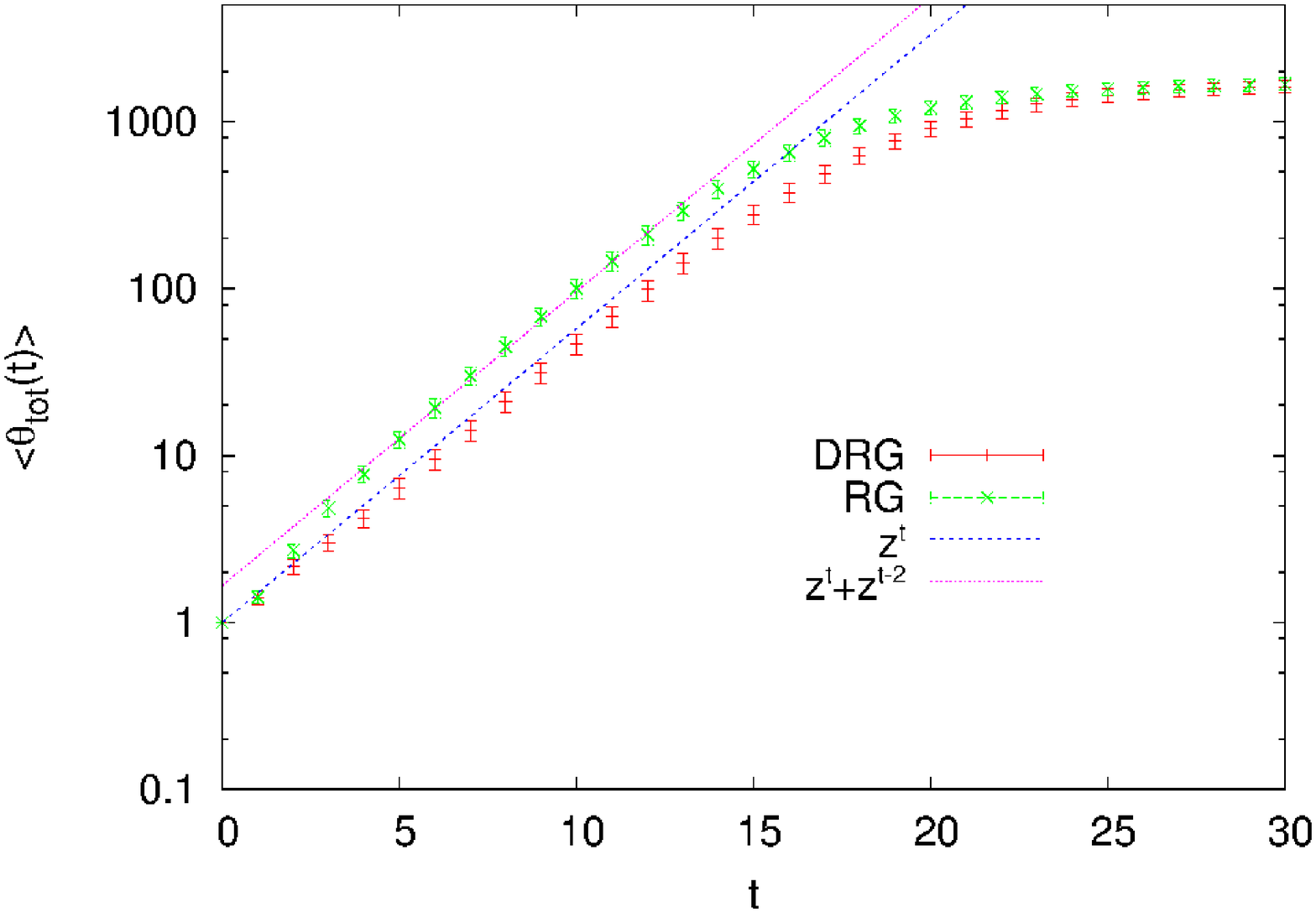}
\includegraphics[width=.45\textwidth,angle=0]{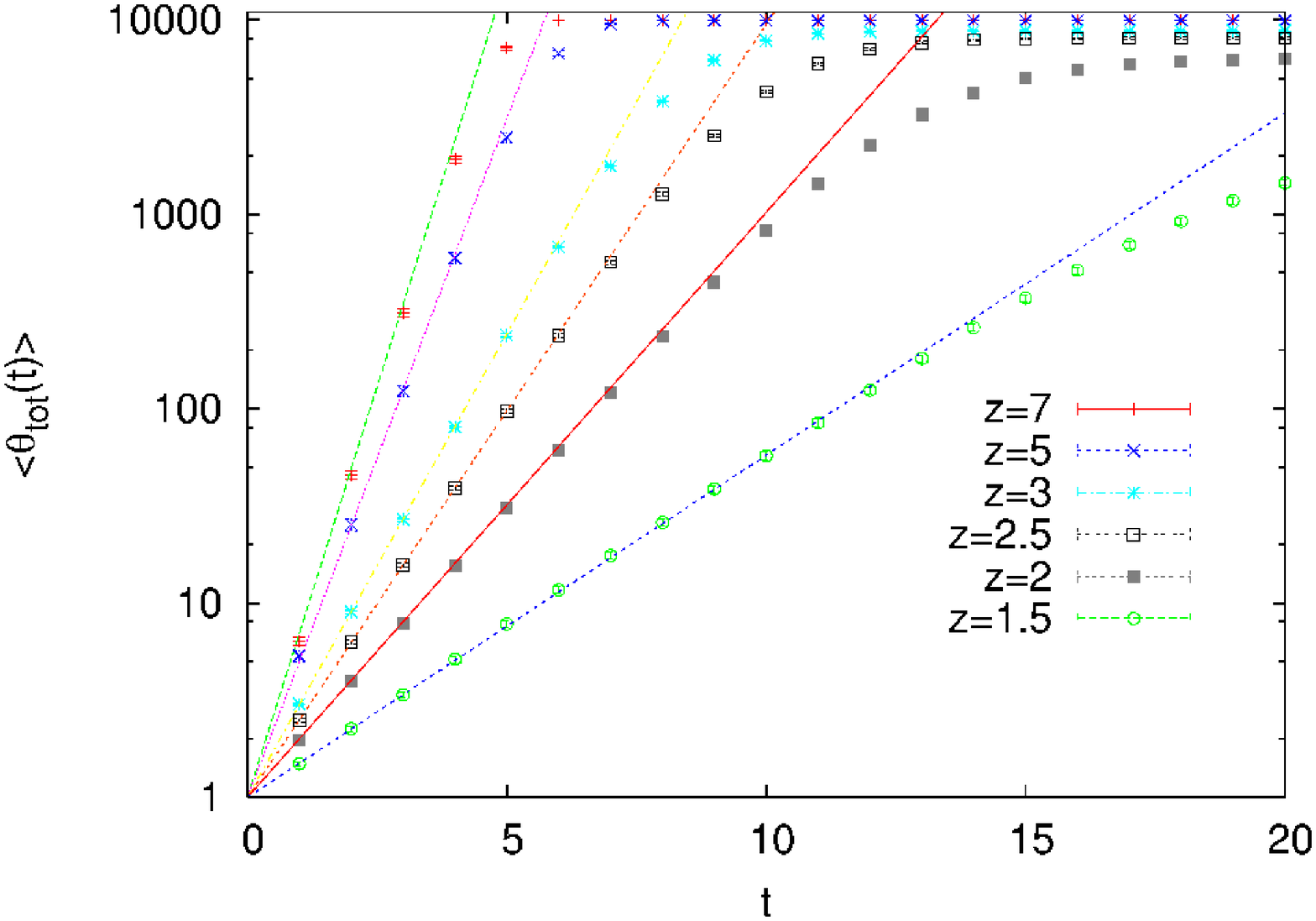}
\end{center}
\caption{Data on free models on random graphs with a large number  
of nodes $N \gg 1$ and deterministically chosen 
delays all equal to $\tau_0=1$ day. The obtained results 
are averaged over at least ${\cal N}_s=200$ different configurations of the 
links. On the left, we present the short time behavior of the average
total number of impaired firms, $\langle \theta_{tot}(t)\rangle$, 
both in the directed random graph, $D(N,p)$, with average in/out-degree 
$z \simeq Np=1.5$ and in the undirected Erd\H{o}s-R\'enyi one, $G(N,p)$, with 
the same $p$ value; here $N=5000$. On the right, we plot
$\langle \theta_{tot}(t)\rangle$ in the directed random graph, $D(N,p)$,
for different $z\simeq pN$ values above the transition point and  
$N=10000$. The obtained curves are compared with the corresponding expected
behaviors, given by Eq.~(\ref{st_drg}) for DRGs, and by an
approximation of Eq.~(\ref{st_rg}) for
RGs, respectively (see the text for details).}
\end{figure}

We present in [Fig. 15] the results of numerical simulations of the
BDE free model with equal delays on random graphs with a
large number of nodes $N=5000\div 10000$, for different values of $z>1$,
above the transition point. On the left, we compare the
short time behavior of  $\langle \theta_{tot}(t) \rangle$ in a 
directed random graph, and in an undirected one with the same $z=1.5$ 
average degree. Here, the regime preceding the attainment of the asymptotic 
constant value is relatively long, and one can appreciate the difference 
between the DRG case, described by Eq.~(\ref{st_drg}), hence
$\langle \theta_{tot}(t) \rangle \simeq z^t$, and the RG one,
where at not too short times the data are better in agreement with
$\langle \theta_{tot}(t) \rangle \simeq z^t +z^{t-2}$, which is an 
approximation of Eq.~(\ref{st_rg}). Notice moreover that one gets compatible 
asymptotic values in both of the cases. On the right we plot 
$\langle \theta_{tot}(t) \rangle$, 
as a function of $t$, in directed random graphs with different $z>1$ 
values, by comparing the curves with the expected $z^t$ short time behaviors.
This figure makes also evident that the asymptotic total number
of impaired firms is an increasing function of the average in/out-degree;
moreover, already for $z$ as small as $z=5$, it practically coincides
with the system size $N$, and the damage spreads in a few steps,
$t_{trans} \sim \log N /\log z $, to all the firms. 

The results on the average asymptotic density, $\langle
\rho_{asym}(z) \rangle$, as a function of the average in/out-degree $z$, 
are presented in [Fig. 16], and they turn out to be in very good agreement 
with the expected curve (see Eq.(\ref{phasetransition})), obtained by 
evaluating numerically the solution $v(z)$ of 
Eq.~(\ref{my_v}). The data shown in the figure correspond to the case of 
directed random graphs, but we checked that in undirected structures with the 
same average degree one gets definitely compatible results. This agreement
does also imply that for the considered number of nodes, $N= 10000$, the
corrections to the behavior in the limit $N \rightarrow \infty$ of 
$\langle \rho_{asym}(z) \rangle$ are practically negligible.

\begin{figure}[htpb]
\begin{center}
\includegraphics[width=.6\textwidth,angle=0]{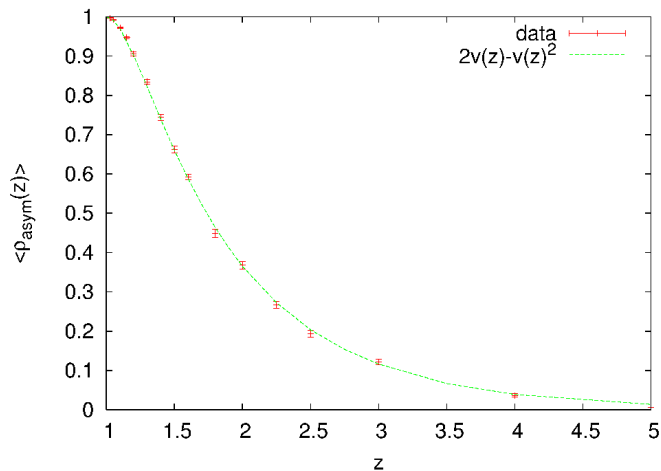}
\end{center}
\caption{Data on free models on directed random graphs, with large network 
size, $N=10000$, and deterministically chosen delays all equal to $\tau_0=1$ 
day. Here the results are averaged over at least ${\cal N}_s=200$ different 
configurations of the 
links. We plot the asymptotic average density, $\langle \rho_{asym}(z) \rangle=
\lim_{t \rightarrow \infty} \langle \rho(t) \rangle$, as 
a function of the mean in/out-degree $z$, compared with the expected
behavior, given by Eq.~(\ref{phasetransition}).}
\end{figure}

To conclude this section, the present results suggest that the study of these 
simple BDE systems on more complex random structures could prove useful 
also for better understanding the topology of the networks themselves, 
and in particular for evaluating the size of their connected components, which
is difficult to be analytically computed in the case of realistic directed 
networks, where the probability distributions of the out and in degree 
do not factorize \cite{Ne07}. A first step in this direction would be 
to consider (directed) ``small-world'' networks \cite{Wa99}, which in
some sense interpolate between the topology of the braid chain and the
one of the Erd\H{o}s-R\'eny random graph. 

It is moreover important to stress that the discussed damage spreading in free
BDE models is not to be confused with other well known examples of
damage spreading, such as in particular the spreading of epidemic diseases
or of the effects of node deletions \cite{AlBa02,Ne03}. In fact, on the 
one hand, here we have assumed that the damage propagates to all the nodes in 
the out-component of the concerned one, though with possibly different 
delays, and, on the other hand, a node attained by the consequences of the 
initial damage is not to be considered removed from the network, since it 
can recover.

\subsection{The free models with random delays}
We now turn onto the study of the free BDE model on a directed random
graph, when the delays are randomly chosen according to Eq.~(\ref{Ptauij}).
The system is described by a set of equations analogous to Eq.~(\ref{newform}):
\begin{equation}
x_i(t)=\mu_i(t) \prod_{j\in {\cal I}(i)} x_j(t-\tau_{i,j}), 
\hspace{.3in} i=1,...,N.
\label{newformij}
\end{equation}
where the $\{ \tau_{i,j} \}$ variables are independently uniformly distributed 
in the interval $[\tau_{min},\tau_{max}]$. Hence the time is in units
of $\tau_{min}=1$ day, and we will take as usual $\tau_{max}=10$ days,
limiting moreover the analysis to the case in which the duration
of the initial damage is $\tau_c=\tau_{min}$.

In order to understand the dynamics, we start by noticing that in the
first time step the signal propagates in average to
$z/\tau_{max}$ other firms. In the next time steps, one still expects
to observe an exponential damage spreading, for networks above the
transition point $z^d_c=1$. Moreover, the structure is locally 
tree-like: this means that, at enough short times, the signal propagates along 
each given branch (and sub-branch and so on) of the tree independently; 
correspondingly, since the propagation paths of different time 
lengths are not concurrent, one expects that the random delays will turn out 
first of all in a global rescaling of the time of a factor 
$1/\tau_{av}$, where $\tau_{av}=\langle \tau_{i,j} \rangle$.

\begin{figure}[htpb]
\begin{center}
\includegraphics[width=.6\textwidth,angle=0]{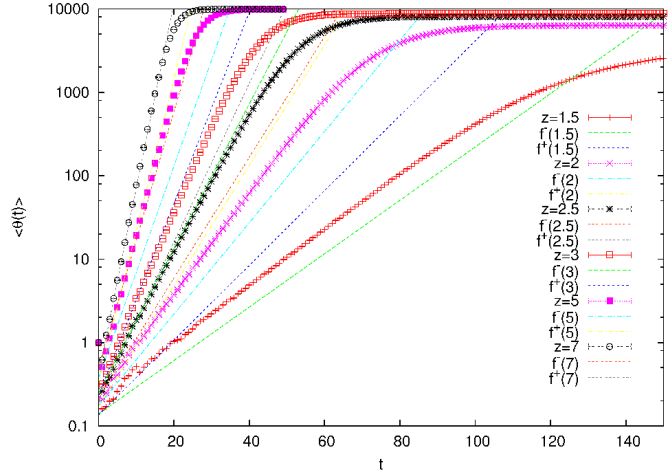}
\end{center}
\caption{The total number of impaired firms, $\theta_{tot}(t)$, as a function 
of time, in the free model on a directed random graph with a large network 
size $N=10000$ and different $z$ values. The data are averaged over at least 
${\cal N}_s=200$ different configurations both of the links and of the random 
delays, the last being chosen uniformly distributed between $\tau_{min}=1$ 
day and $\tau_{max}=10$ days. We compare the curves for each $z$ value
with the expected short time lower limit $f^-(z,t)=f(z^-_{eff},t)$ and 
upper limit $f^+(z,t)=f(z^+_{eff},t)$, where $f$ is given in Eq.~(\ref{effe}).
See the text for details.}
\end{figure}

One obtains:
\begin{equation}
\langle \theta_{tot}(t) \rangle=\frac{z}{\tau_{max}} z_{eff}^
{t/\tau_{av}} \mbox{ for } t \ll \log N/\log z_{eff}.
\end{equation}
The effective average in/out-degree which appears
into this law has to approach $z$ when $z \rightarrow 1^+$, hence
\begin{equation}
z_{eff} \ge z^-_{eff}=z.
\end{equation} 
Nevertheless, since the signal from a given node propagates to its 
customers up to the time $t+\tau_{max}$, one also finds an effective
increasing of the average in/out-degree, which can be easily overestimated:
\begin{equation}
z_{eff} \le z^+_{eff}=z\sum_{l=0}^{\infty}\left ( \frac{z}{\tau_{max}}
\right )^l=\frac{z}{1-z/\tau_{max}}.
\end{equation}
Summarizing, at short times, for $z>1$, we get:
\begin{equation}
f^-(z,t) =f(z_{eff}^-,t)\le \langle \theta_{tot}(t) \rangle \le 
f(z_{eff}^+,t)=f^+(z,t),
\label{limztot}
\end{equation}
where
\begin{equation}
f(z_{eff},t)=\frac{z}{\tau_{max}} z_{eff}^{t/\tau_{av}}.
\label{effe}
\end{equation} 

We compare in [Fig. 17] the data on $\langle \theta_{tot}(t) \rangle$,
for different values of the average in/out-degree $z>1$, and a large system
size $N=10000$, with the expected upper and lower limits on the
short time behaviors, given by Eq.~(\ref{limztot}). The curves 
do always lie between $f^-(z,t)$ and $f^+(z,t)$ in a large time window;
they approach $f^-(z,t)$ for $z \rightarrow 1^+$, and $f^+(z)$ 
for large $z \gg 1$; in detail, the data at short times nearly coincide with 
$f^+(z,t)$ already for $z=3$.

\begin{figure}[htpb]
\begin{center}
\includegraphics[width=.6\textwidth,angle=0]{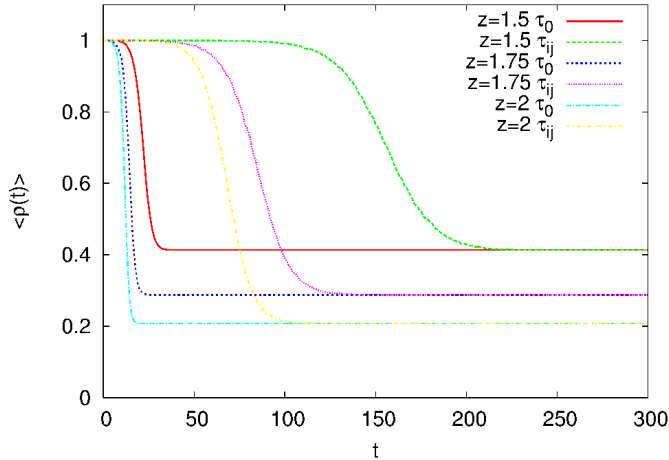}
\end{center}
\caption{The density of fully active firms $\rho(t)$ as a function of time in 
the free model on a directed random graph with different average 
connectivities 
$z>1$. Here we consider single typical configurations of the links. We compare 
the results for delays deterministically taken all equal to $\tau_0=1$ day 
with the ones for a single typical configuration of random delays, 
chosen uniformly distributed between $\tau_{min}=1$ day and $\tau_{max}=10$ 
days. We notice that the densities approach exactly the same values in the two 
cases.}
\end{figure}

In the large time limit, since in the giant strongly connected component
${\cal S}_{gc}$ there are loops, one expects that at the end the damage
will spread to the whole giant out-component ${\cal O}$, with the
same probability $1-v(z)$ that in the model with deterministic delays.
Correspondingly, we verified numerically that one still observes,
for large $N$ values, an asymptotic average density of fully active firms
$\langle \rho_{asym} \rangle =2v(z)-v^2(z)$, in agreement
with Eq.~(\ref{phasetransition}). Moreover, for not too small $N$ values, one 
usually gets the same $\rho_{asym}$ for a given network configuration both for
deterministic equal delays and for randomly chosen $\tau_{i,j}$
(see [Fig. 18]). This can be qualitatively understood because, if the 
initially attained firm is in a connected component where there are no loops,
the economy can recover completely in both of the cases, whereas
if it is in the giant in-component then the damage spreads to the whole
out-component of the particular considered network.

\begin{figure}[htpb]
\begin{center}
\includegraphics[width=.6\textwidth,angle=0]{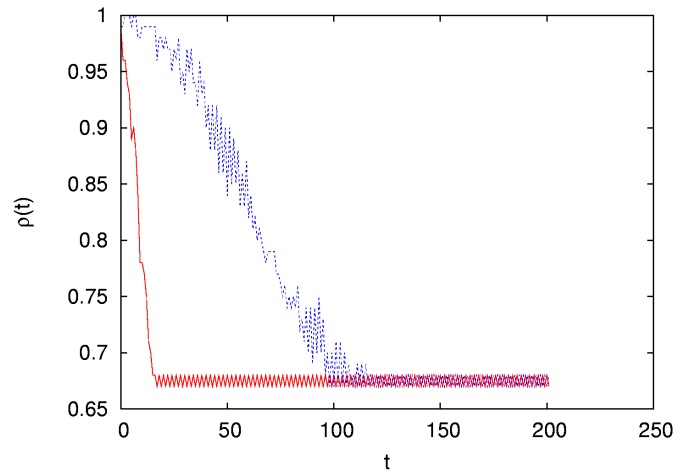}
\caption{The density of fully active firms $\rho(t)$ as a function of time 
in a given, quite peculiar, random directed network configuration with 
$z=0.525$ and $N=100$, after a perturbation of node $i=1$ of duration 
$\tau_c=1$ day. We compare results for deterministically
chosen delays, all equal to $\tau_0=1$ day (lower curves), to the ones for a 
typical configuration of random delays, uniformly distributed 
between $\tau_{min}=1$ day and $\tau_{max}=10$ day (upper curves):
notice that the asymptotic solutions display the same behavior of the
density, which oscillates around a value definitely smaller than one, though
the network is well below the crossover point $z \sim 1$. See the text
for details.}
\end{center}
\end{figure}

Nevertheless, the asymptotic solutions of the considered free BDE systems, for 
the same directed graph configuration, do not need to be coincident when 
considering equal or random delays. This effect becomes evident for relatively 
small $N$, near or below the crossover point, hence for $z\lesssim 1$, since 
it is related to the presence of loops in connected components which contain an
enough large fraction of the nodes, but are not the giant ones. More in detail:
\begin{itemize}
\item when the initially attained firm belongs to a component where
there are no loops, the asymptotic solutions is given by
$x_i \equiv 1 \: \forall i$ in both of the cases; 
\item when it belongs to the giant in-component, hence containing an enough 
large number of loops, the asymptotic solutions is given by 
$x_i \equiv 0 \: \forall i \in {\cal O}$, and $x_i \equiv 1 \: 
\forall i \in {\cal W}\setminus{\cal O}$, hence the solutions are again the 
same in both of the cases, apart possibly for a small fraction of nodes which 
approaches zero in the limit $N \rightarrow \infty$; 
\item in intermediate situations, one usually finds periodic solutions 
which can be largely different.
\end{itemize} 

\begin{figure}[htpb]
\begin{center}
\includegraphics[width=.32\textwidth,angle=0]{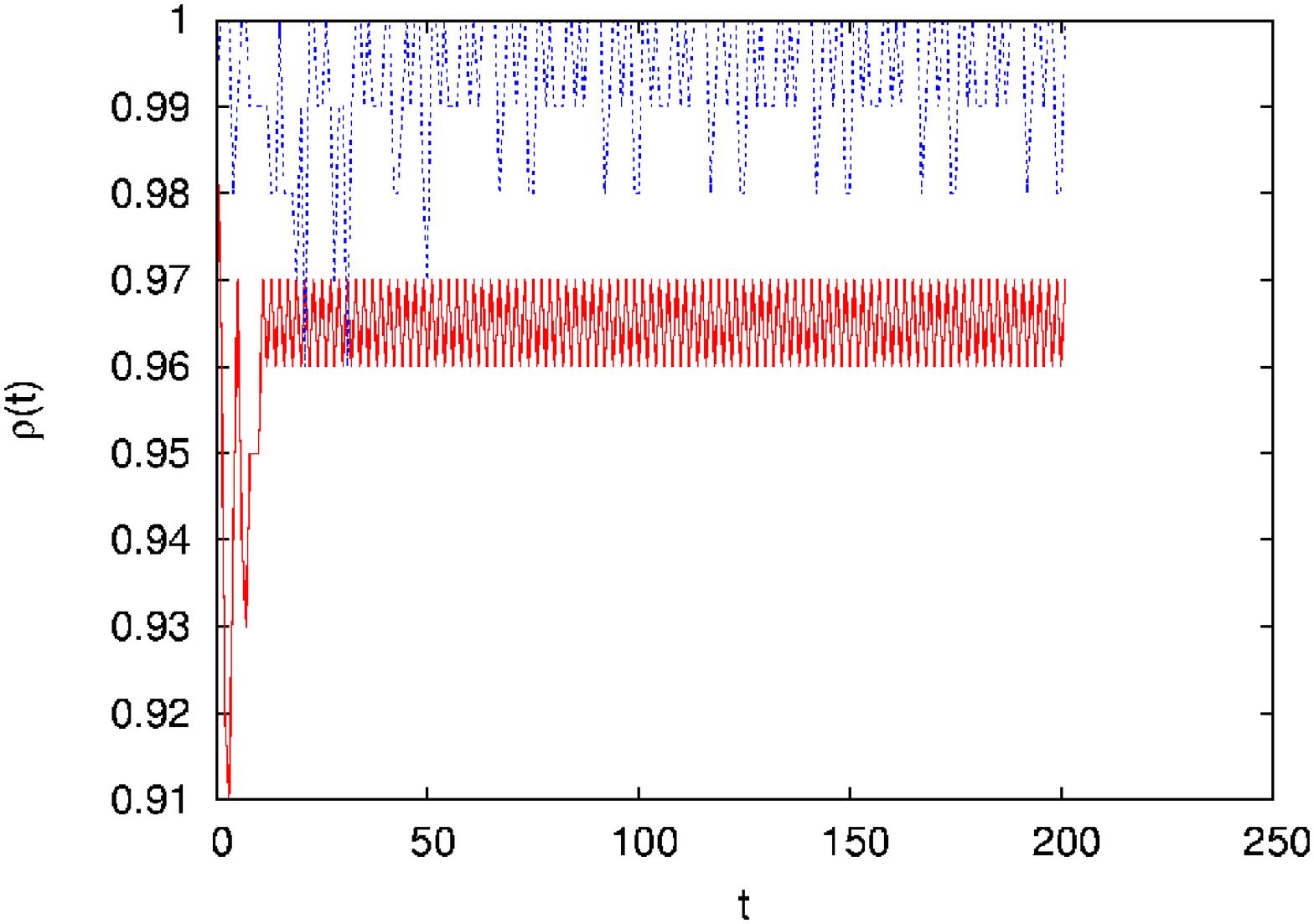}
\includegraphics[width=.32\textwidth,angle=0]{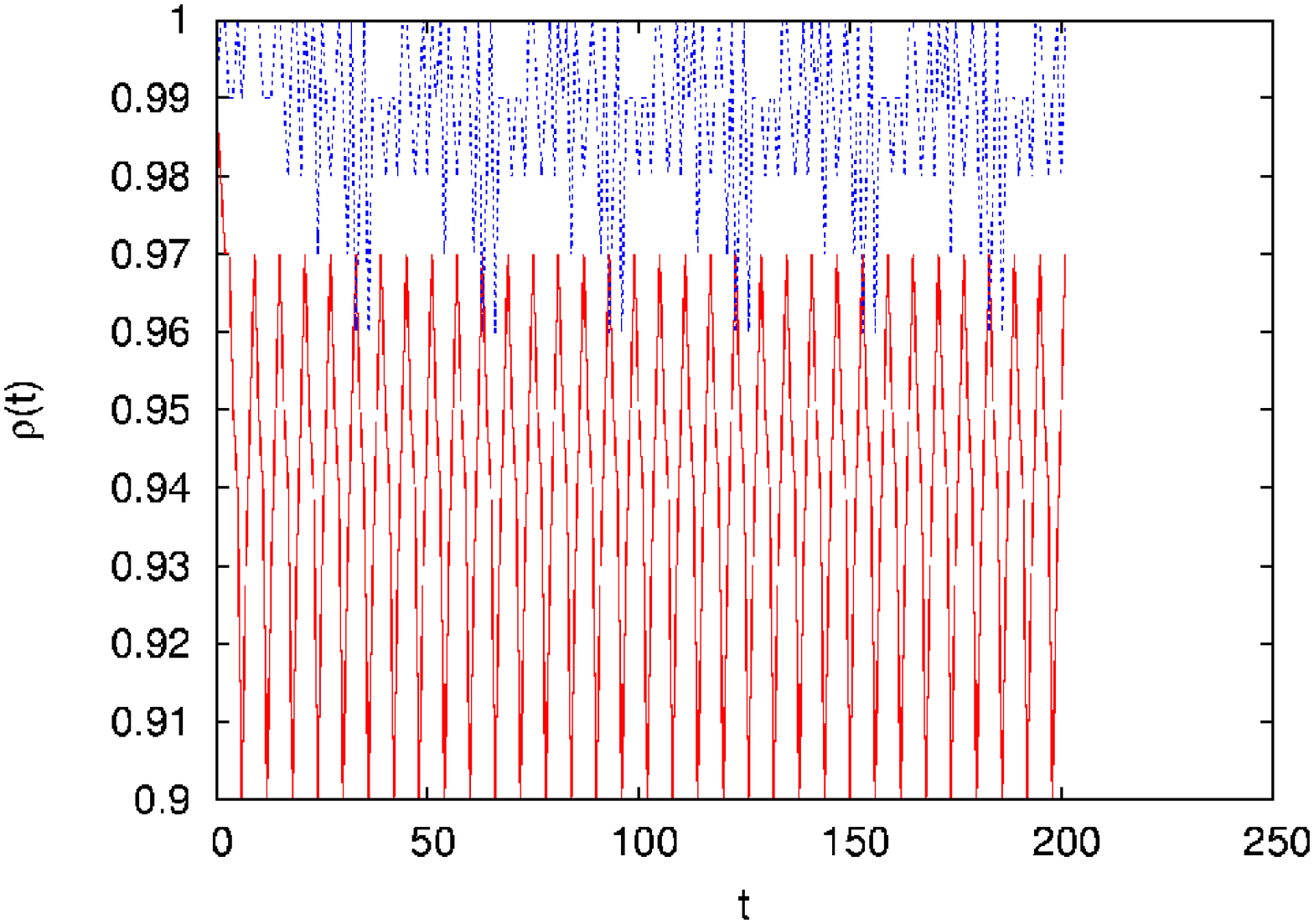}
\includegraphics[width=.32\textwidth,angle=0]{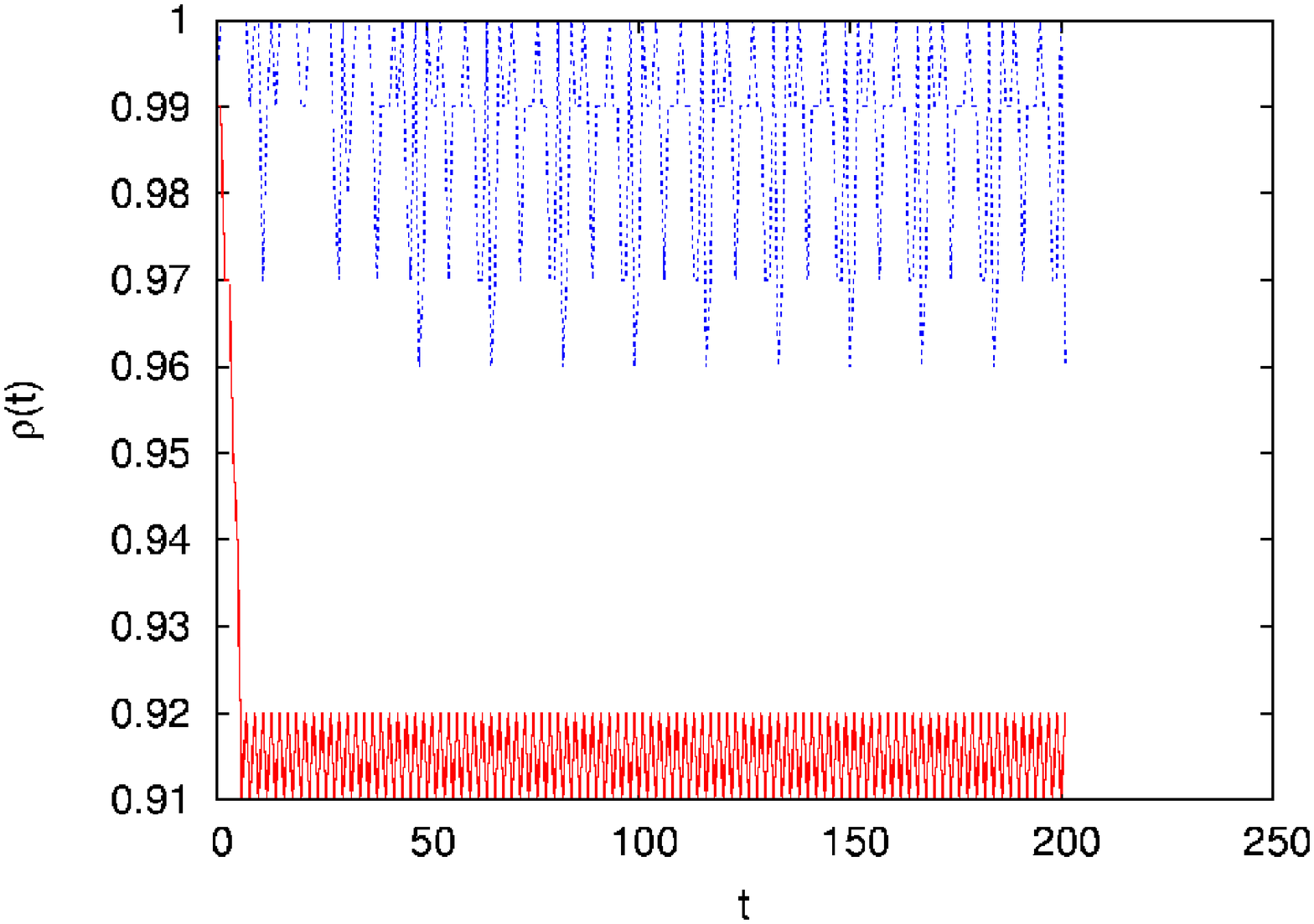}
\includegraphics[width=.32\textwidth,angle=0]{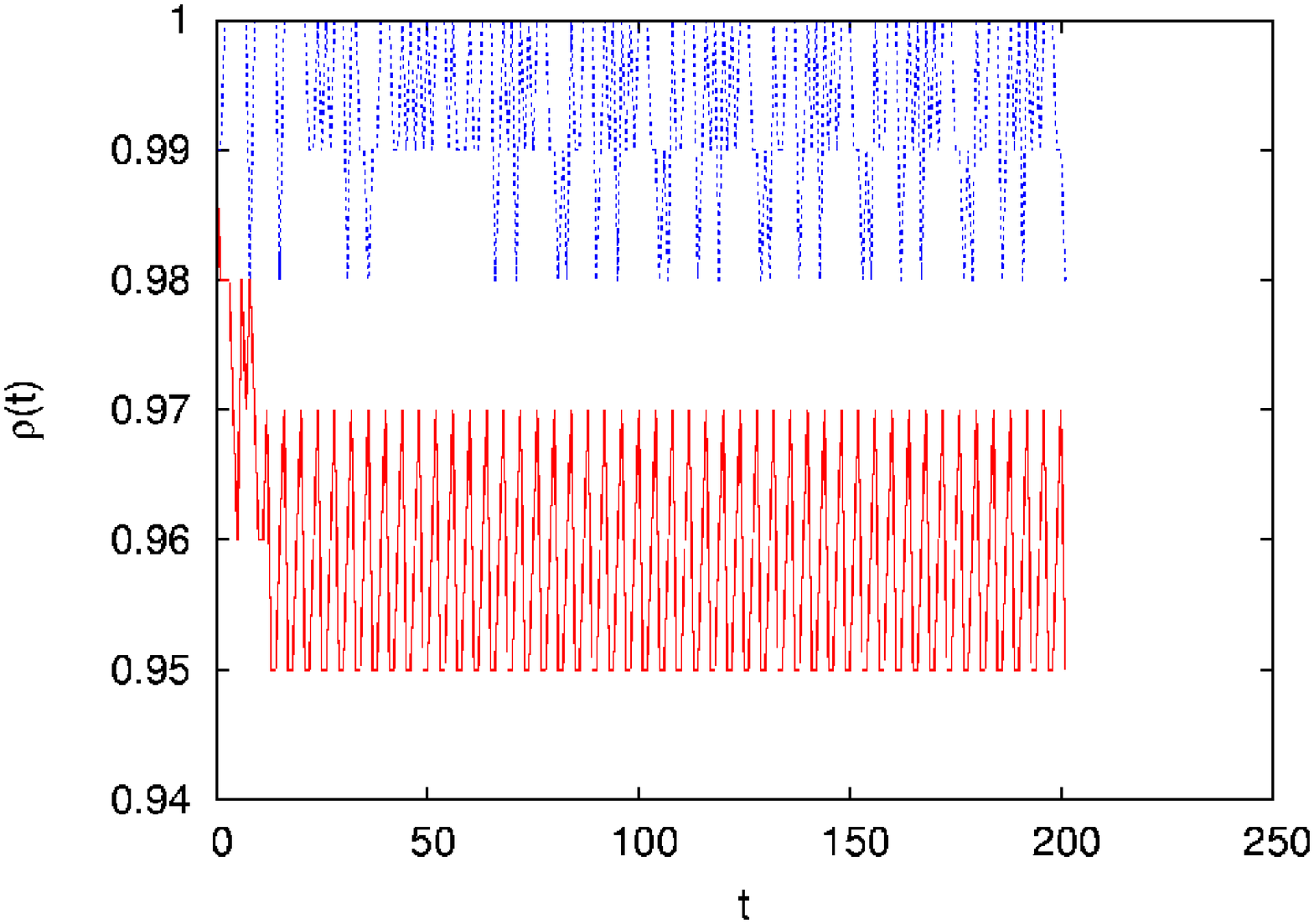}
\includegraphics[width=.32\textwidth,angle=0]{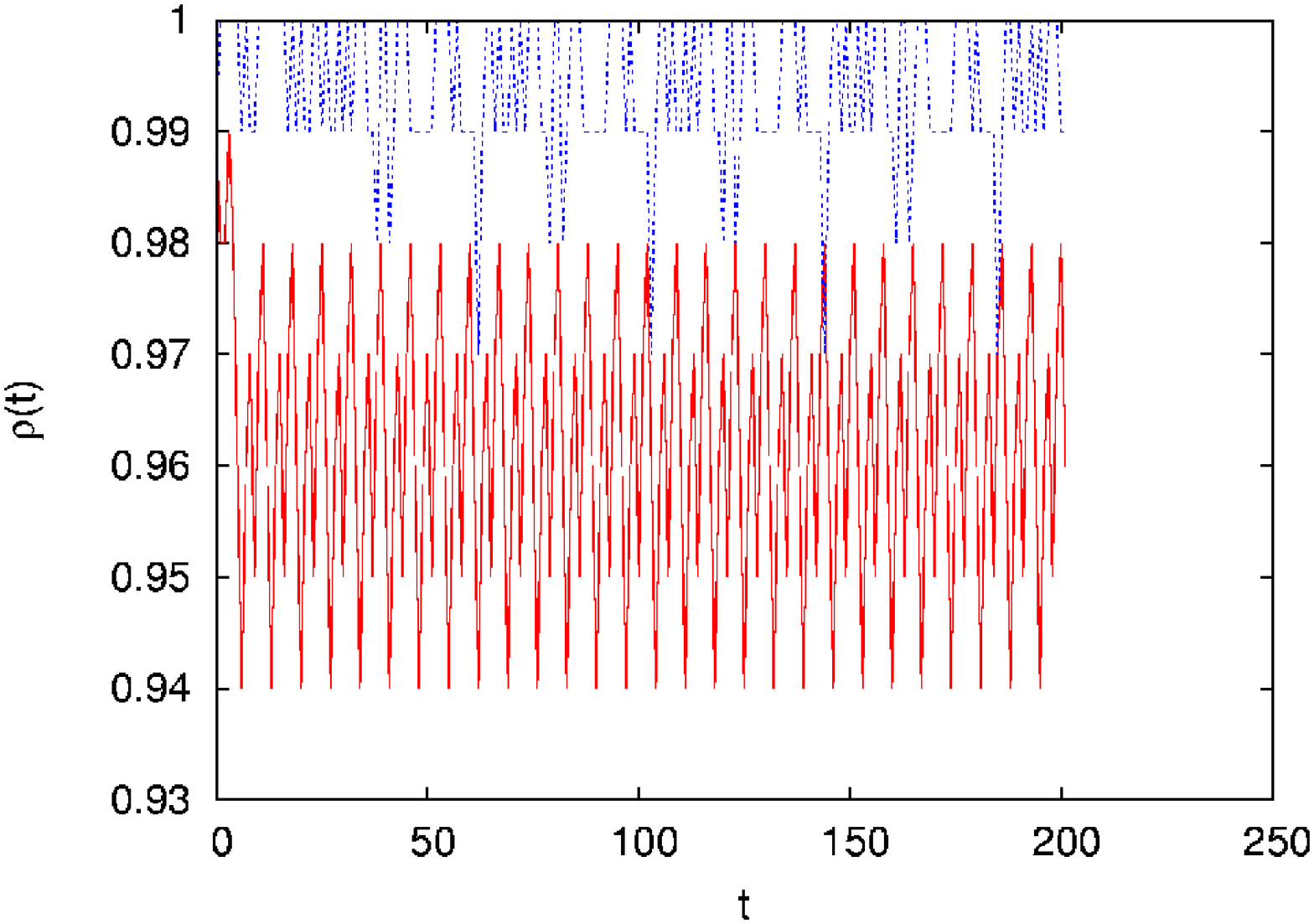}
\includegraphics[width=.32\textwidth,angle=0]{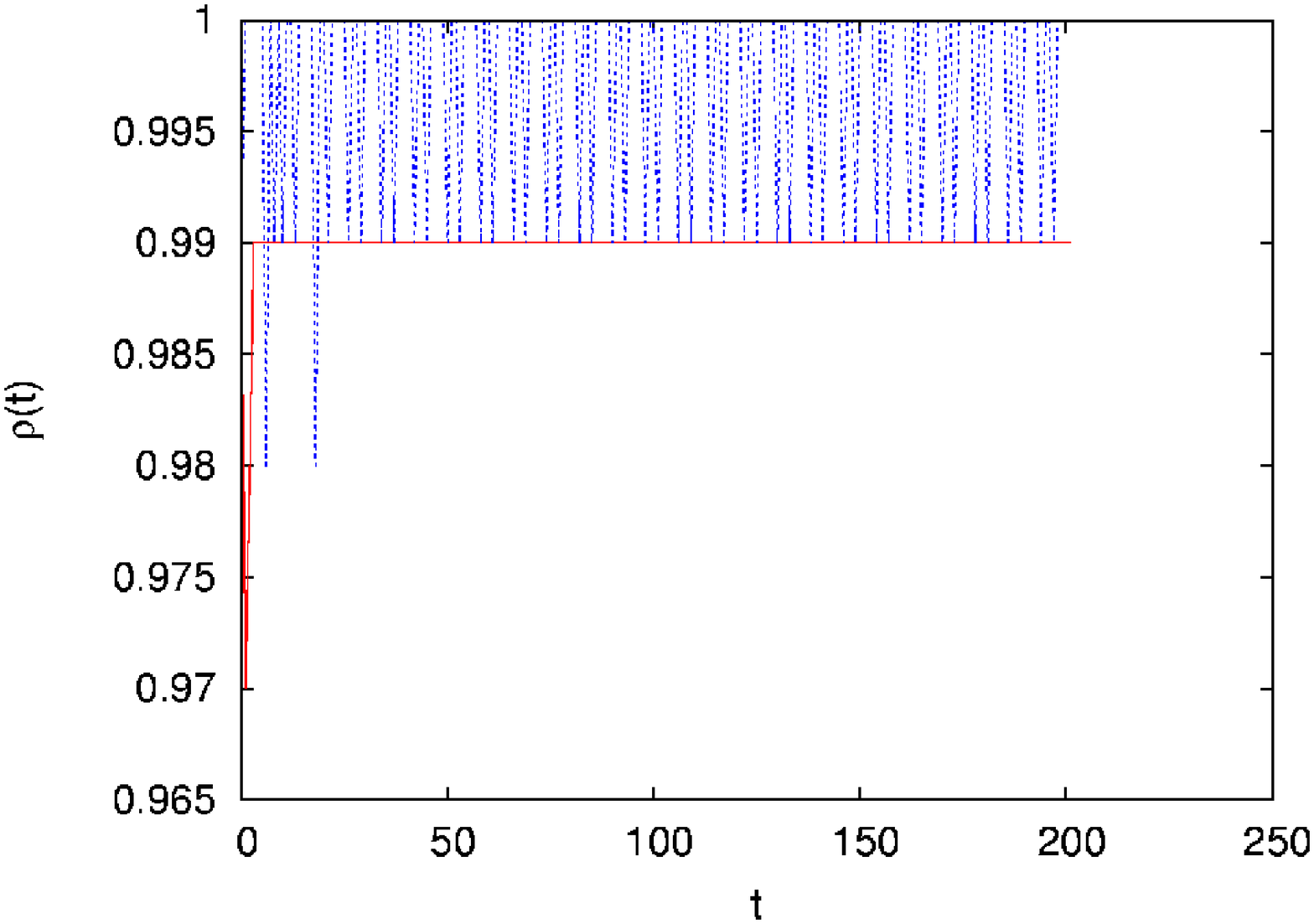}
\includegraphics[width=.32\textwidth,angle=0]{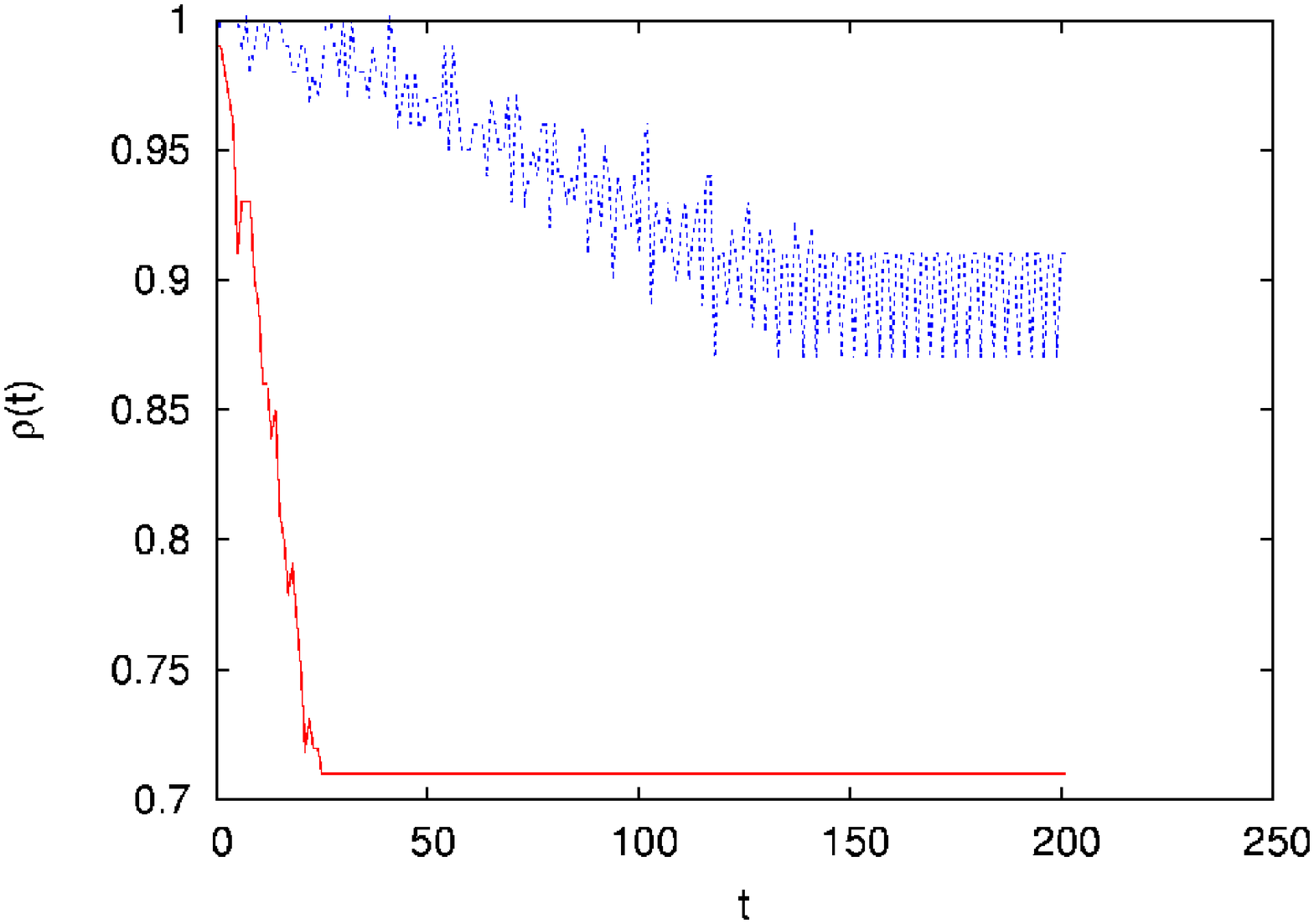}
\includegraphics[width=.32\textwidth,angle=0]{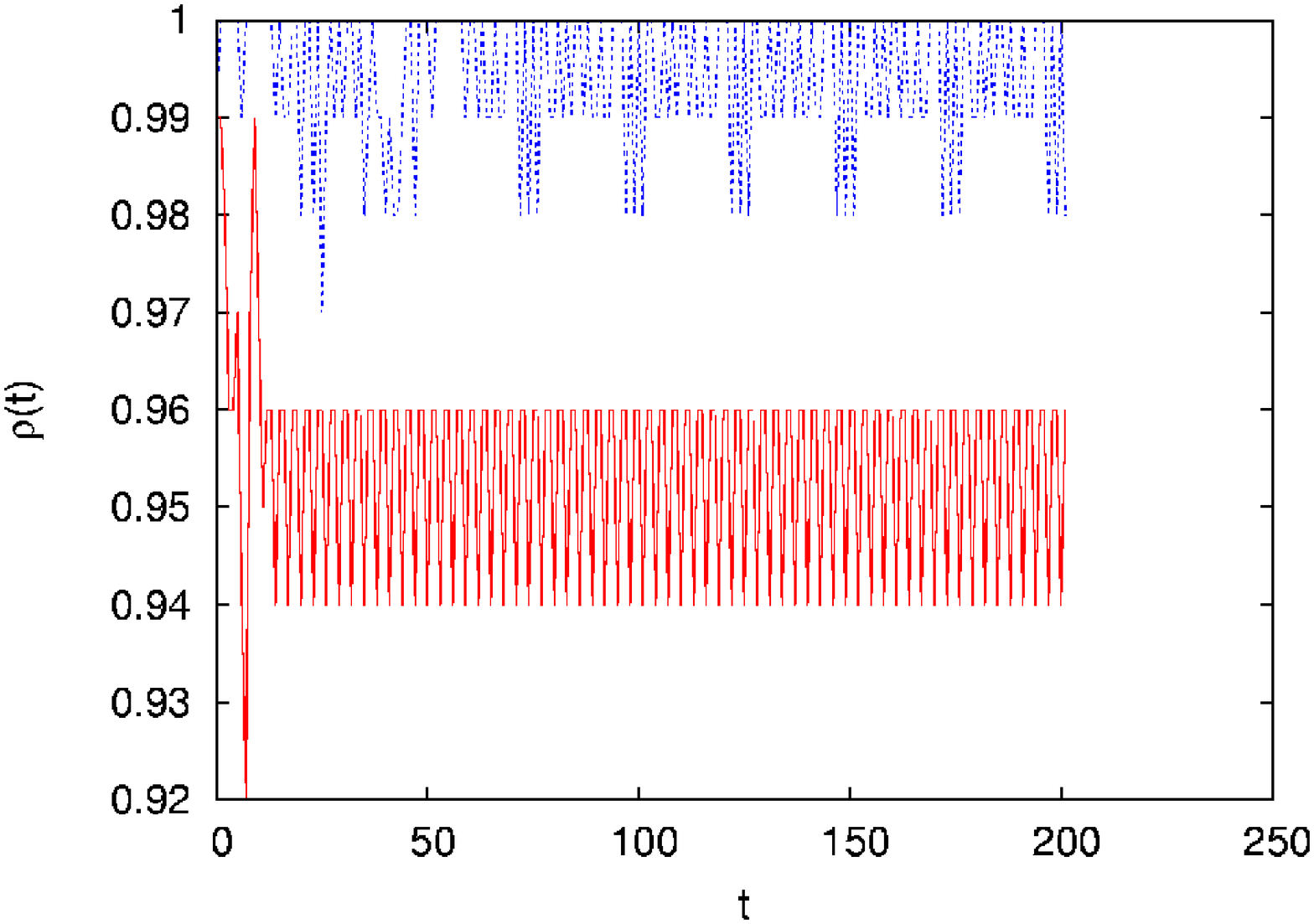}
\includegraphics[width=.32\textwidth,angle=0]{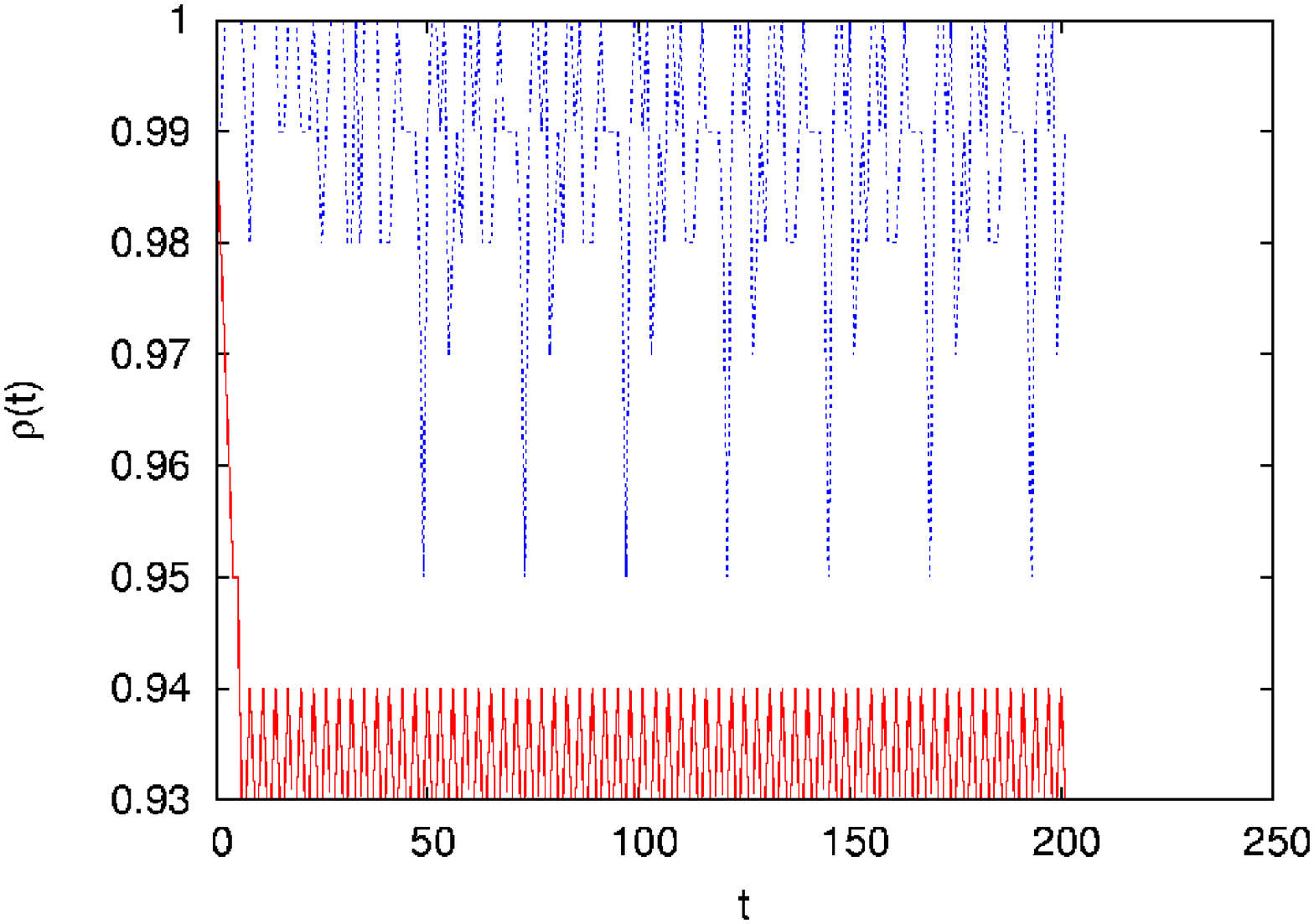}
\caption{The density of fully active firms $\rho(t)$ as a function of time,
after a perturbation of node $i=1$ of duration $\tau_c=1$ day,
in some quite peculiar random network configurations with $z=0.525$, which
probability is not negligible already for the relatively large size $N=100$
that we considered. In all of the cases, we compare results for 
deterministically chosen equal delays (lower curve) with the ones for a 
typical configuration of random unequal delays (upper curve), as in the
previous figure. Notice that here the densities approach definitely different
asymptotic behaviors in the two cases. See the text for details.}  
\end{center}
\end{figure}

Though a detailed analysis of the finite $N$ behavior of free BDE models
on directed random graphs is beyond the scope of the present paper, we
show in [Fig. 19] and [Fig. 20] some examples of peculiar configurations
of $D(N,p)$, with $N=100$ and $z=0.525$, well below the crossover point. 

We note that the observed behavior does also depend on the position of the 
firm which is initially destroyed, here taken to be $i=1$ in all of 
the cases. We compare the data on the density of fully active firms $\rho(t)$ 
for deterministically chosen delays all equal to $\tau_0=1$ day with the
ones for a typical configuration of random unequal delays, uniformly
distributed between $\tau_{min}=1$ day and $\tau_{max}=10$ days; the
duration of the initial damage is $\tau_c=1$ day in both of 
the cases. The results confirm that one can observe different asymptotic
average densities, which implies largely different periodical solutions of the
system; moreover, the probability of these disordered network configurations 
is not negligible, since we have observed not compatible $\rho_{asym}$ values 
in $\sim 1/10$ of the considered samples.

\subsection{The forced models}

We conclude this work by presenting some preliminary results
on the forced models, as defined by Eq.~(\ref{forced}), on a directed random 
graphs structure $D(N,p)$. The equation for the variable $x_i$ of the system 
(\ref{syst}) can be written as:
\begin{equation}
x_i(t)=\mu_i(t) \oo{x}_i(t-\tau_{0})\vee \prod_{j\in {\cal I}(i)} 
x_j(t-\tau_0) 
\hspace{.3in} i=1,...,N.
\label{newformforced}
\end{equation}
when the delays are deterministically taken all equal to $\tau_0=1$ day, and:
\begin{equation}
x_i(t)=\mu_i(t)\prod_{j\in {\cal I}(i)} 
[\oo{x}_i(t-\tau_{i,j}) \vee x_j(t-\tau_{i,j})] 
\hspace{.3in} i=1,...,N.
\label{newformforceds}
\end{equation}
for random delays. We study, as usual, the damage spreading after the initial 
destruction of a single firm, that we take for simplicity of duration 
$\tau_c=1$ day.

As soon as the initial attained firm belongs to a component where there are
no loops, the economy will recover completely as in the previously considered 
free models, hence the asymptotic solution is $x_i \equiv 1 \: \forall i$. 
For large $N$, this happens in particular almost surely below the critical 
point of the directed random graph $z_{c}^d=1$, and with probability 
$v(z)$ for $z>z_c^d$, where $v(z)$ is the solution of Eq.~(\ref{my_v}).

\begin{figure}[htpb]
\begin{center}
\includegraphics[width=.6\textwidth,angle=0]{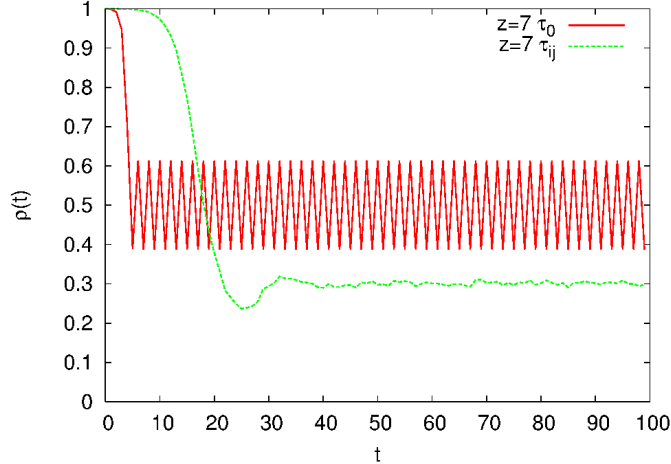}
\end{center}
\caption{The density of fully active firms $\rho(t)$ as a function of time, 
after an initial perturbation of node $i=1$ of duration $\tau_c=1$ day, in 
the forced model on a directed random graph with a large network size 
$N=10000$ and average input/output degree $z=7 \gg z^d_c$. Here we consider a 
single (typical) configuration of the links, and we compare the results for 
the 
case of delays taken all equal to $\tau_0=1$ day with the ones for a 
typical configuration of random unequal delays, chosen uniformly distributed 
between $\tau_{min}=1$ day and $\tau_{max}=10$ days.}
\end{figure}

Nevertheless, because of the external rescue inputs, one expects
that also in the region $z \gg z_c^d$, though the damage spreads 
almost surely ($v(z) \ll 1$) to the whole giant connected out-component, 
whose size is approximately $N$ in this limit, the average fraction of 
fully active firms in the asymptotic solution is larger then zero, {\em i.e.}
the economy can partially recover. More in detail, for 
deterministically taken equal delays, since each firm recovers both
if all of its suppliers did recover and if its activity has been already
impaired for a duration $\tau_0$, one expects that, in the
asymptotic solution, in average the firm is fully active one half of
the time. This means $\langle \rho_{asym} \rangle =1-[1-v(z)]^2/2 \sim 1/2$, 
for $z\gg 1$. In the presence of random delays, since there is a larger number 
of concurrent paths of different time lengths, and the activity of a given
firm can be impaired for a different duration of time depending on the 
good which is lacking, the asymptotic average density is expected to be 
smaller. 

We present in [Fig. 21] our numerical results on the behavior 
of $\rho(t)$, as a function of time, for a single, typical, configuration of 
the directed random graph, of large size $N=10000$, and average in/out-degree 
$z=7 \gg z_{c}^d$. In fact, for deterministically chosen equal delays the 
asymptotic density oscillates around the average value $1/2$, whereas for 
random delays it takes the lower average value $\rho_{asym}\simeq 0.3$. 
Moreover, in the first case the oscillations of the density are quite strong, 
with lower value $\simeq 0.4$ and upper value $\simeq 0.6$; hence a finite 
fraction of the firms, of order $0.2N\simeq 2000$, is involved into them. 

\begin{figure}[htpb]
\begin{center}
\includegraphics[width=.4\textwidth,angle=0]{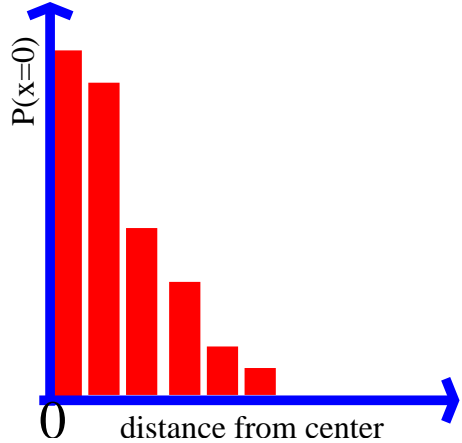}
\end{center}
\caption{A qualitative sketch of the expected behavior of the probability
distribution of impaired firms as a function of the distance of their
position from the center of the connected component of the graph. See the
text for details.}
\end{figure}

One can explain qualitatively this result by arguing that the damage starts 
to spread from some firm in the periphery of the connected component of the 
graph, rapidly propagating into the center. Then, because of the external 
inputs, the most of the economy recovers, apart from a few firms which are 
once 
again in the periphery, and have been reached later by the wave of damage: 
these will be responsible for the forthcoming impulse. Roughly speaking, 
the nodes with larger in/out-degree are the ones occupying the more central 
positions, whereas a small in-degree means that the node is more unlikely
reached by the damage propagation and a small out-degree means that it
is more unlikely to transmit the damage, so that these nodes are to be
considered in the periphery in both of the cases. This picture is sketched in 
[Fig. 22], and it is quite intriguing, since it implies that there is a large 
factor of unpredictability in the behavior of the solution of the BDE system. 
A first step towards a more quantitative analysis could be to look at the
modular structure of the directed random graph \cite{GuAm05}. Notice 
however that the random delays have the effect of smoothening this oscillating 
behavior.
  
\section{Conclusions}
We have studied the propagation of an initial damage, which destroys
a single firm for a given time, in production networks. In the considered free 
models, which represent 
isolated networks, since the local firm dynamics is controlled by a logical 
AND function of the inputs, the damage can invade a finite fraction of the 
nodes or the whole network when these two conditions are fulfilled: i) the 
nodes mean input connectivity is larger than one; ii) the duration of the 
initial damage is larger than the smallest propagation time between two nodes. 
In 
particular, for a braid chain (deterministic and connected) topology, 
these conditions mean that in the asymptotic solution the production of all 
the firms is impaired.

Damage spreading velocity strongly depends upon network topology: we have 
shown that the number of attained firms increases linearly with time 
for the braid chain and exponentially for the directed random graphs. We have 
also introduced a distribution of delays, and we have found that the 
propagation velocity is dominated by the fastest segments (the 
smallest delay) in the braid chain, while the average delay limits speed on 
the directed random graph. Moreover, in random networks the saturation level 
of the fraction of impaired firms does not depend upon the distribution of 
delays, but only upon the particular network topology through the size of the 
connected component, which is in average not negligible as soon as the 
mean in/out-degree is larger than one.

External supplies, modeled here in terms of forced networks, limit damage and 
the asymptotic solutions are periodic: waves of damage move across the 
structure. In the considered forced models with distribution of delays, the 
transient before that these solutions are reached diverges exponentially with 
the network size, though the average density of fully active firms approaches 
a nearly constant value after a definitely shorter effective transient. This 
effective transient corresponds approximately to the duration of the first 
cycle of propagation of the damage across the connected component of the 
networks, which happens similarly to that in the free models, with linear or 
exponential speed depending on the topology. Finally, periodic 
dynamics are obtained also when the duration of the initial damage
is smaller than the smallest propagation time between two nearest neighbor 
nodes.

The models that we have used are extremely simplified with respect to real 
production networks, and firms behavior is represented in a very idealized
way. Still, results from this simple analysis suggest that:
\begin{itemize}
\item Damage spreads in absence of sufficient stocks and flexibility; 
as a result, production shortages are extended and can survive for periods 
much longer than the duration of the initial local damage.
This result suggests that an affected economy can suffer from disaster
consequences even after all physical damages have been repaired.
\item The presence of multiple concurrent production paths with different 
time lengths does not necessarily imply a slowing down of the speed of 
the signal, which can propagate as fast as the shortest path, depending 
on the topology. This result suggests that the most vulnerable supply
chains control the macroeconomic losses and that vulnerability analysis
should focus on the identification of key vulnerabilities.
\end{itemize}

\section*{Appendix~A}
The average total number of impaired firms $\langle \theta_{tot}(t)\rangle$, 
in the free model on a braid chain network with $n=1$, defined accordingly 
to Eq.~(\ref{defztot}) and Eq.~(\ref{defrho1}), can be expressed as
function of the probability ${\cal P}({\cal T}_k,k)$, where ${\cal T}_k$ is 
the sum of $k$ random variables $\tau_{i,j}$:
\begin{eqnarray}
\langle \theta_{tot}(t) \rangle &\equiv& 
\langle \sum_{i=1}^N \oo{x}_i(t) \rangle
\equiv \sum_{i=1}^N \int \prod d\tau 
{\cal P}(\tau) \oo{x}_{i-k}(t-{\cal T}_k)= \nonumber \\
&=&\sum_{k=[t/\tau_{max}]+1}^{[t]\tau_{max}} {\cal P}([t],k) 
\sum_{i=1}^N \delta_{i-k,1}=\sum_{k=[t/\tau_{max}]+1}^{[t]\tau_{max}} 
{\cal P}([t],k). 
\label{defztotav}
\end{eqnarray}
Here $[v]$ means the integer part of $v$, and we are assuming for simplicity
that the duration of the initial damage is $\tau_c=\tau_{min}=1$
day, hence we are using the conditions on the initial state of the system,
$\{ x_i(t) = \delta_{i,1}(0) \}$ for $t \in [0,1]$.
  
One has, from Eq.~(\ref{Ptauij}):
\begin{eqnarray}
\langle \tau_{i,j} \rangle &=& 
\int_{1}^{\tau_{max}} d\tau_{i,j}{\cal P}(\tau_{i,j}) \tau_{i,j}=
\frac{(\tau_{max}+1)}{2} \\
\sigma^2_{\tau_{i,j}}&=&\frac{(\tau_{max}+1)(2\tau_{max}+1)}{6}
-\frac{(\tau_{max}+1)^2}{4},
\end{eqnarray}
for the mean value and the variance of the distribution of the delays, 
respectively. For $k \ge 2$, assuming that the variables are independent
(which is clearly an approximation for $k>N$), one can apply the central limit 
theorem:
\begin{equation}
{\cal P}({\cal T}_k,k) \simeq {\cal P}_{G}
(T_k,k)=\frac{1}{\sqrt{2\pi k \sigma^2_{\tau_{i,j}}}}
\exp \left( - \frac{({\cal T}_k-k\langle \tau_{i,j} \rangle)^2}
{2 k \sigma^2_{\tau_{i,j}}} \right ).
\label{Gaussian}
\end{equation}
By using this approach, we find:
\begin{equation}
\langle \theta_{tot}(t)\rangle \simeq
\left \{
\begin{array}{ll}
 1 & t \in [0,1) \\
{\cal P}(1,1)={1}/{\tau_{max}} & t \in [1,2) \\
{1}/{\tau_{max}} +\sum_{k=2}^{([t] \tau_{max})} 
{\cal P}_{G}([t],k) & t \in [[t],[t]+1), \:
2\le t \le \tau_{max} \\
\sum_{k=[t/\tau_{max}]+1}^{([t] \tau_{max})} 
{\cal P}_{G}([t],k) & t \in [[t],[t]+1),\: 
t > \tau_{max} 
\end{array}
\right.
\label{ztotcmn1}
\end{equation}  
where the sums are clearly dominated by the terms corresponding to 
$ k^* \langle \tau_{i,j} \rangle \simeq [t]$, {\em i.e.} the same that are 
best approximated by the Gaussian distribution. The corresponding average 
density can be straightaway obtained from Eq.~(\ref{rdaz}).

In the large $t$ limit, $\langle \theta_{tot}(t) \rangle$ approaches quite 
rapidly a nearly constant value, that in the numerically studied case of 
$\tau_{max}=10$ is found to be:
\begin{equation}
\lim_{t \rightarrow \infty} \langle \theta_{tot}(t) \rangle \simeq
\lim_{t \rightarrow \infty} \sum_{k=[[t]/\tau_{max}]+1}^{[t] \tau_{max}} 
{\cal P}_{G}([t],k)\simeq 0.182,
\label{estimation}
\end{equation}
which gives, again from Eq.~(\ref{rdaz}):
\begin{equation}
\lim_{t \rightarrow \infty} \langle \rho(t) \rangle=1-\frac{0.182}{N}.
\label{asym}
\end{equation}

\section*{Appendix~B}
By using the Boolean rule $\overline{a\wedge b}=\overline{a}\vee
\overline{b}$, Eq.~(\ref{freecirc}), defining the
free model on the circulant matrix, turns out to be equivalent to:
\begin{equation}
\overline{x}_i(t)=\sum_{j=1}^n \overline{x}_{i-j}(t-\tau_{i,i-j}),
\end{equation}
which, in the case of equal delays $\{ \tau_{i,j}=\tau_0 \: \forall i,j \}$,
allows the simplification:
\begin{eqnarray}
\overline{x}_i(t)&=&\sum_{j_1=1}^n \sum_{j_2=1}^n
\overline{x}_{i-j_1-j_2}(t-2\tau_0)= \nonumber \\
&=& \sum_{j_1=1}^n \sum_{j_2=1}^n \cdots \sum_{j_k=1}^n
\overline{x}_{i-j_1-j_2-...-j_k}(t-k\tau_0)= \nonumber \\
&=& \sum_{j=k}^{kn} \overline{x}_{i-j}(t-k \tau_0),
\end{eqnarray}
since the different terms containing the same argument are  
redundant. Hence, by choosing $k=[t/\tau_0]=[t]$, one gets:
\begin{equation}
\theta_{tot}(t)=
\sum_{i=1}^N \overline{x}_i(t)=
\sum_{j=[t]}^{[t]n} \overline{x}_{i-j}(t-[t]) =[t](n-1)+1 \mbox{ for }
t\in[[t],[t]+1),
\end{equation}
where we are assuming $[t](n-1)+1\le N$, and we used the conditions on the 
initial state of the system, 
$\{ \overline{x}_{i-j}(t)=\delta_{i-j,1} \mbox{ for }
t\in[0,\tau_c)\}$, by considering for simplicity a duration of the
initial damage $\tau_c=\tau_0$. This analysis confirms that, 
as soon as $n>1$, the system is dissipative, and in particular
the total number of impaired firms (a constant function
in each time step) is linearly increasing with time, with
slope $n-1$. Correspondingly, for a finite size $N$, the
asymptotically stable solution $\{ x_i \equiv 0 \: \forall i\}$, 
and the zero limit value of the density, are attained after the time 
$t_{trans}\simeq(N-1)/(n-1)$, in agreement with Eq.~(\ref{ttransndet}). One 
can moreover analogously show that, for $\tau_c \neq \tau_0$, the density is 
still linearly decreasing with time, and the asymptotic solution
is the same for $\tau_c > \tau_0$, whereas it is periodic of period
$\tau_0$ for $\tau_c < \tau_0$, as described in Section~3.4. 

In order to work out the same approach for randomly distributed delays,
we label $l_h=\sum_{v=1}^h j_v$. One finds:
\begin{eqnarray}
\overline{x}_i(t)&=&
\sum_{j_1=1}^n \sum_{j_2=1}^n \cdots \sum_{j_k=1}^n
\overline{x}_{i-l_k}\left(t-\sum_{h=0}^{k} \tau_{i-l_h,i-l_{h+1}}\right) = 
\nonumber \\
&=&\sum_{j=k}^{kn} \sum_{\{ l_1,l_2,\dots,l_k \}: l_k=j}
\overline{x}_{i-j}\left(t-\sum_{h=0}^{k} \tau_{i-l_h,i-l_{h+1}}\right). 
\end{eqnarray}
Since the $\{ \tau_{i,j} \}$ are independent identically
distributed random variables which take integer values (in unities
of $\tau_{min}$) in the interval $[1,\tau_{max}]$, it follows
that their sums ${\cal T}_k=\sum_{h=0}^{k} \tau_{i-l_h,i-l_{h+1}}$ take
integer values in the interval $[k,k\tau_{max}]$ with probability
${\cal P}({\cal T}_k,k)$; moreover, terms 
$\overline{x}_{i-j}$ with the same argument are redundant.
The average over the disorder can be therefore computed as:
\begin{eqnarray}
\langle \overline{x}_i(t) \rangle&=&
\int \prod d\tau {\cal P}(\tau) \overline{x}_i(t)= \nonumber \\
&=&\sum_{j=k}^{kn} \sum_{{\cal T}_k=k}^{k\tau_{max}} 
{\cal P}({\cal T}_k,k) 
\overline{x}_{i-j}(t-{\cal T}_k).
\end{eqnarray}
Then, one can choose in particular $k$ values in the different terms 
such that $t-{\cal T}_k$ is always in the initial interval; in detail, looking 
for simplicity only at $t$ values which are integer multiple of $\tau_{max}$, 
one has:
\begin{eqnarray}
\langle \overline{x}_i(t) \rangle &=& \left [ \sum_{j=t}^{nt}{\cal P}_t(t,t) 
+\sum_{j=t-1}^{n(t-1)} {\cal P}_t(t,t-1) +...\right. \nonumber \\
... &+& \left. 
\sum_{j=t/\tau_{max}}^{nt/\tau_{max}} {\cal P}_t(t,t/\tau_{max}) \right ] 
\overline{x}_{i-j}(0)= \nonumber \\
&=& \sum_{j=t/\tau_{max}}^t \sum_{k=j}^{nj} {\cal P}_t(t,j) \delta_{i-j,1},
\end{eqnarray}
and therefore, up to the time $t_{trans}$ at which 
$\langle \theta_{tot}(t_{trans}) \rangle \simeq N$, one has:
\begin{equation}
\langle \theta_{tot}(t) \rangle =
\sum_{k=t/\tau_{max}}^{t} [(n-1)k+1] {\cal P}_t(t,k).
\label{bla}
\end{equation}
Here, ${\cal P}_t(t,k)$ is the probability for the signal to 
have propagated $k$ steps along the chain at time $t$, which can
be approximated by a Gaussian with mean value $k \langle \tau_{i,j} \rangle$ 
and variance $k \sigma^2_{i,j}$ (see Eq.~(\ref{Gaussian})), but needs to be
correctly normalized in order to get:
\begin{equation}
\sum_{k=t/\tau_{max}}^t {\cal P}_t(t,k)=1 \hspace{.2in} \forall t.
\label{normalization}
\end{equation}
In detail, we take ${\cal P}_t(t,k) \simeq C(t) {\cal P}_G(t,k)$, where the 
normalization constant $C(t)$ approaches rapidly its large time limit 
$C_{\infty}\simeq 1/0.182$, in agreement with Eq.~(\ref{estimation}). 

In fact, we notice that one would get straightaway a linear behavior 
of $\langle \theta_{tot}(t) \rangle$ from
Eq.~(\ref{bla}), by taking ${\cal P}_t(t,k) \equiv 1$; nevertheless,
the Gaussian approximation and the use of the correct normalization constant 
$C(t)$ are essential in order evaluate accurately enough the slope. These 
results therefore confirm, from a different point of view, the 
analysis of Section~3.5; in particular, for the considered case of 
$n=20$ and $\tau_{max}=10$, the density 
$ \langle \rho(t) \rangle =1-\langle \theta_{tot}(t) \rangle/N$ obtained with 
this approach is in very good agreement with the numerical data, as shown in 
[Fig. 6].

\section*{Appendix~C}
Here we briefly recall the framework of probability generating functions
\cite{NeStWa01,DoMeSa01}: this allows to simply evaluate the
size of connected components in Erd\H{o}s-R\'enyi random graphs, and 
it is moreover well suitable to be generalized to directed networks and to more
complex probability distribution ${\cal P}(k)$ of the node degrees.

One defines:
\begin{equation}
{\cal G}_0(y)=\sum_{k=0}^{\infty} {\cal P}(k) y^k,
\end{equation}
which implies:
\begin{equation}
{\cal P}(k)=\frac{1}{k!} \left . \frac{d^k}{dy^k} {\cal G}_0(y) \right |_{y=0}.
\end{equation}
Analogously, the generating function for the probability distribution
${\cal P}_1(k)$, to have $k$ edges leaving a node apart the one from which
the signal arrived, is given by:
\begin{equation}
{\cal G}_1(y)=\sum_{k=0}^{\infty} {\cal P}_1(k) y^k=
\frac{\sum_{k=0}^{\infty}(k+1){\cal P}(k+1)y^k}
{\sum_{k=0}^{\infty} k {\cal P}(k)}=\frac{{\cal G'}_0(y)}{z},
\end{equation}
and in the Erd\H{o}s Renyi random graph one has:
\begin{equation}
{\cal G}_0(y)={\cal G}_1(y)=e^{z(y-1)}.
\end{equation}

One can show in particular that, if ${\cal G}$ is the generating
function for the probability of some property of an object, then
the probability of the sum of the same property on $l$ independent objects 
is generated by ${\cal G}^l$. Hence, if one defines ${\cal H}_0(x)$ 
(respectively ${\cal H}_1(x)$) as the generating function for the distribution 
probability of the sizes of the connected components (respectively of the
clusters), {\em i.e.} of the nodes which can be reached from a randomly
chosen one (respectively from the end of one 
of the edges of a randomly chosen one), it is:
\begin{eqnarray}
{\cal H}_1(y)&=&y \sum_{k=0}^{\infty} {\cal P}_1(k) 
\left [ {\cal H}_1(y) \right ]^k=y{\cal G}_1[{\cal H}_1(y)] \nonumber \\
{\cal H}_0(y)&=&\sum_{k=0}^{\infty}
{\cal P}(k)\left [ {\cal H}_1(y) \right ]^k= 
y{\cal G}_0[{\cal H}_1(y)].
\end{eqnarray}
Therefore, one obtains a set of equations which can be in principle
solved self-consistently. Most importantly from our present point of view, 
one gets immediately the mean connected component size:
\begin{itemize}
\item below the transition one can show that
\begin{equation}
\langle s \rangle ={\cal H'}_0(1)=1+{\cal G'}_0(1) {\cal H'}_1(1)=
1+\frac{{\cal G'}_0(1)}{1-{\cal G'}_1(1)},
\end{equation}
which means, in the Erd\H{o}s-R\'enyi random graph,
\begin{equation}
\langle s \rangle = \frac{1}{1-z},
\end{equation}
as expected.
\item above the transition, ${\cal H}_0(x)$ and ${\cal H}_1(x)$ can be
defined as the generating function for the distribution probability
of the finite size connected components, which have still a tree-like
structure; correspondingly $H_0(1)=v<1$, and one has
\begin{equation}
\left \{
\begin{array}{lcl}
s_{gc}&=&1-v=1-{\cal G}_0(v) \\
v&=&{\cal G}_{1}(v)\\
\end{array}
\right.,
\end{equation}
which for Erd\H{o}s-R\'enyi random graphs gives $v=e^{z(v-1)}$, 
{\em i.e.} Eq.~(\ref{my_v}).
\end{itemize}

\end{document}